\documentclass[10pt]{article}
\usepackage{ucs} 
\usepackage[utf8x]{inputenc}
\usepackage{amsmath,amsfonts,amssymb,amsthm,amsopn,amscd}
\usepackage{bbm}
\usepackage{graphicx}
\usepackage[sort&compress]{natbib} 
\usepackage{epsfig}
\usepackage[dvips]{color}
\usepackage{subfigure}

\theoremstyle{plain}
\newtheorem{theorem}{Theorem}[section]

\theoremstyle{definition}
\newtheorem{lemma}[theorem]{Lemma}

\newtheorem{proposition}[theorem]{Proposition}

\newtheorem{definition}{Definition}[section]
\newtheorem{Assump}{Assumption}[section]
\newtheorem*{remark}{Remark}

\theoremstyle{remark}

\newcommand{\bd}{\begin{document}}
\newcommand{\ed}{\end{document}}
\newcommand{\beq}{\begin{equation}}
\newcommand{\eeq}{\end{equation}}
\newcommand{\nid}{\noindent}
\newcommand{\ben}{\begin{enumerate}}
\newcommand{\een}{\end{enumerate}}
\newcommand{\bit}{\begin{itemize}}
\newcommand{\eit}{\end{itemize}}
\newcommand{\baR}{\begin{array}}
\newcommand{\eaR}{\end{array}}
\newcommand{\su}{\section}\usepackage{graphicx}
\newcommand{\ssu}{\subsection}
\newcommand{\sssu}{\subsubsection}
\newcommand{\deq}{\stackrel {\rm def}{=}}
\newcommand{\bea}{\begin{eqnarray}}
\newcommand{\eea}{\end{eqnarray}}
\newcommand{\mZ}{\mathbf{Z}}
\newcommand{\mg}{\mathbf{g}}
\newcommand{\mm}{\mathbf{m}}

\newcommand {\VV} {\mathbf{V}}
\newcommand {\V} {\mathcal{V}}
\newcommand {\G} {\mathcal{G}}
\newcommand{\R}{\mathbbm{R}}
\newcommand{\N}{\mathbbm{N}}
\newcommand{\Lop}{\mathcal{L}}
\newcommand{\F}{\mathcal{F}}
\newcommand{\Pro}{\mathbbm{P}}
\newcommand{\f}{\varphi}
\newcommand{\Diag}{{\mathbf{\rm Diag}}}
\newcommand{\ty}{\tilde{y}}
\newcommand{\Jb}{\bar{J}}
\newcommand{\Jbab}{\bar{J}_{\alpha\beta}}
\newcommand{\Jbcd}{\bar{J}_{\gamma\delta}}
\newcommand{\Jab}{\sigma_{\alpha\beta}}
\newcommand{\Jdab}{\sigma^2_{\alpha\beta}}
\newcommand{\ma}{m_\alpha}
\newcommand{\Ma}{M_\alpha}
\newcommand{\mua}{\mu_\alpha}
\newcommand{\muax}{\mua^X}
\newcommand{\muay}{\mua^Y}
\newcommand{\va}{v_\alpha}
\newcommand{\vax}{v_\alpha^X}
\newcommand{\Ca}{C_{\alpha\alpha}}
\newcommand{\Cax}{C_{\alpha\alpha}^X}
\newcommand{\Cbx}{C_{\beta\beta}^X}
\newcommand{\Cay}{C_{\alpha\alpha}^Y}
\newcommand{\Ia}{I_\alpha}
\newcommand{\Da}{\Delta_\alpha}
\newcommand{\Deab}{\Delta_{\alpha\beta}}
\newcommand{\Deabx}{\Delta_{\alpha\beta}^X}
\newcommand{\Debx}{\Delta_{\beta}^X}
\newcommand{\sa}{s_\alpha}
\newcommand{\ta}{\tau_\alpha}
\newcommand{\mub}{\mu_\beta}
\newcommand{\mb}{m_{\alpha\beta}}
\newcommand{\vb}{v_\beta}
\newcommand{\vbx}{v_\beta^X}
\newcommand{\Cb}{C_{\beta\beta}}
\newcommand{\Db}{\Delta_\beta}
\newcommand{\Mb}{M_\beta}
\newcommand{\qb}{q_\beta}
\newcommand{\gb}{\gamma_\beta(\mub,\vb;\Cb)}
\newcommand{\gbt}{\gamma_\beta(\mub,\vb;\Cb(t))}
\newcommand{\tb}{\tau_\beta}
\newcommand{\Sb}{S_\beta}
\newcommand{\Ha}{H_{\alpha}}
\newcommand{\Hax}{H_{\alpha}^X}
\newcommand{\Hab}{H_{\alpha\beta}}
\newcommand{\Habx}{H_{\alpha\beta}^X}
\newcommand{\Dab}{D_{\alpha\beta}}
\newcommand{\Dabx}{D_{\alpha\beta}^X}
\newcommand{\cJ}{{\cal J}}

\newcommand\bbbr{{\sf I\!R}}

\newcommand{\ind}[1]{\mathbbm{1}_{#1}}          

\newcommand{\derpart}[2]{ \frac{\partial #1}{\partial #2} } 
\newcommand{\der}[2]{ \frac{\text{d} #1}{\text{d} #2} }  
\newcommand{\derparts}[2]{ \frac{\partial^2 #1}{\partial #2 ^2}}
\newcommand{\derpartsxy}[3]{ \frac{\partial^2 #1}{\partial #2 \partial #3}} 
\newcommand{\argmin}{\mathop{\mathrm{Arg\,Min}}}
\newcommand{\eps}{\varepsilon}

\DeclareMathOperator{\eqlaw}{\stackrel{\mathcal{L}}{=}}         

\newcommand{\Herm}[1]{\mathcal{H}_{#1}}            
\newcommand{\Hermnu}{\mathcal{H}_{\nu}}            
\newcommand{\erf}{\mathrm{erf}}              
\newcommand{\Cov}{\mathrm{Cov}}

\newcommand{\m}{\mathcal}
\newcommand{\Ie}{I_{\text{ext}}}

\newcommand {\image} {\includegraphics}

\newcommand{\mC}{\mathbf{C}}
\newcommand{\mD}{\mathbf{D}}
\newcommand{\mE}{\mathbf{E}}
\newcommand{\mX}{\mathbf{X}}
\newcommand{\mx}{\mathbf{x}}
\newcommand{\mY}{\mathbf{Y}}
\newcommand{\mL}{\mathbf{L}}
\newcommand{\mQ}{\mathbf{Q}}
\newcommand{\mI}{\mathbf{I}}
\newcommand{\mJ}{\mathbf{J}}
\newcommand{\mW}{\mathbf{W}}
\newcommand{\mU}{\mathbf{U}}
\newcommand{\mone}{\mathbf{1}}

\newcommand{\msig}{\boldsymbol{\sigma}}
\newcommand{\mtheta}{\boldsymbol{\theta}}
\newcommand{\mdelta}{\boldsymbol{\Delta}}
\newcommand{\mla}{\boldsymbol{F}}
\newcommand{\mga}{\boldsymbol{\Gamma}}
\newcommand{\mxi}{\boldsymbol{\xi}}

\newcommand{\X}{\mathcal{X}}
\newcommand{\proc}{\mathcal{M}_1^+(C([t_0,T], \R)}
\newcommand{\procP}[1]{\mathcal{M}_1^+(C([t_0,T], \R^{#1}))}
\newcommand{\proci}{\mathcal{M}_1^+(C((-\infty,T], \R)}
\newcommand{\prociP}[1]{\mathcal{M}_1^+(C((-\infty,T], \R^{#1})}
\newcommand{\procs}[1]{\mathcal{M}_1^+(C([0,#1], \R)}
\newcommand{\Exp}[1]{\mathbb{E}\left [ #1 \right]}
\newcommand{\ExpSimple}{\mathbb{E}}
\newcommand{\nvar}[1]{\left \| #1 \right\|_{\mathrm{var}}}
\newcommand{\nvarP}[2]{\left \| #1 \right\|_{\mathrm{var}, #2}}
\newcommand{\abs}[1]{\left \vert #1 \right \vert }
\newcommand{\pente}{\|S_{\beta}'\|_{\infty}}
\newcommand{\Fs}{\mathcal{F}_{\text{stat}}}
\newcommand{\triplebar}[1]{\vert\vert\vert #1 \vert\vert\vert}
\newcommand{\procsP}[2]{\mathcal{M}_1^+(C([t_0,#1], \R^{#2})}
\newcommand{\dvar}[2]{d_{\rm{var}}\left(#1,\; #2\right)}
\newcommand{\dt}[2]{d_t\left(#1,\; #2\right)}
\newcommand{\norm}[1]{\left \| #1 \right \|}
\newcommand{\Si}{S}
\newcommand{\T}{\mathcal{T}}
\newcommand{\IT}{\mathcal{IT}}
\newcommand{\tV}{\mathbf{\widetilde{V}}}

\newcommand{\mA}{\mathbf{B}}

\newenvironment{heuriproof}{\noindent \emph{Heuristic proof.}}{\begin{flushright} \qed \end{flushright}}
\newcommand{\eqdef}{\overset{\rm def}{=}}

\usepackage{ifthen}
\newboolean{DisplaySOS}
\setboolean{DisplaySOS}{true} 
\newcommand{\SOS}[1]{\ifthenelse{\boolean{DisplaySOS}}{{\bf[#1]}}{}}

\begin{document}

\title{A constructive mean-field analysis of multi-population neural networks with random synaptic weights and stochastic inputs}

\author{Olivier Faugeras \thanks{Correspondence: Odyssee Laboratory, INRIA/ENS/ENPC, 2004 Route des Lucioles, 06902 Sophia-Antipolis, France, email: olivier.faugeras@sophia.inria.fr} \thanks{Odyss\'ee Laboratory, INRIA/ENS/ENPC, France} \and Jonathan Touboul \footnotemark[2] \and Bruno Cessac \footnotemark[2] \thanks{Laboratoire Jean-Alexandre Dieudonn\'e, France} \thanks{Universit\'e de Nice, France}}

\date{\today}

\maketitle

\section*{Abstract}
We deal with the problem of bridging the gap between two scales in neuronal modeling. At the first (microscopic) scale, neurons are considered individually and their behavior described by
stochastic differential equations that govern the time variations of their membrane potentials. They are coupled by synaptic connections acting on their resulting activity, 
a nonlinear function of their membrane potential. At the second (mesoscopic) scale, interacting populations of neurons are described individually by similar equations. 
The equations describing the dynamical and the stationary mean field behaviors are considered as functional equations on a set of stochastic processes. Using this new point of view allows us to prove that these equations are well-posed on any finite time interval and to provide a constructive method for effectively computing their unique solution. This method is proved to converge to the unique solution and we characterize its complexity and convergence rate. We also provide partial results for the stationary problem on infinite time intervals.
These results shed some new light on such neural mass models as the one of Jansen and Rit \cite{jansen-rit:95}: 
their dynamics appears as a coarse approximation of  the much richer dynamics that emerges from our analysis.
Our  numerical experiments confirm that the framework we propose and the numerical methods we derive from it  provide a new and powerful tool for the exploration of neural behaviors at different scales.

{\noindent \bf Keywords:} mean-field analysis, stochastic processes, stochastic differential equations, stochastic networks, stochastic functional equations, random connectivities, multi populations networks, neural mass models.

\section{Introduction}
Modeling neural activity at scales integrating the effect
of thousands of neurons is of central importance for several reasons. First, most
imaging techniques are not able to measure individual neuron activity (``microscopic'' scale),
but are instead measuring mesoscopic effects 
resulting from the activity of several hundreds to several hundreds of thousands of
neurons. Second, anatomical data recorded in the cortex reveal the existence
of structures, such as the cortical columns, with a diameter of about
$50 \mu m$ to $1 mm$, containing of the order of one hundred to one hundred thousand neurons
belonging to a few different species. These columns
have specific functions. For example, in the visual cortex V1, they respond to preferential
orientations of bar-shaped visual stimuli. In this case,
information processing does not occur at the  scale of individual neurons
but rather corresponds to an  activity integrating the collective dynamics
of many interacting neurons and resulting in a mesoscopic signal. The description of this collective dynamics
requires models which are different from individual neurons models.
In particular, if the accurate description of one neuron requires
``$m$'' parameters (such as sodium, potassium, calcium conductances,
membrane capacitance, etc...), it is not necessarily true that an accurate mesoscopic
description of an assembly of $N$ neurons requires $Nm$ parameters.
Indeed, when $N$ is large enough averaging
effects  appear, and the collective dynamics is well described by an effective
mean-field, summarizing the effect of the interactions of a neuron with the
other neurons, and  depending on a few effective control parameters.
This vision, inherited from statistical physics requires
that the space scale be large enough to include a large number of microscopic
components (here neurons) and small enough so that the region considered
is homogeneous. This is in effect for instance the case of cortical columns.

However, obtaining the evolution equations of the effective mean-field
 from microscopic dynamics is far from being evident. In simple physical models this can
be achieved via the law of large numbers and the central limit theorem, provided
that time correlations decrease sufficiently fast.
This type of approach has been generalized to such fields
 as quantum field theory or non equilibrium 
statistical mechanics. To the best of our knowledge,
the idea of applying mean-field methods to neural networks
dates back to Amari \cite{amari:72,amari-yoshida-etal:77}. In his approach,
the author uses an assumption that he called the ``local chaos hypothesis'',
reminiscent of Boltzmann's ``molecular chaos hypothesis'',
that postulates the vanishing of individual
correlations between neurons, when the number
$N$ of neurons  tends to infinity.
Later on, Crisanti, Sompolinsky and coworkers \cite{sompolinsky-crisanti-etal:88} used a dynamic mean-field
approach to conjecture the existence of chaos in an homogeneous neural network
with random independent synaptic weights. This approach was
formerly developed by Sompolinsky and coworkers for spin-glasses 
\cite{sompolinsky-zippelius:82,crisanti-sompolinsky:87,crisanti-sompolinsky:87b}, where complex effects
such as aging or coexistence of a diverging number of metastable states,
renders the mean-field analysis delicate in the long time limit \cite{houghton-jain-etal:83}.

On the opposite, these effects do not appear in the neural network considered
in \cite{sompolinsky-crisanti-etal:88} because the synaptic weights are independent \cite{cessac:95} (and especially
non  symmetric, in opposition to spin glasses). In this case, the Amari approach
and the dynamic mean-field approach lead to the same mean-field equations.
Later on, the mean-field equations derived by Sompolinsky and Zippelius \cite{sompolinsky-zippelius:82} for spin-glasses were rigorously obtained by Ben Arous and Guionnet \cite{ben-arous-guionnet:95,ben-arous-guionnet:97,guionnet:97}. The application of their method to a discrete time version of the neural network considered in  \cite{sompolinsky-crisanti-etal:88}
and in \cite{molgedey-schuchardt-etal:92} was done by Moynot and Samuelides \cite{moynot-samuelides:02}.

Mean-field methods are often used in the neural network community but there
are only a few  rigorous results using the dynamic mean-field method. The main advantage of dynamic mean-field techniques
is that they allow one to consider neural networks where synaptic weights
are random (and independent). The mean-field approach allows one to state general and generic results about the dynamics as a function of the statistical parameters controlling the probability distribution of the synaptic
weights \cite{samuelides-cessac:07}.  It does not only provide
the evolution of the mean activity of the network but, because it is an equation on the law of the mean-field, it also provides informations on the fluctuations around the mean and their correlations.
These correlations are of crucial importance as revealed in the paper by Sompolinsky
and coworkers \cite{sompolinsky-crisanti-etal:88}. Indeed, in their work, the analysis of correlations 
allows them to discriminate between two distinct regimes:
a dynamics with a stable fixed point and a chaotic dynamics,
while the mean is identically zero in the two regimes.

However, this approach has also several drawbacks explaining why it is 
so seldom used. First, this method  uses a generating function approach
that requires heavy computations and some ``art'' for obtaining
the mean-field equations. Second, it  is hard to  generalize to models including  several populations. 
Finally,  dynamic mean-field equations are usually supposed to
characterize \textit{in fine} a  stationary process. 
It is then natural to search for stationary solutions. This considerably simplifies the 
dynamic mean-field equations by reducing them to a set of differential equations (see section \ref{sect:numerics}) 
but the price to pay is the unavoidable occurrence in the equations of a non free parameter, the initial condition, 
that can only be characterized through the investigation of the non stationary case. 

Hence  it is not clear whether such a stationary solution exists, and, if it is the case, how to characterize it. To the best of our knowledge, this difficult question has only been investigated for neural networks in one  paper  by Crisanti and coworkers \cite{crisanti-sommers-etal:90}. 

Different alternative approaches have been used to get a mean-field description of a given neural network and to find its solutions. In the neuroscience community, a static mean-field study of multi population network activity was developed by Treves in \cite{treves:93}.  This author did not consider external inputs but incorporated dynamical synaptic currents and adaptation effects. His analysis was completed in \cite{abbott-van-vreeswijk:93}, where the authors considered a unique population of nonlinear oscillators subject to a noisy input current. They proved, using a stationary Fokker-Planck formalism, the stability of an asynchronous state in the network. Later on, Gerstner in \cite{gerstner:95} built a new approach to characterize the mean-field dynamics for the Spike Response Model, via the introduction of suitable kernels propagating the collective activity of a neural population in time.

Brunel and Hakim considered a network composed of integrate-and-fire neurons connected with constant synaptic weights \cite{brunel-hakim:99}. In the case of sparse connectivity, stationarity, and considering a regime where individual neurons emit spikes at low rate, they were able to study analytically the dynamics of the network and to show that the network exhibited a sharp transition between a stationary regime and a regime of fast collective oscillations weakly synchronized. Their approach was based on a perturbative analysis of the Fokker-Planck equation. A similar formalism was used in \cite{mattia-del-giudice:02} which, when complemented with self-consistency equations, resulted in the dynamical description of the mean-field equations of the network, and was extended to a multi population network.

Finally, Chizhov and Graham \cite{chizhov-graham:07} have recently proposed
a new method based on a population density approach allowing to characterize the mesoscopic
behaviour of neuron populations in conductance-based models. We shortly discuss their approach and
compare it to ours in the discussion section \ref{section:discussion}.\\
 
In the present paper, we investigate the problem of deriving
the equations of evolution of neural masses at mesoscopic scales from neurons dynamics,
using a new and rigorous approach based on stochastic analysis.

The article is organized as follows. In section \ref{sect:equations} we derive from first principles the equations relating the membrane potential of each of a set of neurons as function of the external injected current and noise and of the shapes and intensities of the postsynaptic potentials in the case where these shapes depend only on the post-synaptic neuron (the so-called voltage-based model). Assuming that the shapes of the postsynaptic potentials can be described by linear (possibly time-dependent) differential equations we express the dynamics of the neurons as a set of stochastic differential equations. Assuming that the synaptic connectivities between neurons satisfy statistical relationship only depending on the population they belong to, we obtain the mean-field equations summarizing the interactions of the $P$ populations in the limit where the number of neurons tend to infinity. These equations can be derived in several ways, either heuristically following the lines of  Amari \cite{amari:72,amari-yoshida-etal:77}, Sompolinsky \cite{sompolinsky-crisanti-etal:88,crisanti-sommers-etal:90}, and Cessac \cite{cessac:95,samuelides-cessac:07}, or rigorously as in the work of Benarous and Guionnet  \cite{ben-arous-guionnet:95,ben-arous-guionnet:97,guionnet:97}. The purpose of this article is not the derivation of these mean-field equations but to prove that they are well-posed and to provide an algorithm for computing their solution. Before we do this we provide the reader with two important examples of such mean-field equations. The first example is what we call the simple model, a straightforward generalization of the case studied by Amari and Sompolinsky. The second example is a neuronal assembly model, or neural mass model, as introduced by Freeman \cite{freeman:75} and examplified in Jansen and Rit's cortical column model \cite{jansen-rit:95}.

In section \ref{sect:existUniq} we consider the problem of solutions over a finite time interval $[t_0,T]$. We prove, under some mild assumptions, the existence and uniqueness of a solution of the dynamic mean-field equations given an initial condition at time $t_0$. The proof consists in showing that a nonlinear equation defined on the set of multidimensional Gaussian random processes defined on $[t_0,T]$ has a fixed point.
We extend this proof in section \ref{sect:station} to the case of stationary solutions over the time interval $[-\infty,T]$ for the simple model. 
Both proofs are constructive and provide an algorithm for computing numerically the solutions of the mean-field equations. 

We then study in section \ref{sect:numerics} the complexity and the convergence rate of this algorithm and 
put it to good use:  We first compare our numerical results to the theoretical results of Sompolinsky and coworkers \cite{sompolinsky-crisanti-etal:88,crisanti-sommers-etal:90}. We then provide an example of numerical experiments in the case of two populations of neurons where the role of the mean-field fluctuations is emphasized. 

Along the paper we introduce several constants. To help the reader we have collected in table \ref{table:summ} the most important ones and the place where they are defined in the text.

\section[Mean-Field Equations]{Mean-field equations for multi-populations neural network models}\label{sect:equations}
In this section we introduce the classical neural mass models and compute the related mean-field equations they satisfy in the limit of an infinite
number of neurons.

\subsection{The general model}

\subsubsection{General framework}\label{subsubsection:general}
\begin{figure}
 \begin{center}
  \includegraphics[width=.5\textwidth]{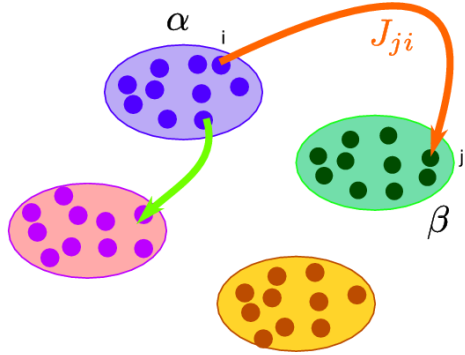}
 \end{center}
 \caption{General network considered: $N$ neurons belonging to $P$ populations are interconnected with random synaptic weights whose probability distributions only depend upon the population indexes, see text.}
 \label{fig:MFNet}
\end{figure}

We consider a network composed of $N$ neurons indexed by $i \in \{1,\,\ldots,\,N\}$ belonging to $P$ populations indexed by $\alpha \in \{1,\,\ldots,\, P\}$
such as those shown in figure \ref{fig:MFNet}.  Let $N_{\alpha}$ be the number of neurons in population $\alpha$. We have $N=\sum_{\alpha=1}^P N_{\alpha}$. We define the population  which the neuron $i$, $i=1,\cdots,N$ belongs to.

\begin{definition}\label{def:population}
The function $p: \{1,\,\ldots,\,N\} \to \{1,\,\ldots,\, P\}$ associates to each neuron $i \in \{1,\cdots,N\}$, the population $\alpha=p(i) \in \{1,\cdots,P\}$, it belongs to.
\end{definition}
We consider that each neuron $i$ is described by its membrane potential $V_i(t)$, and the related instantaneous firing rate is deduced from it through a relation 
of the form $\nu_i(t)=S_i(V_i(t))$ \cite{gerstner-kistler:02,dayan-abbott:01}, where $S_i$ is a sigmoidal function.

\noindent
A single action potential from neuron  $j$ generates a
post-synaptic potential $PSP_{ij}(u)$ on the postsynaptic neuron  $i$,
where $u$ is the time elapsed after the spike is received. We neglect the delays due to the distance
travelled down the axon by the spikes. 

Assuming that the post-synaptic potentials sum linearly, the average membrane potential
of neuron $i$ is\quad
\[
V_i(t)=\sum_{j,k,t_k > t_0} PSP_{ij}(t-t_k)+V_i(t_0),
\]
where the sum is taken over the arrival times of the spikes produced
by the neurons $j$ after some reference time $t_0$. The number of spikes arriving
between $t$ and $t+dt$ is $\nu_j(t)dt$. Therefore we have
\begin{equation}\label{eq:vi}
V_i(t)=\sum_j \int_{t_0}^t PSP_{ij}(t-s) \nu_j(s)\,ds+V_i(t_0)=
\sum_j \int_{t_0}^t PSP_{ij}(t-s) S_j(V_j(s))\,ds+V_i(t_0),
\end{equation}
or, equivalently
\begin{equation}\label{eq:ui}
\nu_i(t)=S_i\left(\sum_j \int_{t_0}^t PSP_{ij}(t-s) \nu_j(s)\,ds +V_i(t_0)\right).
\end{equation}
The $PSP_{ij}$s can depend on several variables in order to account for instance for 
adaptation or learning.\\

We now make the simplifying assumption that the shape of the postsynaptic potential $PSP_{ij}$ only depends on the postsynaptic population, which
corresponds to the voltage based models in Ermentrout’s classification \cite{ermentrout:98}.\\

\paragraph{The voltage-based model}\ \\
The
assumption, made in  \cite{hopfield:84}, is that the post-synaptic potential
has the same shape no matter which presynaptic population caused it,
the sign and amplitude may vary though. This leads to the relation
\[
PSP_{ij}(t)=\bar{J}_{ij} g_i(t).
\]
$g_{i}$ represents the unweighted shape (called a g-shape) of the postsynaptic potentials and $\bar{J}_{ij}$ is the strength of the postsynaptic potentials elicited by neuron  $j$ on neuron $i$. At this stage of the discussion, these weights are supposed to be deterministic. This is reflected in the notation $\bar{J}_{ij}$ which indicates an average value. \footnote{When we come to the mean-field equations they will be modeled as random variables.} From equation \eqref{eq:vi} we have
\[
 V_i(t)=\int_{t_0}^t g_i(t-s) \left(\sum_j \bar{J}_{ij} \nu_j(s)\right)\,ds+V_i(t_0).
\]
So far we have only considered the synaptic inputs to the neurons. We enrich our model by assuming that the neuron $i$ receives also an external current density composed of a deterministic part, noted $I_i(t)$, and a stochastic part, noted $n_i(t)$, so that\quad
\begin{equation}\label{eq:V}
V_i(t)=\int_{t_0}^t g_i(t-s) \left(\sum_j \bar{J}_{ij} \nu_j(s)+I_i(s)+n_i(s)\right)\,ds+V_i(t_0).
\end{equation}
We assume, and this is essential for deriving the mean-field equations below, that all indexed quantities depend only upon the $P$ populations of neurons (see definition \ref{def:population}), i.e.
\begin{multline}\label{eq:population}
 g_i(t) \eqdef g_{p(i)}(t) \quad \bar{J}_{ij} \eqdef \bar{J}_{p(i)p(j)} \quad I_i(t) \eqdef I_{p(i)}(t) \\ n_i(t) \overset{\rm def}{\sim} n_{p(i)}(t) \quad S_j(\cdot)=S_{p(j)}(\cdot),
\end{multline}
where $x \sim y$ indicates that the two random variables $x$ and $y$ have the same probability distribution.
In other words, all neurons in the same population are described by {\em identical} equations (in law).

The g-shapes describe the shape of the postsynaptic potentials and can reasonably well be approximated by smooth functions.

In detail we
assume that $g_{\alpha}$, $\alpha=1,\cdots,P$ is the Green function of a linear differential equation of order $k$, i.e. satisfies
\begin{equation}\label{eq:PSP}
\sum_{l=0}^k b_{l\alpha}(t) \frac{d^l g_\alpha}{d t^l}(t)=\delta(t),
\end{equation}
where $\delta(t)$ is the Dirac delta function.

The functions $b_{l \alpha}(t)$, $l=0,\cdots,k$, $\alpha=1,\cdots,P$, are assumed to be continuous. We also assume for simplicity that 
\begin{equation}\label{eq:calpha}
b_{k\alpha}(t)\equiv c_\alpha \neq 0,
\end{equation} 
for all $t\in \R$, $\alpha=1,\cdots,P$. We note $D_\alpha^k$ the corresponding differential operator: 
\begin{equation}\label{eq:Dk}
 D_\alpha^k g_\alpha(t) \eqdef \sum_{l=0}^k b_{l\alpha}(t) \frac{d^l g_\alpha}{d t^l}(t)=\delta(t)
\end{equation}
Applying $D_\alpha^k$ to both sides of \eqref{eq:V}, using \eqref{eq:Dk} and the fact that $\nu_j(s)=S_j(V_j(s))$, we obtain a $k$th order differential equation for $V_i$
\begin{equation}\label{eq:DV}
 D_i^k V_i(t)=\sum_{j=1}^N \bar{J}_{ij} S_j(V_j(t))+I_i(t)+n_i(t).
\end{equation}
With a slight abuse of notation, we split the sum with respect to $j$ into $P$ sums:
\[
 D_i^k V_i(t)=\sum_{\beta=1}^P \sum_{j=1}^{N_\beta}\bar{J}_{ij} S_j(V_j(t))+I_i(t)+n_i(t)
\]
We classically turn the $k$th-order differential equation \eqref{eq:DV} into a $k$-dimensional system of coupled 1st-order differential equations (we divided both sides of the last equation by $c_i$, see equation \eqref{eq:calpha}):
\begin{equation}\label{eq:odeVi}
\begin{array}{lcl}
 dV_{li}(t)&=&V_{l+1\,i}(t) dt \quad l=0,\cdots,k-2\\
 dV_{k-1\,i}(t)&=&\left(-\sum_{l=0}^{k-1} b_{lp(i)}(t) V_{li}(t)+\sum_j \bar{J}_{ij}S_{p(j)}(V_j(t))+I_{p(i)}(t)+n_i(t)\right)dt
\end{array}
\end{equation}

A well-known example  of g-shapes, see section \ref{ssect:JansenRit} below or \cite{gerstner-kistler:02}, is
\begin{equation}\label{eq:g1}
 g(t)=Ke^{-t/\tau} Y(t),
\end{equation}
where $Y(t)$ is the Heaviside function. This is an exponentially decaying postsynaptic potential corresponding to 
\[
 k=1 \quad b_1(t)= \frac{1}{K} \quad \text{and} \quad  b_0(t) = \frac 1 {K\, \tau}
\]
in equation \eqref{eq:PSP}.

Another well-known example is 
\begin{equation}\label{eq:g2}
 g(t)=Kte^{-t/\tau} Y(t).
\end{equation}
This is a somewhat smoother function corresponding to
\[
 k=2 \quad b_2(t) = \frac{1}{K}  \quad b_1(t) = \frac{2}{\tau}  \quad b_0(t) = \frac 1 {\tau^2}
\]
in equation \eqref{eq:PSP}.

\paragraph{The dynamics}\ \\
We modify the equations \eqref{eq:odeVi} by perturbing the first $k-1$ equations with Brownian noise and assuming that $n_i(t)$ is white noise. This has the effect that the quantities that appear in equations \eqref{eq:odeVi} are {\em not} anymore the derivatives up to order $k-1$ of $V_i$. This becomes true again only in the limit where the added Brownian noise is null.
This may seem artificial at first glance but 1) it is a technical assumption that is necessary in the proofs of the well-posedness of the mean-field equations, see assumptions \ref{assumption:all} below, and 2) it generates a rich class of external stochastic input, as shown below. With this in mind, the equations \eqref{eq:odeVi} now read
\begin{equation}\label{eq:sdeVi}
\begin{array}{lcl}
 dV_{li}(t)&=&V_{l+1\,i}(t) dt+f_{li}(t) dW_{li}(t) \quad l=0,\cdots,k-2\\
 dV_{k-1\,i}(t)&=&\left(-\sum_{l=0}^{k-1} b_{lp(i)}(t) V_{li}(t)+\sum_j \bar{J}_{ij}S_{p(j)}(V_j(t))+I_{p(i)}(t)\right)dt+\\
		& & \quad \quad f_{k-1\,i}(t) dW_{k-1\,i}(t)
\end{array}
\end{equation}
$W_{li}(t)$, $l=0,\cdots,k-1$, $i=1,\cdots,N$,  are $kN$ independent standard Brownian processes. 
Because we want the neurons in the same class to be essentially identical we also assume that the functions $f_{li}(t)$ that control the amount of noise on each derivative satisfy
\[
 f_{li}(t)=f_{lp(i)}(t),\quad l=0,\cdots,k-1,\quad i=1,\cdots,N
\]

Note that in the limit $f_{l\alpha}(t)=0$ for $l=0,\cdots,k-1$ and $\alpha=1,\cdots,P$, the components $V_{li}(t)$ of the vector $\tV_i(t)$ are the derivatives of the membrane potential $V_i$, for $l=0,\cdots,k-1$ and the equations \eqref{eq:sdeVi} turn into equations \eqref{eq:odeVi}. The system of differential equations \eqref{eq:sdeVi} implies that the class of admissible external stochastic input $n_i(t)$ to the neuron $i$ are Brownian noise integrated through the filter of the synapse, i.e. involving the $l$th primitives of the Brownian motion for $l\leq k$. 

We now introduce the $k-1$ $N$-dimensional vectors $\VV_l(t)=[V_{l1},\cdots,V_{lN}]^T$, $l=1,\cdots,k-1$ of the $l$th-order derivative (in the limit of $f_{lp(i)}(t)=0$) of $\VV(t)$, and concatenate them with $\VV(t)$ into the $Nk$-dimensional vector 
\begin{equation}\label{eq:Vtilde}
 \widetilde{\VV}(t)=\left[ \begin{array}{c}
                            \VV(t)\\
         \VV_1(t)\\
            \vdots\\
         \VV_{k-1}(t)
                           \end{array}
\right].
\end{equation}
The $N$-neurons network is described by the $Nk$-dimensional vector $\widetilde{\VV}(t)$. By definition the $l$th $N$-dimensional component$\tV_l$ of $\tV$ is equal to $\VV_l$. In the limit $f_{l\alpha}(t)=0$ we have
\[
 \widetilde{\VV}_l=\VV_l=\frac{d^l \VV}{dt^l} \quad l=0,\cdots,k-1, \quad \text{with} \quad \widetilde{\VV}_0=\VV
\]

We next write the equations governing the time variation of the  $k$ $N$-dimensional sub-vectors of $\widetilde{\VV}(t)$, i.e. the derivatives of order 0, \ldots $k-1$ of $\VV(t)$. These are vector versions of equations \eqref{eq:sdeVi}. We write
\begin{equation}\label{eq:MicroGMMd}
d\tV_l(t)=\tV_{l+1}(t)\,dt+ \mla_l(t) \cdot d\mW_l(t) \quad l=0,\cdots,k-2.
\end{equation}
$\mla_l(t)$ is the $N \times N$ diagonal matrix ${\rm diag}(\underset{N_1}{\underbrace{f_{l1}(t), \cdots, f_{l1}(t)}},\cdots,\underset{N_P}{\underbrace{f_{lP}(t), \cdots, f_{lP}(t)}})$, where $f_{l\alpha}(t)$, $\alpha=1,\cdots,P$ is repeated $N_\alpha$ times, and the $\mW_l(t)$, $l=0,\cdots,k-2$, are $k-1$ $N$-dimensional independent standard Brownian processes. 

The equation governing the $(k-1)$th differential of the membrane potential has a linear part determined by the differential operators $D_\alpha^{k}$, $\alpha=1,\cdots,P$ and accounts for the external inputs (deterministic and stochastic) and the activity of the neighbors. We note $\boldsymbol{\mathcal L}(t)$ the $N \times Nk$ matrix describing the relation between the neurons membrane potentials and their derivatives up to the order $k-1$ and the $(k-1)$th derivative of $\VV$. This matrix  is defined as the concatenation of the $k$ $N \times N$ diagonal matrixes $\mA_l(t)={\rm diag}(\underset{N_1}{\underbrace{b_{l1}(t),\cdots,b_{l1}(t)}},\cdots,\underset{N_P}{\underbrace{b_{lP}(t),\cdots,b_{lP}(t)}})$ for $l=0,\cdots,k-1$:
\[
\boldsymbol{\mathcal L}(t)=[\mA_0(t),\cdots,\mA_{k-1}(t)]
\]
We have:
\begin{multline}\label{eq:MicroGMM}
 d\tV_{k-1}(t) = \Big ( -\boldsymbol{\mathcal L}(t) \cdot \tV(t) +  \big(\mathbf{\bar{J}} \cdot S(\tV_0(t))\big) + \mathbf{I}(t)\Big) \, dt  \\
 + \mla_{k-1}(t) \cdot d\mathbf{W}_{k-1}(t),
\end{multline}
where $\mathbf{W}_{k-1}(t)$ is an $N$-dimensional standard Brownian process independent of $\mW_l(t)$, $l=0,\cdots,k-2$. The coordinates of the $N$-dimensional vector $\mI(t)$ are the external deterministic input currents, $\mI(t)=[\underset{N_1}{\underbrace{I_1(t),\cdots,I_1(t)}},\cdots,\underset{N_P}{\underbrace{I_P(t),\cdots,I_P(t)}}]^T$, $\mathbf{\bar{J}}$ the $N \times N$ matrix of the weights $\bar{J}_{ij}$ which are equal to $\bar{J}_{p(i)p(j)}$ (see equations \eqref{eq:population}), and $S$ is a mapping from $\R^N$ to $\R^N$ such that 
\begin{equation}\label{eq:population-S}
 S(\VV)_i=S_{p(i)}(V_i) \quad \text{for} \quad  i=1,\cdots,N.
\end{equation}

We  define
\[
 \mL(t)=\left[ 
      \begin{array}{cccc}
       0_{N\times N}       &     {\rm Id}_N  &     \cdots   &  0_{N\times N}\\
      0_{N\times N}        &  0_{N\times N}     &     \ddots   &  0_{N\times N}\\
         \vdots      &  \vdots      &        &  {\rm Id}_N\\
      \mA_0(t)  & \mA_1(t) &  \cdots   & \mA_{k-1}(t) 
      \end{array}
   \right],
\]
where ${\rm Id}_N$ is the $N \times N$ identity matrix and $0_{N \times N}$ the $N \times N$ null matrix. We also define the two $kN$-dimensional vectors:
\[ 
\widetilde{\mU}_t = \left[
   \begin{array}{c}
    0_N\\
   \vdots \\
   0_N\\
   \mathbf{\bar{J}} \cdot S(\tV_0(t))
   \end{array}
\right]=
\left[
   \begin{array}{c}
    0_N\\
   \vdots \\
   0_N\\
   \mathbf{\bar{J}} \cdot S(\VV(t))
   \end{array}
\right]\ 
\textrm{ and } \widetilde{\mI}_t = \left[
   \begin{array}{c}
    0_N\\
   \vdots \\
   0_N\\
   \mI(t)
   \end{array}
\right],
\]
where $0_N$ is the $N$-dimensional null vector.

Combining equations \eqref{eq:MicroGMMd} and \eqref{eq:MicroGMM} the full equation satisfied by $\tV$ can be written:
\begin{equation}\label{eq:VNtilde}
  d \tV(t)=\left(-\mL(t)\tV(t)+ \widetilde{\mU}_t + \widetilde{\mI}_t \right) dt+
\mla(t) \cdot d\mathbf{W}_t,
\end{equation}
where the $kN \times kN$ matrix $\mla(t)$ is equal to ${\rm diag}(\mla_0,\cdots,\mla_{k-1})$ and $\mW_t$ is an $kN$-dimensional standard Brownian process.

\subsection{The Mean-Field equations}
One of the central goals of this paper is to analyze what happens when we let the total number $N$ of neurons grow to infinity. 
Can we ``summarize'' the $kN$ equations \eqref{eq:VNtilde} with a smaller number of equations that would account for the populations activity? We show that the answer to this question is yes and that the populations activity can indeed be represented by $P$ stochastic differential equations of order $k$. Despite the fact that their solutions are Gaussian processes, these equations turn out to be quite complicated because these processes are non-Markovian.

We assume that the proportions of neurons in each population are non-trivial, i.e. :
\[\lim\limits_{N\to\infty} \frac{N_{\alpha}}{N} = n_{\alpha} \;\in \;(0,1)\; \forall \alpha \in \{1,\,\ldots,\,P\},\quad \text{and} \quad \sum_\alpha n_\alpha=1.\]
If it were not the case the corresponding population would not affect the global behavior of the system, would not contribute to the mean-field equation, and could be neglected.

\subsubsection{General derivation of the mean-field equation}\label{subsubsection:mean-field}

When investigating the structure of such mesoscopic neural assemblies as cortical columns, experimentalists are able to provide the average value $\bar{J}_{ij}$ of the synaptic efficacy  $J_{ij}$ of neural population $j$ to population $i$. These values are obviously subject to some uncertainty which can be modeled as Gaussian random variables. We also impose that the distribution of the $J_{ij}$s depends only on the population pair $\alpha=p(i),\beta=p(j)$, and on the total number of neurons  $N_\beta$ of population $\beta$:

\begin{equation}\label{eq:lawJij}
J_{ij} \sim \m N \Big(\frac{\Jbab}{N_{\beta}}, \frac{\Jab} {\sqrt{N_{\beta}}}\Big).
\end{equation}

We also make the additional assumption that the $J_{ij}$'s are independent. This is a reasonable assumption as far as modeling cortical columns from experimental data is concerned.
 Indeed, it is already
difficult for experimentalists to provide the average value of the synaptic strength 
$\Jbab$ from  population
$\beta$ to population $\alpha$ and to estimate the corresponding error bars ($\Jab$),
but measuring synaptic efficacies correlations 
in a large assembly of neurons seems currently out of reach.
Though, it is known that synaptic weights are indeed correlated (e.g. 
via synaptic plasticity mechanisms),  these correlations are built by dynamics 
via a complex interwoven  evolution between neurons and synapses dynamics and postulating
the form of synaptic weights correlations requires, on theoretical grounds,
 a detailed investigation of
the whole history of neurons-synapses dynamics.

Let us now discuss the scaling form of the probability distribution (\ref{eq:lawJij})
of the $J_{ij}$'s, namely the division by $N_\beta$ for the mean and variance of the Gaussian
distribution. This scaling ensures that the ``local interaction field'' $\sum_{j=1}^{N_\beta}
J_{ij}S(V_j(t))$, summarizing the effects of the neurons in population $\beta$ on neuron $i$,
has a mean and variance which do not depend on $N_\beta$ and is only controlled
by the phenomenological parameters $\Jbab,\Jab$.

We are interested in the limit law when $N \to \infty$ of the $N$-dimensional vector $\VV$ defined in equation \eqref{eq:V} under the joint law of the connectivities and the Brownian motions, which we call the mean-field limit. This law can be described by a set of $P$ equations, the mean-field equations. As mentioned in the introduction these equations can be derived in several ways, either heuristically as in the work of Amari \cite{amari:72,amari-yoshida-etal:77}, Sompolinsky \cite{sompolinsky-crisanti-etal:88,crisanti-sommers-etal:90}, and Cessac \cite{cessac:95,samuelides-cessac:07}, or rigorously as in the work of Benarous and Guionnet  \cite{ben-arous-guionnet:95,ben-arous-guionnet:97,guionnet:97}. We derive them here in a pedestrian way, prove that they are well-posed, and provide an algorithm for computing their solution.

The effective description of the network population by population is possible because the neurons in each population are interchangeable, i.e. have the same probability distribution under the joint law of the multidimensional Brownian motion and the connectivity weights. This is the case because of the relations \eqref{eq:population} and \eqref{eq:population-S} which imply the form of equation \eqref{eq:VNtilde}.

\paragraph{The mean ideas of dynamic mean-field equations.}\ \\
Before diving into the mathematical developments let us comment briefly what are the basic ideas and conclusions of the mean-field
approach.
Following equation \eqref{eq:DV}, the evolution of the membrane
potential of some neuron $i$ in population $\alpha$ is given
by:

\beq\label{eq:VBM}
\sum_{l=0}^k b_{l\alpha}(t) 
\frac{d^l V_i}{d t^l}(t)=\sum_{j=1}^N J_{ij} S_j(V_j(t))+ I_i(t)+n_i(t), \, p(i)=\alpha.
\eeq

Using the assumption that
$S_i$, $I_i$, $n_i$ depend only on neuron population, this gives:

\beq\label{eq:VBMF}
\sum_{l=0}^k b_{l\alpha}(t) 
\frac{d^l V_i}{d t^l}(t)=\sum_{\beta=1}^P\eta_{i\beta}(V(t)) + 
I_\alpha(t)+n_i(t), \, i \in \alpha,
\eeq
where we have introduced the local interaction field
$\eta_{i\beta}(V(t))=\sum_{j=1}^{N_\beta} J_{ij}S_\beta(V_j(t))$, summarizing the effects of neurons in population $\beta$ on neuron $i$ and whose probability distribution only depends on the pre- and postsynaptic populations $\alpha$ and $\beta$.

In the simplest situation where the $J_{ij}$'s have no fluctuations ($\Jab=0$) this field
reads $\eta_{i\beta}(V(t))=\Jbab\Phi_\beta(V(t))$. The term $\Phi_\beta(V(t))=\frac{1}{N_\beta}\sum_{j=1}^{N_\beta} S_\beta(V_j(t))$ 
is the frequency rate of
neurons in population $\beta$, averaged over this population. Introducing in the 
same way the average membrane potential in population $\beta$, 
$V_\beta(t)=\frac{1}{N_\beta}\sum_{j=1}^{N_\beta} V_j(t)$, one obtains:

\beq
\sum_{l=0}^k b_{l\alpha}(t) 
\frac{d^l V_\alpha}{d t^l}(t)=\sum_{\beta=1}^P \Jbab\Phi_\beta(V(t))
+ I_\alpha(t)+n_\alpha(t).
\eeq
 
This equation ressembles very much equation (\ref{eq:VBM}) if one
makes the following reasoning: ``Since  $\Phi_\beta(V(t)$ 
is the frequency rate of neurons in population $\beta$, averaged over this population,
and since, for one neuron, the frequency rate is $\nu_i(t)=S_i(V_i(t))$ let us
write $\Phi_\beta(V(t))=S_\beta(V_\beta(t))$''. This leads to:

\beq\label{eq:NMF}
\sum_{l=0}^k b_{l\alpha}(t) 
\frac{d^l V_\alpha}{d t^l}(t)=\sum_{\beta=1}^P \Jbab S_\beta(V_\beta(t))
+ I_\alpha(t)+n_\alpha(t),
\eeq
which has exactly the same form as (\ref{eq:VBM}) but
at the level of a neuron population.

Equations such as (\ref{eq:NMF}), which are obtained via a very strong
assumption:
\beq\label{eq:NMFA}
\frac{1}{N_\beta}\sum_{j=1}^{N_\beta} S_\beta(V_j(t))
=S_\beta\left(\frac{1}{N_\beta}\sum_{j=1}^{N_\beta}V_j(t)\right),
\eeq
are typically those obtained by Jansen-Rit \cite{jansen-rit:95}.
Surprisingly, they are correct and can be rigorously derived,
as discussed below, provided $\Jab=0$.

However, they cannot remain true, as soon as the
synaptic weights fluctuate. Indeed, the transition
from equation (\ref{eq:VBM}) to equation (\ref{eq:NMF}) corresponds 
to a projection from a $NP$-dimensional space
to a $P$-dimensional one, which holds because
the $NP \times NP$ dimensional synaptic weights matrix
has in fact only $P$ linearly independent rows.
This does not hold anymore if the $J_{ij}$'s are random
and the synaptic weights matrix has generically full rank.
Moreover, the effects of the nonlinear dynamics
on the synaptic weights variations about their mean,
is not small even if the $\Jab$s are and the real trajectories
of (\ref{eq:VBM}) can depart strongly from the trajectories
of (\ref{eq:NMF}). This is the main message of this paper.

To finish this qualitative description, let us say in a few words
what happens to the mean-field equations when $\Jab \neq 0$.
We show below that the local interaction fields $\eta_{\alpha\beta}(V(t))$
becomes, in the limit $N_\beta \to \infty$, a time dependent Gaussian field $U_{\alpha\beta}(t)$.
One of the main results is that this field is  \textit{non Markovian}, i.e.
it integrates the whole history, via the synaptic responses $g$
which are convolution products. Despite the fact that  the  evolution equation
for the membrane potential averaged over a population writes
in a very simple form:
\beq\label{eq:RMF}
\sum_{l=0}^k b_{l\alpha}(t) 
\frac{d^l V_\alpha}{d t^l}(t)=\sum_{\beta=1}^P U_{\alpha\beta}(t)
+ I_\alpha(t)+n_\alpha(t),
\eeq
it hides a real difficulty, since $U_{\alpha\beta}(t)$
depends on the whole past. Therefore, \textit{the introduction of synaptic
weights variability leads to a drastic change in neural mass models,
as we now develop.}

\paragraph{The Mean-Field equations}
We note $C([t_0,T],\R^P)$ (respectively 
$C((-\infty,T],\R^P)$) the set of continuous functions from the real interval $[t_0,T]$ (respectively $(-\infty,T]$) to $\R^P$. By assigning a probability to subsets of such functions, a continuous stochastic process $X$ defines a positive measure of unit mass on $C([t_0,T],\R^P)$ (respectively 
$C((-\infty,T],\R^P)$). This set of positive measures of unit mass is noted $\procP{P}$ (respectively $\prociP{P}$).

We now define a process of particular importance for describing the limit process: the effective interaction process. 
\begin{definition}[Effective Interaction Process]\label{def:effectiveInteractionProcess}
 Let $X\in \procP{P}$ (resp. $\prociP{P}$) be a given Gaussian stochastic process. The effective interaction term is the Gaussian process $\mU^{X} \in \procP{P \times P}$, (resp. $\prociP{P \times P}$) defined by:
 \begin{equation}\label{eq:effectiveInteractionProcessParams}
  \begin{cases}
   \Exp{U^{X}_{\alpha\beta}(t)} = \Jbab m^X_{\beta}(t) \\
   \Cov(U^{X}_{\alpha\beta}(t), U^{X}_{\gamma \delta}(s))  = \left\{
	\begin{array}{ll}
	\Jdab  \Debx(t,s) & \text{if} \quad \alpha=\gamma \; \text{and} \; \beta=\delta\\
     	0 & \text{otherwise} 
	\end{array} \right.
  \end{cases},
 \end{equation}
where 
\[
 m^X_{\beta}(t) \eqdef \mathbb{E}[S_{\beta}(X_{\beta}(t))],
\]
and
\[
 \Debx(t,s) \eqdef \mathbb{E}\Big[S_{\beta}(X_{\beta}(t))S_{\beta}(X_{\beta}(s))\Big]
\]
\end{definition}
In order to construct the solution of the mean-field equations (see section \ref{sect:existUniq}) we will need more explicit expressions for $m_\beta^X(t)$ and $\Debx(t,s)$ which we obtain in the next proposition.
\begin{proposition}\label{prop:muCexpressions}
Let $\mu(t)=\Exp{X_t}$ be the mean of the process $X$ and $C(t,s)=\Exp{(X_t-\mu(t))(X_s-\mu(s))^T}$ be its covariance matrix. The vectors $\mathbf{m}^X(t)$ and $\Delta^X(t,s)$ that appear in the definition of the effective interaction process $\mU^X$ are defined by the following expressions:
\begin{equation}\label{eq:mean-interaction}
 m_\beta^X(t)=\int_\R S_\beta\left(x \sqrt{C_{\beta \beta}(t,t)}+\mu_\beta(t)\right) \,Dx,
\end{equation}
and
\begin{multline}\label{eq:covariance-interaction}
 \Debx(t,s)=\int_{\R^2} S_{\beta}\Bigg(\frac{\sqrt{C_{\beta \beta}(t,t)C_{\beta \beta}(s,s)-C_{\beta \beta}(t,s)^2}}{\sqrt{C_{\beta \beta}(t,t)}}x+
\frac{C_{\beta \beta}(t,s)}{\sqrt{C_{\beta \beta}(t,t)}}y+\mu_\beta(s) \Bigg)\\
S_{\beta}\left(y \sqrt{C_{\beta \beta}(t,t)}+ \mu_\beta(t)\right)\,Dx\,Dy,
\end{multline}
where
\[
 Dx=\frac{1}{\sqrt{2\pi}}e^{-\frac{x^2}{2}}\,dx.
\]
is the probability density of a zero-mean, unit variance, Gaussian variable.
\end{proposition}
\begin{proof}
The results follow immediatly by a change of variable from the fact that $X_\beta(t)$ is a univariate Gaussian random variable of mean $\mu_\beta(t)$ and variance $C_{\beta \beta}(t,t)$ and the pair $(X_\beta(t),X_\beta(s))$ is bivariate Gaussian random variable with mean $(\mu_\beta(t),\mu_\beta(s))$ and covariance matrix
\[
 \left[
\begin{array}{cc}
 C_{\beta \beta}(t,t) & C_{\beta \beta}(t,s)\\
C_{\beta \beta}(t,s) & C_{\beta \beta}(s,s)
\end{array}
\right]
\]
\end{proof}

Choose $P$ neurons $i_1,\ldots, i_P$, one in each population (neuron $i_\alpha$ belongs to the population $\alpha$). We define the $kP$-dimensional vector $\widetilde{\mathcal{V}}^{(N)}(t)$ by choosing, in each of the $k$ $N$-dimensional components $\tV_l(t)$, $l=0,\cdots,k-1$, of the vector $\tV(t)$ defined in equation \eqref{eq:Vtilde} the coordinates of indexes $i_1,\cdots,i_P$. Then it can be shown, using either a heuristic argument or  large deviations techniques (see appendix \ref{appendix:MFECalculus}), that the sequence of $kP$-dimensional processes $\Big(\widetilde{\mathcal{V}}^{(N)}_{t\geq t_0}\Big)_{N\geq 1}$ converges in law to the process $\widetilde{\mathcal{V}}(t)=[\mathcal{V}(t)^T,\mathcal{V}_1(t)^T,\cdots,\mathcal{V}_{k-1}(t)^T]^T$ solution of the following mean-field equation:
\begin{equation}\label{eq:MFEwithMatrix}
 d \widetilde{\mathcal{V}}(t)=\left(-\mL(t)\widetilde{\mathcal{V}}(t)+ \widetilde{\mathbf{U}}^{\mathcal{V}}_t + \widetilde{\mathbf{I}}(t)\right) dt+
\mla(t) \cdot d\mathbf{W}_t.
\end{equation}
$\mL$ is the $Pk \times Pk$ matrix
\[
 \mL(t)=\left[ 
      \begin{array}{cccc}
       0_{P \times P}         &     {\rm Id}_P  &     \cdots   &  0_{P \times P}\\
      0_{P \times P}       &  0_{P \times P}    &     \ddots   &  0_{P \times P}\\
         \vdots      &  \vdots      &        &  {\rm Id}_P\\
      \mA_{0}(t)  & \mA_{1}(t)&  \cdots   &\mA_{k-1}(t)  
      \end{array}
   \right].
\]
The $P \times P$ matrixes  $\mA_l(t)$, $l=0,\cdots,k-1$ are, with a slight abuse of notations, equal to ${\rm diag}(b_{l1}(t),\cdots,b_{lP}(t))$.
$(\mW_t)_{t \geq t_0}$ is a $kP$-dimensional standard Brownian process. $\widetilde{\mathbf{U}}^{\mathcal{V}}$ has the law of the $P$-dimensional effective interaction vector associated to the vector $\mathcal{V}$ (first $P$-dimensional component of $\widetilde{\mathcal{V}}$)  and is statistically independent of the external noise $(\mW_t)_{t\geq t_0}$ and of the initial condition $\widetilde{\mathcal{V}}(t_0) $ (when $t_0>-\infty$):
\[
 \widetilde{\mathbf{U}}^{\mathcal{V}}_t =\left[
   \begin{array}{c}
    0_P\\
   \vdots \\
   0_P\\
   \mathbf{U}^{\mathcal{V}}_t \cdot \mathbf{1} 
   \end{array}
\right] \quad  \widetilde{\mathbf{I}}(t)=\left[
   \begin{array}{c}
    0_P\\
   \vdots \\
   0_P\\
   \mathbf{I}(t)
   \end{array}
\right] \quad \mla(t)={\rm diag}(
    \mla_0(t),\cdots,
   \mla_{k-1}(t)
).
\]
We have used for the matrixes $\mla_l(t)$, $l=0.\cdots,k-1$ the same abuse of notations as for the matrixes $\mA_l(t)$, i.e. $\mla_l(t)={\rm diag}(f_{l1}(t),\cdots,f_{lP}(t))$ for $l=0,\cdots,k-1$. $\mI(t)$ is the $P$-dimensional external current $[I_1(t)\cdots,I_P(t)]^T$.

The process $(\mU^{\mathcal{V}}_t)_{t\geq t_0}$ is a $P \times P$-dimensional process and is applied, as a matrix, to the $P$-dimensional vector $\mathbf{1}$ with all coordinates equal to 1, resulting in the $P$-dimensional vector $\mathbf{U}^{\mathcal{V}}_t  \cdot \mathbf{1}$ whose mean and covariance function can be readily obtained from definition \ref{def:effectiveInteractionProcess}:
\begin{equation}\label{eq:mean-meanf}
 \Exp{\mathbf{U}^{\mathcal{V}}_t  \cdot \mathbf{1}}=\sum_{\beta=1}^P \Jbab m^{\mathcal{V}}_{\beta}(t), \quad m^{\mathcal{V}}_{\beta}(t)=\Exp{S_\beta\left(\mathcal{V}_\beta(t)\right)}
\end{equation}
and
\begin{equation}\label{eq:cov-meanf}
 \Cov((\mathbf{U}^{\mathcal{V}}_t \cdot \mathbf{1})_\alpha (\mathbf{U}^{\mathcal{V}}_s  \cdot \mathbf{1})_\gamma)= \left\{
		\begin{array}{ll}
		\sum_{\beta=1}^P \sigma_{\alpha\beta}^2 \Delta_\beta^{\mathcal{V}}(t,s) & \text{if} \quad \alpha=\gamma\\
		&\\
		0 & \text{otherwise}
		\end{array}\right.
\end{equation}
We have of course
\[
 \Delta_\beta^{\mathcal{V}}(t,s)=\Exp{S_\beta\left(\mathcal{V}_\beta(t)\right)S_\beta\left(\mathcal{V}_\beta(s)\right)}
\]
Equations \eqref{eq:MFEwithMatrix} are formally very similar to equations \eqref{eq:VNtilde} but there are some very important differences. The first ones are of dimension $kP$ whereas the second are of dimension $kN$ which grows arbitrarily large when $N \to \infty$. The interaction term of the  second, $\mathbf{\bar{J}} \cdot S(\VV(t))$, is simply the synaptic weight matrix applied to the activities of the $N$ neurons at time $t$. The interaction term of the first equation, $\widetilde{\mathbf{U}}^{\mathcal{V}}_t$, though inocuous looking, is in fact quite complex (see equations \eqref{eq:mean-meanf} and \eqref{eq:cov-meanf}). In fact the stochastic process $\widetilde{\mathbf{U}}^{\mathcal{V}}_t$, putative solution of equations \eqref{eq:MFEwithMatrix}, is in general non Markovian.

To proceed further we formally integrate the equation using the flow, or resolvent, of the equation \eqref{eq:MFEwithMatrix}, noted $\Phi_L(t,t_0)$ (see appendix \ref{appendix:TOop}), and we obtain, since we assumed $\mL$ continuous, an implicit representation of $\widetilde{\mathcal{V}}(t)$:

\begin{multline}\label{eq:IntegratedGMM}
 \widetilde{\mathcal{V}}(t) = \Phi_L(t,t_0) \widetilde{\mathcal{V}}(t_0) + \int_{t_0}^t \Phi_L(t,s)\cdot \left(  \widetilde{\mU}^{\mathcal{V}}_s + \widetilde{\mathbf{I}}(s)\right) \, ds   \\
 + \int_{t_0}^t \Phi_L(t,s) \cdot \mla(s) \cdot d\mathbf{W}_s
\end{multline}

We now introduce for future reference a simpler model which is quite frequently used in the description on neural networks and has been formally analyzed by Sompolinski and colleagues, \cite{sompolinsky-crisanti-etal:88,crisanti-sommers-etal:90} in the case of one population ($P=1$).

\subsubsection{Example I: The Simple Model}\label{ssect:SimpleModel}

In the Simple Model, each neuron membrane potential decreases exponentially to its rest value if it receives no input, with a time constant $\ta$ depending only on the population. In other words, we assume that the g-shape describing the shape of the PSPs is equation \eqref{eq:g1}, with $K=1$ for simplicity. The noise is modeled by an independent Brownian process per neuron whose standard deviation is the same for all neurons belonging to a given population. 

Hence the dynamics of a given neuron $i$ from population $\alpha$ of the network reads:

\begin{multline}\label{eq:NetworkBasicEquations}
dV_i(t) = \left[ -\frac{V_i(t)}{ \tau_{p(i)} } + \sum_{\beta=1}^P \sum_{j=1}^{N_{\beta}} J_{ij} S_{p(j)}\left(V_j(t)\right) + I_{p(i)}(t)\right] \, dt  \\
 + f_{p(i)} dW_i(t).
\end{multline}
This is a special case of equation \eqref{eq:sdeVi} where $k=1$, $b_{0\alpha}(t)=1/\tau_\alpha$, $ b_{1\alpha}(t)=1$ for $\alpha=1,\cdots,P$. 
The corresponding mean-field equation reads:
 
 \begin{equation}\label{eq:V_MFE}
   d\mathcal{V}_{\alpha}(t) = \Big( -\frac{\mathcal{V}_{\alpha}(t)}{\tau_{\alpha}} + \sum_{\beta=1}^P U_{\alpha\beta}^{\mathcal{V}}(t) + I_{\alpha}(t)\Big)\, dt  + f_{\alpha} dW_{\alpha}(t), \;\;  \forall \alpha \in \{1,\,\ldots,\, P\},
 \end{equation}
 
\noindent where the processes $(W_{\alpha}(t))_{t\geq t_0}$ are independent standard Brownian motions,  $\mU^{\mathcal{V}}(t) = (U_{\alpha\beta}^{\mathcal{V}}(t);\; \alpha, \beta \in \{1,\, \ldots,\, P\})_{t}$ is the effective interaction term, see definition \ref{def:effectiveInteractionProcess}. This is a special case of equation \eqref{eq:MFEwithMatrix} with $\mL={\rm diag}(\frac{1}{\tau_1},\cdots,\frac{1}{\tau_P})$,  and $\mla={\rm diag}(f_1,\cdots,f_P)$.

Taking the expected value of both sides of equation \eqref{eq:V_MFE} and using we obtain equation \eqref{eq:mean-interaction} that the mean $\mu_\alpha(t)$ of $\mathcal{V}_{\alpha}(t)$ satisfies the differential equation
\[
 \frac{d \mu_\alpha(t)}{dt}= 
-\frac{\mu_\alpha(t)}{\tau_{\alpha}}+\sum_{\beta=1}^P \Jbab 
\int_\R S_\beta\left(x \sqrt{C_{\beta \beta}(t,t)}+\mu_\beta(t)\right) \,Dx +I_\alpha(t),
\]
If $C_{\beta \beta}(t,t)$ vanishes for all $t\geq t_0$
this equation reduces to:
\[
 \frac{d \mu_\alpha(t)}{dt}= 
-\frac{\mu_\alpha(t)}{\tau_{\alpha}}+\sum_{\beta=1}^P \Jbab
S_\beta\left(\mu_\beta(t)\right) +I_\alpha(t),
\] 
which is precisely the ``naive'' mean-field equation (\ref{eq:NMF})
obtained with the assumption (\ref{eq:NMFA}). We see that  equations (\ref{eq:NMF})
are indeed correct, provided that $C_{\beta \beta}(t,t)=0, \ \forall t \geq t_0$.

Equation \eqref{eq:V_MFE} can be formally integrated implicitly and we obtain the following integral representation of the process $\mathcal{V}_{\alpha}(t)$:

\begin{multline}\label{eq:IntegratedV}
 \mathcal{V}_{\alpha}(t) = e^{-(t-t_0)/\tau_{\alpha}} \mathcal{V}_{\alpha}(t_0) + \int_{t_0}^t e^{-(t-s)/\tau_{\alpha}} \Big( \sum_{\beta=1}^P U_{\alpha\beta}^{\mathcal{V}}(s) + I_{\alpha}(s) \Big) \, ds  \\
 + f_{\alpha} \int_{t_0}^t e^{-(t-s)/\tau_{\alpha}} dW_{\alpha}(s)
\end{multline}

\noindent where $t_0$ is the initial time. It is an implicit equation on the probability distribution of $\mathcal{V}(t)$, a special case 
 of \eqref{eq:IntegratedGMM}, with $\Phi_L(t,t_0)={\rm diag}(e^{-(t-t_0)/\tau_1},\cdots,e^{-(t-t_0)/\tau_P})$.

The variance  $C_{\alpha \alpha}(t,t)$ of $\mathcal{V}_{\alpha}(t)$ can easily be obtained from equation \eqref{eq:IntegratedV}. It reads
\begin{multline*}
 C_{\alpha \alpha}(t,t)= e^{-2t/\ta}\Big[C_{\alpha \alpha}(t_0,t_0)+ \frac{\ta f_\alpha^2}{2}\left(e^{\frac{2t}{\ta}}-1\right) \\ 
+ \sum_{\beta=1}^P \Jdab \int_{t_0}^t\int_{t_0}^t e^{(u+v)/\ta}\Delta_\beta(u,v)dudv\Big],
\end{multline*}
where $\Delta_\beta(u,v)$ is given by equation \eqref{eq:covariance-interaction}.

If $\Jab=0$ and if $\sa=0$ then $C_{\alpha \alpha}(t,t)=0, \forall t \geq t_0$ is
a solution of this equation. Thus, mean-field equations for the simple model
reduce to the naive mean-field equations (\ref{eq:NMF}) in this case. This conclusion
extends as well to all models of synaptic responses, ruled by equation (\ref{eq:PSP}). 

However, the equation of $ C_{\alpha \alpha}(t,t)$ shows that, in the
general case, 
in order to solve the differential equation for $\mu_\alpha(t)$,
we need to know the whole past of the process $\mathcal{V}$. 
This examplifies a previous statement 
on the non Markovian nature of the solution of the mean-field equations.

\subsubsection{Example II: The model of Jansen and Rit}\label{ssect:JansenRit}

\begin{figure}[!h]
\begin{center}
 \subfigure[Populations involved in Jansen's model]{\includegraphics[width=.5\textwidth]{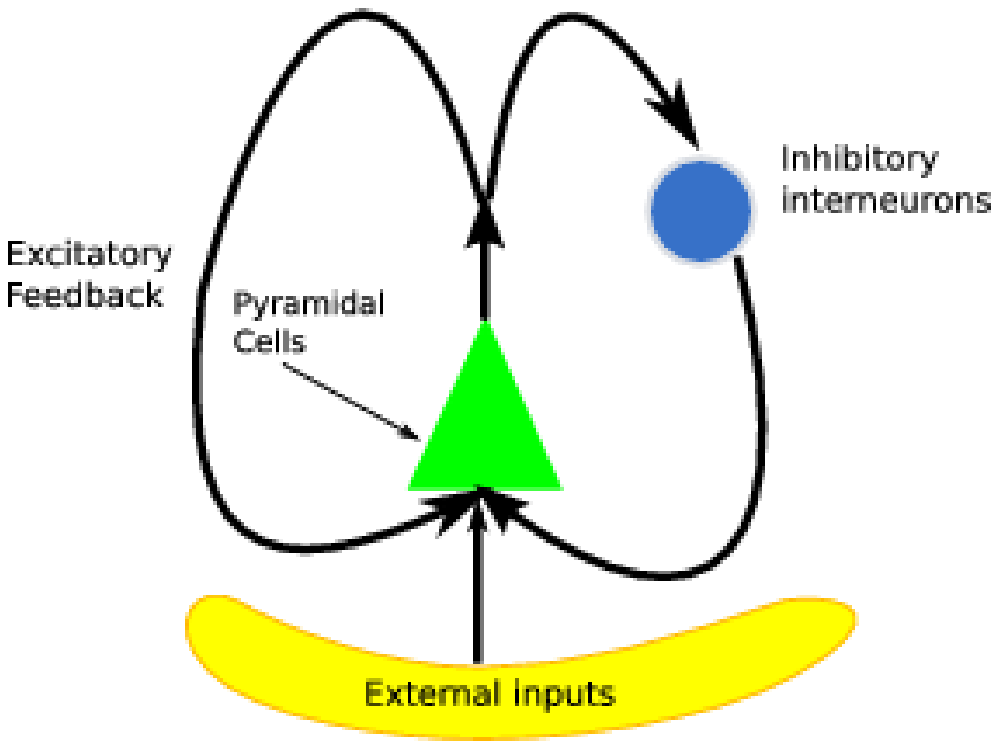}\label{sfig:JansenPopsMF}}
 \subfigure[Block diagram]{\includegraphics[width=.45\textwidth]{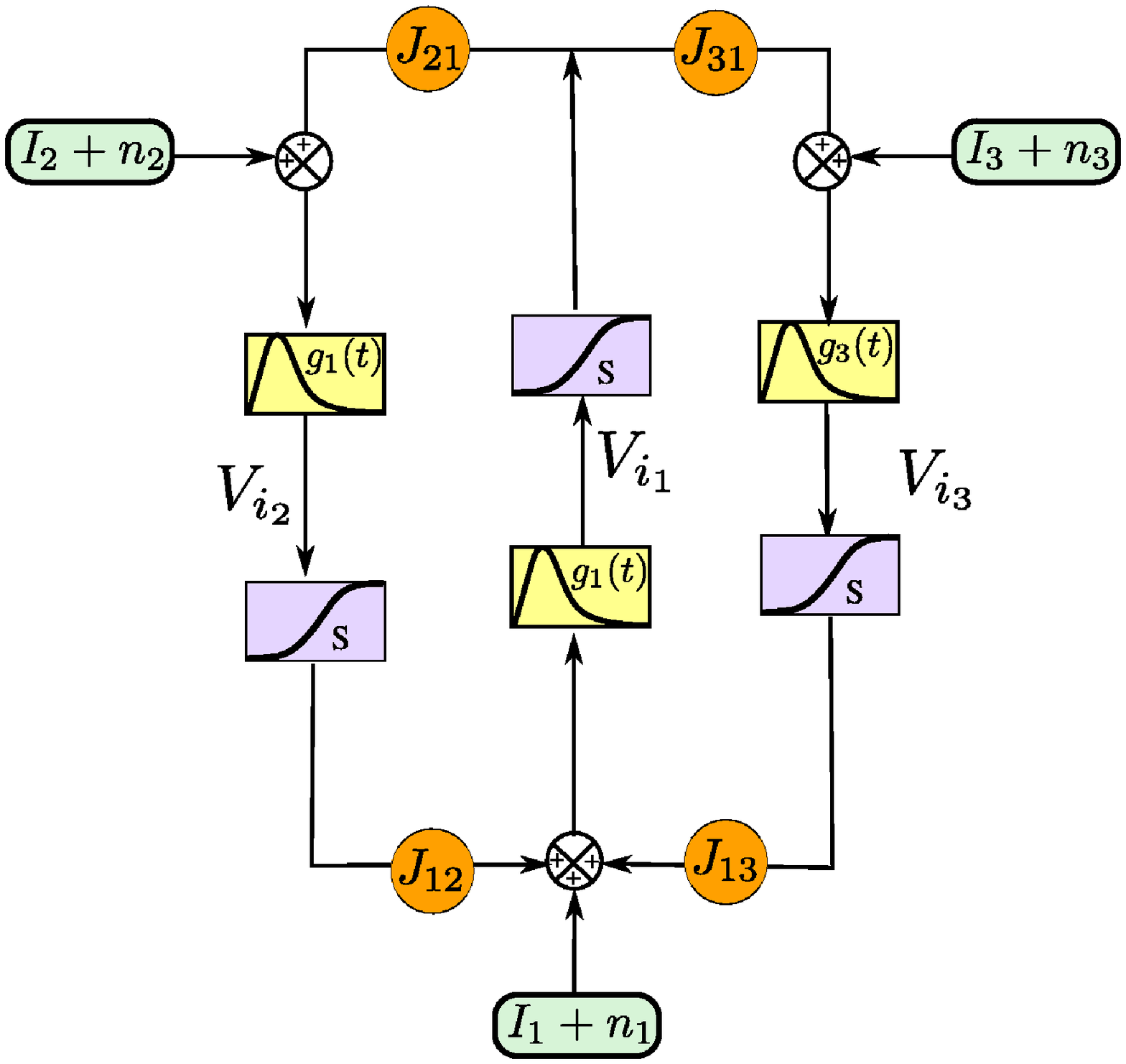}\label{sfig:JansenblocksMF}}
\end{center}
\caption[Jansen and Rit's model]{ a. Neural mass model: a
      population of pyramidal cells interacts with itself in an excitatory mode and with an inhibitory population of inter-neurons.
      b. Block representation of the model. The \emph{g} boxes
      account for the synaptic integration between neuronal populations. $S$ boxes
      simulate cell bodies of neurons by transforming the membrane potential 
      of a population into an output firing rate. The coefficients $J_{\alpha \beta}$
      are the random synaptic efficiency of population $\beta$ on population $\alpha$ 
      (1 represents the pyramidal population, 2 the excitatory feedback, and 3 the inhibitory inter-neurons).
    }
\label{fig:jansenDiagram}
\end{figure}

One of the motivations of this study is to characterize the global behavior of an assembly of neurons in particular to get a better understanding of recordings of cortical signals like EEG or MEG. 
One of the classical models of neural masses is Jansen and Rit's mass model \cite{jansen-rit:95}, in short the JR model (see figure \ref{fig:jansenDiagram}).

The model features a population of pyramidal neurons
 that receives inhibitory inputs from local inter-neurons, 
excitatory feedbacks, and excitatory inputs from neighboring cortical units
and sub-cortical structures such as the thalamus. The excitatory input is represented by an external firing rate
that has a deterministic part $I_1(t)$ accounting for specific activity of other cortical units and a stochastic 
part $n_1(t)$ accounting for a non specific background activity. We formally consider that the excitatory feedback of the 
pyramidal neurons is a new neural population, making the number $P$ of populations equal to 3. We also represent the external inputs to the
other two populations by the sum of a deterministic part $I_j(t)$ and a stochastic part $n_j(t)$, $j=2,3$, see figure~\ref{fig:jansenDiagram}.

In the model introduced originally by Jansen and Rit, the connectivity weights were assumed to be constant, i.e. 
equal to their mean value. Nevertheless, there exists a variability of these coefficients, and as we show in the 
sequel, the effect of the connectivity variability impacts the solution at the level of the neural mass. Statistical 
properties of the connectivities have been studied in details for instance in \cite{braitenberg-schuz:98}.

We consider a network of $N$ neurons, $N_\alpha$, $\alpha=1,2,3$ belonging to population $\alpha$. We index by 1 (respectively 2, and 3) the pyramidal (respectively excitatory feedback, inhibitory interneuron) populations. We choose in each population a particular  neuron  indexed by $i_\alpha$, $\alpha=1,2,3$. The evolution equations of the network can be written for instance in terms of the potentials $V_{i_1}$, $V_{i_2}$ and $V_{i_3}$ labelled in figure \ref{fig:jansenDiagram} and these equations read:

\[
\begin{cases}
 V_{i_1} &= \displaystyle{g_1 \ast \left (\sum_{j=1}^{N_2} J_{i_1j}\Si(V_{j})+\sum_{j=1}^{N_3} J_{i_1j}\Si(V_{j})+I_1+n_1\right) } \\
 V_{i_2} &= \displaystyle{g_1 \ast \left (\sum_{j=1}^{N_1} J_{i_2 j} \Si( V_{j}) + I_2+n_2 \right)}\\
 V_{i_3} &= \displaystyle{g_3 \ast \left (\sum_{j=1}^{N_1} J_{i_3j} \Si(V_{j})+I_3+n_3 \right)} \\
 \end{cases}
\]

In the mean-field limit, denoting by $\mathcal{V}_\alpha$, $\alpha=1,2,3$ the average membrance potential of each class, we obtain the following equations:

\begin{equation}\label{eq:MeanFieldJR}
\begin{cases}
 \mathcal{V}_{1} &= \displaystyle{g_1 \ast \left(U_{12}^{{\mathcal V}}+U_{13}^{{\mathcal V}}+I_1+n_1\right)} \\
 \mathcal{V}_{2} &= \displaystyle{g_1 \ast \left (U_{21}^{{\mathcal V}}+ I_2+n_2 \right)}\\
 \mathcal{V}_{3} &= \displaystyle{g_3 \ast \left(U_{31}^{{\mathcal V}}+I_3+n_3 \right)} \\
 \end{cases}
\end{equation}

\noindent where $\mU^{\mathcal{V}}=(U_{\alpha \beta}^{\mathcal{V}})_{\alpha,\,\beta=1,2,3}$ is the effective interaction process associated with this problem, i.e. a Gaussian process of mean:

\[
\begin{cases}
 \Exp{U_{12}^{\mathcal{V}}} &= \bar{J}_{12} \Exp{\Si(\mathcal{V}_2)} \\
\Exp{U_{13}^{\mathcal{V}}} &= \bar{J}_{13} \Exp{\Si(\mathcal{V}_3)} \\
 \Exp{U_{21}^{\mathcal{V}}} &= \bar{J}_{21} \Exp{\Si(\mathcal{V}_1)} \\
 \Exp{U_{31}^{\mathcal{V}}} &= \bar{J}_{31} \Exp{\Si(\mathcal{V}_1)},
\end{cases}
\]
All other means correspond to the non-interacting populations and are equal to zero.
The covariance matrix can be deduced from \eqref{eq:effectiveInteractionProcessParams}:
\[
 {\rm Cov}\left(U_{\alpha \beta}^{\mathcal{V}}(t),\, U_{\gamma \delta}^{\mathcal{V}}(s)  \right)=\left\{
	\begin{array}{l}
	 \sigma_{\alpha \beta}^2 \Delta_\beta^{\mathcal{V}}(t,s)  \quad \text{if} \quad \alpha=\beta \quad \text{and} \quad \gamma=\delta\\
	 0 \quad \text{otherwise}
	\end{array}
\right.
\]
where
\[
 \Delta_\beta^{\mathcal{V}}(t,s)=\Exp{S\left(\mathcal{V}_\alpha(t)\right)S\left(\mathcal{V}_\beta(s)\right)}
\]
This model is  a voltage-based model in the sense of Ermentrout \cite{ermentrout:98}. Let us now instantiate the synaptic dynamics and compare the mean-field equations with Jansen's population equations\footnote{We have modified the original model which is not voltage-based.} (sometimes improperly called also mean-field equations). 

The simplest model of synaptic integration is a first-order integration, which yields exponentially decaying post-synaptic potentials:
\[
g(t)=\left\{
\begin{array}{ll}
K e^{-\frac{t}{\tau}} & t \geq 0\\
0 & t<0
\end{array}
\right.
\]

\noindent Note that this is exactly equation \eqref{eq:g1}. The corresponding g-shape satisfies the following 1st-order differential equation
\[
 \dot{g}(t)=-\frac{1}{\tau}g(t)+K \delta(t),
\]

In this equation $\tau$ is the time constant of the synaptic integration and $K$ the synaptic efficiency. The coefficients $K$ and $\tau$ are the same for the pyramidal and the excitatory feedback population (characteristic of the pyramidal neurons and defining the g-shape $g_1$), and different for the inhibitory population (defining the g-shape $g_3$). In the pyramidal or excitatory (respectively the inhibitory) case we have $K=K_1$, $\tau=\tau_1$ (respectively $K=K_3$, $\tau=\tau_3$). Finally, the sigmoid functions $\Si$ is given by
\[\Si(v)=\frac{\nu_{\max}}{1+e^{r(v_0-v)}},\]
\noindent where $\nu_{\max}$ is the maximum firing rate, and $v_0$ is a voltage reference.

With this synaptic dynamics we obtain the first-order Jansen and Rit's equation:

\begin{equation}\label{eq:JRMF1}
 \begin{cases}
  \frac{d\mathcal{V}_{1}}{dt} &= -\frac{1}{\tau_1} \mathcal{V}_1 + K_1 \left(U_{12}^{{\mathcal V}}+U_{13}^{{\mathcal V}}+I_1+n_1\right) \\
  \frac{d\mathcal{V}_{2}}{dt} &= -\frac{1}{\tau_1} \mathcal{V}_2 + K_1 \left (U_{21}^{{\mathcal V}}+ I_2+n_2 \right) \\
  \frac{d\mathcal{V}_{3}}{dt} &= -\frac{1}{\tau_3} \mathcal{V}_3 + K_3 \left(U_{31}^{{\mathcal V}}+I_3+n_3 \right)
 \end{cases}.
\end{equation}
The ``original'' Jansen and Rit's equation \cite{jansen-rit:95,grimbert-faugeras:06} amount considering only the mean of the process $\mathcal{V}$ and assuming that  $\Exp{\Si_i(\mathcal{V}_j)}=\Si_i(\Exp{\mathcal{V}_j})$ for $i,j \in \{1,2,3\}$, i.e. that the expectation commutes with the sigmoidal function $\Si$. This is a very strong assumption, and that the fluctuations of the solutions of the mean-field equation around the mean imply that the sigmoid cannot be considered as linear in the general case.

A higher order model was introduced by van Rotterdam and colleagues \cite{rotterdam-lopes-da-silva-etal:82} to better account for the synaptic integration and to better reproduce the characteristics of real postsynaptic potentials. In this model the g-shapes satisfy a second order differential equation:
\[
g(t)=\left\{
\begin{array}{ll}
K te^{- \frac{t}{\tau}} & t\geq0\\
0 & t<0
\end{array}
\right.,
\]
\noindent We recognize the g-shape defined by equation \eqref{eq:g2}
 solution of the second-order differential equation $\ddot{y}(t)+\frac{2}{\tau} \dot{y}(t)+\frac{1}{\tau^2}y(t) = K\delta(t)$.
With this type of synaptic integration, we obtain the following mean-field equations: 
\begin{equation}\label{eq:JRMF2}
  \begin{cases}
    \frac{d^2{\mathcal{V}}_1}{dt^2} &=  -\frac{2}{\tau_1} \frac{d\mathcal{V}_1}{dt}-\frac{1}{\tau_1^2} \mathcal{V}_1+K_1 \left(U_{12}^{{\mathcal V}}+U_{13}^{{\mathcal V}}+I_1+n_1\right)\\
    \frac{d^2\mathcal{V}_2}{dt^2} & =-\frac{2}{\tau_1} \frac{d\mathcal{V}_2}{dt}-\frac{1}{\tau_1^2}\mathcal{V}_2+ K_1 \left (U_{21}^{{\mathcal V}}+ I_2+n_2 \right)\\
    \frac{d^2\mathcal{V}_3}{dt^2} & =- \frac{2}{\tau_3}\frac{d\mathcal{V}_3}{dt}-\frac{1}{\tau_3^2}\mathcal{V}_3+K_3 \left(U_{31}^{{\mathcal V}}+I_3+n_3 \right)
  \end{cases}
\end{equation}
Here again, going from the mean-field equations \eqref{eq:JRMF2} to the original Jansen and Rit's neural mass model consists in studying the equation of the mean of the process given by \eqref{eq:JRMF2} and commuting the sigmoidal function with the expectation. 

Note that the introduction of higher order synaptic integrations results in richer behaviors. For instance, Grimbert and Faugeras \cite{grimbert-faugeras:06} showed that some bifurcations can appear in the second-order JR model giving rise to epileptic like oscillations and alpha activity, that do not appear in the first order  model.

\section{Existence and uniqueness of solutions in finite time}\label{sect:existUniq}
The mean-field equation \eqref{eq:IntegratedGMM}
is an implicit equation of the stochastic process $(V(t))_{t \geq t_0}$. We prove in this section that under some mild assumptions this implicit equation has a unique solution. These assumptions are the following.
\begin{Assump}\label{assumption:all}\ \\
 \begin{enumerate}
  	\item The matrix $\mL(t)$ is $C^0$ and satisfies $\norm{\mL(t)} \leq k_L$ for all $t$ in $[t_0,T]$, for some matrix norm $\norm{\ }$ and some strictly positive constant $k_L$.
	\item The matrix $\mla(t)$ has all its singular values lowerbounded (respectively upperbounded) by the strictly positive constant\footnote{We note $\Gamma(t)$ the matrix $\mla(t)\mla(t)^T$.} $\lambda_{\rm min}^\Gamma$ (respectively $\lambda_{\rm max}^\Gamma$) for all $t$ in $[t_0,T]$.
	\item The deterministic external input vector $\mI(t)$ is bounded and we have $\norm{\mI(t)}_\infty \leq I_{\rm max}$ for all $t$ in $[t_0,T]$ and some strictly positive constant $I_{\rm max}$
 \end{enumerate}
\end{Assump}

This solution is the fixed point in the set $\procP{kP}$ of $kP$-dimensional processes of an equation that we will define from the mean-field equations. We will construct a sequence of Gaussian processes and prove that it converges in distribution toward this fixed point.

We first recall some results on the convergence of
random variables and stochastic processes.

\subsection{Convergence of Gaussian processes}

We recall the following result from \cite{bogachev:98} which formalizes the intuition that a sequence of Gaussian processes converges toward a Gaussian process if and only if the means and covariance functions converge. In fact in order for this to be true, it is only necessary to add one more condition, namely that the corresponding sequence of measures (elements of $\procP{kP}$) do not have ``any mass at infinity''. This property is called uniform tightness \cite{billingsley:99}. More precisely we have
\begin{definition}[Uniform tightness]
 Let $\{X_n\}_{n=1}^{\infty}$ be a sequence of $kP$-dimensional processes defined on $[t_0,T]$  and $P_n$ be the associated elements of $\procP{kP}$. The sequence $\procP{kP}$ is called uniformly tight if and only if for all $\varepsilon > 0$ there exists a compact set $K$ of $C([t_0,T], \R^{kP})$ such that $P_n(K) > 1-\varepsilon$, $n \geq 1$.
\end{definition}

\begin{theorem}\label{theo:convergence}
 Let $\{X_n\}_{n=1}^{\infty}$ be a sequence of $kP$-dimensional Gaussian  processes defined on $[t_0,T]$ or on an unbounded interval\footnote{In \cite[Chapter 3.8]{bogachev:98}, the property is stated whenever the mean and covariance are defined on a separable Hilbert space.} of $\R$. The sequence converges
to a Gaussian process $X$ if and only if the following three conditions are satisfied:
\begin{itemize}
   \item The sequence $\{X_n\}_{n=1}^{\infty}$ is uniformly tight.
   \item The sequence $\mu^n(t)$ of the mean functions converges for the uniform norm.
   \item The sequence $C^n$ of the covariance operators converges for the uniform norm.
\end{itemize}
\end{theorem}
\noindent
We now, as advertised, define such a sequence of Gaussian processes. 

Let us fix $Z_0$, a $kP$-dimensional Gaussian random variable, independent of the Brownian and of the process $((X)_t)_{t \in [t_0,T]}$. 
\begin{definition}\label{def:F1Fk}
Let $X$ be an element of $\procP{kP}$ and 
 $\mathcal{F}_k$ be the function $\procP{kP} \to \procP{kP}$ such that
 \begin{multline*}
  \mathcal{F}_k(X)_t = \Phi_L(t,t_0) \cdot Z_0 + \int_{t_0}^t \Phi_L(t,s) \cdot \big(\widetilde{\mathbf{U}}^{X}_s + \widetilde{\mathbf{I}}(s)\big) \, ds   \\
 + \int_{t_0}^t \Phi_L(t,s) \cdot \mla(s) d\mathbf{W}_s
 \end{multline*}
  \noindent where $\widetilde{\mathbf{U}}^{X}_s$ and $\widetilde{\mathbf{I}}(s)$ are defined\footnote{For simplicity we abuse notations and identify $\tilde{X}$ and $X$.} in section \ref{sect:equations}.
\end{definition}

Note that, by definition \ref{def:effectiveInteractionProcess} the random process $(\mathcal{F}_k(X))_{t\in [t_0,T]}$, $k \geq 1$ is the sum of a deterministic function (defined by the external current) and three independent random processes defined by $Z_0$, the interaction between neurons, and the external noise. These three processes being Gaussian processes, so is $(\mathcal{F}_k(X))_{t\in [t_0,T]}$. Also note that $\mathcal{F}_k(X)_{t_0}=Z_0$. It should be clear that a solution $\mathcal{V}$  of the mean-field equation \eqref{eq:IntegratedGMM} satisfies $\mathcal{V}(t_0)=Z_0$ and is a fixed point of $\mathcal{F}_k$, i.e. $\mathcal{F}_k(\mathcal{V})_t=\mathcal{V}(t)$.

\noindent
Let $X$ be a given stochastic process of $\procP{kP}$ such that $X_{t_0}=Z_0$ (hence $X_{t_0}$ is independent of the Brownian). 
We define the sequence of Gaussian processes $\{X_n\}_{n=0}^\infty \in \procsP{T}{kP}$ by:

\begin{equation}\label{eq:DefSequenceXn}
\begin{cases}
   X_0 &= X \\
   X_{n+1} & = \mathcal{F}_k(X_n) = \mathcal{F}_k^{(n)}(X_0).\quad n \geq 0, \quad \mathcal{F}_k^{(0)}={\rm Id}
  \end{cases}
\end{equation}

In the remaining of this section we show that the sequence of processes $\{\mathcal{F}_k^{(n)}(X)\}_{n=0}^\infty$ converges in distribution toward the unique fixed-point $Y$ of $\mathcal{F}_k$ which is also the unique solution of the mean-field equation \eqref{eq:IntegratedGMM}.

\subsection{Existence and uniqueness of a solution for the mean-field equations}

The following upper and lower bounds are used in the sequel.
\begin{lemma}\label{lemma:upper}
Consider the Gaussian process $((\mU^X_t \cdot \mone)_t)_{t \in [t_0,T]}$. $\mU^X$ is defined in \ref{def:effectiveInteractionProcess} and $\mone$ is the $P$-dimensional vector with all coordinates equal to 1.
We have
\begin{equation}\label{eq:mu}
 \norm{\Exp{\mU^X_t \cdot \mone}}_\infty \leq \mu \eqdef \max_\alpha \sum_\beta |\Jbab|\, \|S_{\beta}\|_\infty
\end{equation}
for all $t_0 \leq t \leq T$.
The maximum eigenvalue of its covariance matrix is upperbounded by $\sigma_{\rm max}^2\overset{\rm def}{=}\max_\alpha \sum_\beta \Jdab \, \|S_{\beta}\|_\infty^2$ where $\|S_{\beta}\|_\infty$ is the supremum of the absolute value of $S_{ \beta}$. We also note $\sigma_{\rm min}^2 \eqdef \min_{\alpha,\beta} \sigma_{\alpha\beta}^2$.
\end{lemma}
\begin{proof}
 The proof is straightforward from definition \ref{def:F1Fk}.
\end{proof}

The proof of existence and uniqueness of solution, and of the convergence of the sequence \eqref{eq:DefSequenceXn} is in two main steps. We first prove that the sequence of Gaussian processes $\{\mathcal{F}_k^{(n)}(X)\}_{n=0}^\infty,\,k\geq 1$ is uniformly tight by proving that it satisfies Kolmogorov's criterion for tightness. This takes care of condition 1) in theorem \ref{theo:convergence}.  We then prove that the sequences of the mean functions and covariance operators are Cauchy sequences for the uniform norms, taking care of conditions 2) and 3).

\subsubsection{Uniform tightness}
We first recall the following theorem due to Kolmogorov \cite[Chapter 4.1]{kushner:84}.
\begin{theorem}[Kolmogorov's criterion for thightness]
 Let $\{X_n\}_{n=1}^{\infty}$ be a sequence of $kP$-dimensional processes defined on $[t_0,T]$. If  there exist $\alpha,\,\beta,\,C > 0$ such that
\[
 \Exp{\norm{X_n(t)-X_n(s)}^\beta} \leq C|t-s|^{1+\alpha} \quad \forall s,\,t\,\in [t_0,T] \quad n \geq 1,
\]
then the sequence is uniformly tight.  
\end{theorem}
Using this theorem we prove that the sequence $\{\mathcal{F}_k^{(n)}(X)\}_{n=0}^\infty,\,k\geq 1$ satisfies Kolmogorov's criterion for $\beta=4$ and $\alpha \geq 1$. The reason for choosing $\beta=4$ is that, heuristically, $dW \simeq (dt)^{1/2}$. Therefore in order to upperbound $\Exp{\norm{X_n(t)-X_n(s)}^\beta}$ by a power of $|t-s|$ greater than or equal to 2 (hence strictly larger than 1) we need to raise $\norm{X_n(t)-X_n(s)}$ to a power at least equal to 4. The proof itself is technical and uses standard inequalities (Cauchy-Schwarz' and Jensen's), properties of Gaussian integrals, elementary properties of the stochastic integral, and lemma  \ref{lemma:upper}. It also uses the fact that the input current is bounded, i.e. that $\max_{\alpha=1,\cdots,P} \sup_{t \in [t_0,T]} |I_\alpha(t)| \leq I_{\rm max} <\infty$, this is assumption 3 in \ref{assumption:all}.
\begin{theorem}\label{theo:tight}
The sequence of processes $\left\{\mathcal{F}_k^{(n)}(X)\right\}_{n=0}^\infty$, $k \geq 1$ is uniformly tight.
\end{theorem}

\begin{proof}
 We do the proof for $k=1$, the case $k > 1$ is similar. If we assume that $n \geq 1$ and $s < t$ we can rewrite the difference $\mathcal{F}_1^{(n)}(X)_t-\mathcal{F}_1^{(n)}(X)_s$ as follows, using property i) in proposition \ref{prop:ETO} in appendix \ref{appendix:TOop}.

\begin{align*}
 &\mathcal{F}_1^{(n)}(X)_t-\mathcal{F}_1^{(n)}(X)_s=(\Phi_L(t,t_0)-\Phi_L(s,t_0)) X_{t_0}\\
 &\qquad + (\Phi_L(t,s)-Id) \int_{t_0}^s \Phi_L(s,u)  \mU^{\mathcal{F}_1^{(n-1)}(X)}_u \cdot \mone\,du
 + \int_s^t \Phi_L(t,u) \mU^{\mathcal{F}_1^{(n-1)}(X)}_u \cdot \mone\,du\\
 &\qquad + (\Phi_L(t,s)-Id) \int_{t_0}^s   \Phi_L(s,u) \mla(u)\,d\mW_u+
 \int_s^t \Phi_L(t,u) \mla(u)\,d\mW_u\\
 & \qquad + (\Phi_L(t,s)-Id)\int_{t_0}^s \Phi_L(s,u) \mI(u)\,du+\int_s^t \Phi_L(t,u) \mI(u)\,du
\end{align*}

The righthand side is the sum of seven terms
and therefore (Cauchy-Schwarz inequality):

\begin{align*}
 & \frac{1}{7}\|\mathcal{F}_1^{(n)}(X)_t-\mathcal{F}_1^{(n)}(X)_s\|^2 \leq \|\Phi_L(t,t_0)-\Phi_L(s,t_0)\|^2 \| X_{t_0} \|^2\\ \\
 &\qquad + (s-t_0) \|\Phi_L(t,s)-Id\|^2 \int_{t_0}^s  \|\Phi_L(s,u)\|^2 \|\mU^{\mathcal{F}_1^{(n-1)}(X)}_{u} \cdot \mone\|^2\,du\\
 &\qquad + (t-s)\int_s^t \|\Phi_L(t,u)\|^2 \| \mU^{\mathcal{F}_1^{(n-1)}(X)}_{u} \cdot \mone\|^2\,du\\
 &\qquad + \norm{\Phi_L(t,s)-Id}^2 \left\| \int_{t_0}^s \Phi_L(s,u)  \mla(u)\,d\mW_u \right\|^2+ \left\| \int_s^t \Phi_L(t,u) \mla(u)\,d\mW_u \right\|^2 \\
 &\qquad + (s-t_0)^2 \|\Phi_L(t,s)-Id\|^2 I_{\max}^2 \sup_{u\in [t_0,s]}  \|\Phi_L(s,u)\|^2 \\
 &\qquad + (t-s)^2 I_{\max}^2 \sup_{u\in [s,t]}  \|\Phi_L(t,u)\|^2.
\end{align*}

Because $\|\Phi_L(t,t_0)-\Phi_L(s,t_0)\| \leq |t-s| \|\mL\|$ 
we see that all terms in the righthand side of the inequality but the second one involving the Brownian motion are of the order of $(t-s)^2$. We raise again both sides to the second power, use the Cauchy-Schwarz inequality, and take the expected value: 

\begin{align}
 \nonumber & \frac{1}{7^3}\Exp{\|\mathcal{F}_1^{(n)}(X)_t-\mathcal{F}_1^{(n)}(X)_s\|^4} \leq \|\Phi_L(t,t_0)-\Phi_L(s,t_0)\|^4 \Exp{\| \mX_{t_0} \|^4}\\
 \nonumber &\qquad + (s-t_0)^3 \|\Phi_L(t,s)-Id\|^4 \int_{t_0}^s  \|\Phi_L(s,u)\|^4 \Exp{\|\mU^{\mathcal{F}_1^{(n-1)}(X)}_{u} \cdot \mone\|^4}\,du\\
 \nonumber &\qquad + (t-s)^3\int_s^t \|\Phi_L(t,u)\|^4 \Exp{\| \mU^{\mathcal{F}_1^{(n-1)}(X)}_{u} \cdot \mone\|^4}\,du\\
 \nonumber &\qquad + \norm{\Phi_L(t,s)-Id}^4\Exp{\left\| \int_{t_0}^s \Phi_L(s,u)  \mla(u)\,d\mW_u \right\|^4}\\
 \label{eq:kolmo} &\qquad + \Exp{\left\| \int_s^t \Phi_L(t,u) \mla(u)\,d\mW_u \right\|^4}\\
 \nonumber &\qquad + (s-t_0)^4 \|\Phi_L(t,s)-Id\|^4 I_{\max}^4 \sup_{u\in [t_0,s]}  \|\Phi_L(s,u)\|^4  \\
 \nonumber &\qquad + (t-s)^4 I_{\max}^4 \sup_{u\in [s,t]}  \|\Phi_L(t,u)\|^4.
\end{align}

Remember that $\mU^{\mathcal{F}_1^{(n-1)}(X)}_u \cdot \mone$ is a $P$-dimensional diagonal Gaussian process, noted $\mY_u$ in the sequel, therefore:
\[
 \Exp{\|\mY_u\|^4}=\sum_\alpha \Exp{Y_\alpha(u)^4}+\sum_{\alpha_1 \neq \alpha_2} \Exp{Y_{\alpha_1}^2(u)}\Exp{Y_{\alpha_2}^2(u)}.
\]
The second order moments are upperbounded by some regular function of $\mu$ and $\sigma_{\rm max}$ (defined in lemma \ref{lemma:upper})
 and, because of the properties of Gaussian integrals, so are
the fourth order moments.

We now define $\mA(u)=\Phi_L(s,u)  \mla(u)$ and evaluate $\Exp{\left\|\int_{t_0}^s \mA(u) \,d\mW_u\right\|^4}$. We have

\begin{align*}
 & \Exp{\left\|\int_{t_0}^s \mA(u) \,d\mW_u\right\|^4}=\Exp{\left(\left\|\int_{t_0}^s \mA(u) \,d\mW_u\right\|^2\right)^2}\\
 &\qquad = \Exp{\left( \sum_{i=1}^P \left( \sum_{j=1}^P \int_{t_0}^s B_{ij}(u)\,dW_u^j\right) \left( \sum_{k=1}^P \int_{t_0}^s B_{ik}(u)\,dW_u^k\right) \right)^2}\\
 &\qquad = \sum_{i_1,i_2,j_1,j_2,k_1,k_2} \Exp{\int_{t_0}^s B_{i_1j_1}(u)\,dW_u^{j_1}\int_{t_0}^s B_{i_1k_1}(u)\,dW_u^{k_1}\int_{t_0}^s B_{i_2j_2}(u)\,dW_u^{j_2}\int_{t_0}^s B_{i_2k_2}(u)\,dW_u^{k_2}}.
\end{align*}

Because $W_u^i$ is by construction independent of $W_u^j$ if $i \neq j$ and $\Exp{\int_{t_0}^s B_{ij}(u) dW_u^j}=0$ for all $i,\,j$ (property of the Itô integral),  the last term is the sum of only three types of terms:
\begin{enumerate}
\item If $j_1=k_1=j_2=k_2$ we define
\[
T_1=\sum_{i_1,i_2} \Exp{\left(\int_{t_0}^s B_{i_1j}(u)\,dW_u^{j}\right)^2\left(\int_{t_0}^s B_{i_2j}(u)\,dW_u^{j}\right)^2},
\]
and, using Cauchy-Schwarz:
\[
 T_1 \leq 
\sum_{i_1,i_2} \Exp{\left(\int_{t_0}^s B_{i_1j}(u)\,dW_u^{j}\right)^4}^{1/2}\,\Exp{\left(\int_{t_0}^s B_{i_2j}(u)\,dW_u^{j}\right)^4}^{1/2}
\]
\item If $j_1=k_1$ and $j_2=k_2$ but $_1 \neq j_2$ we define
\[
 T_2=\sum_{i_1,i_2,j_1 \neq j_2} \Exp{\left(\int_{t_0}^s B_{i_1j_1}(u)\,dW_u^{j_1}\right)^2 \left(\int_{t_0}^s B_{i_2j_2}(u)\,dW_u^{j_2}\right)^2},
\]
which is equal, because of the independence of $W_u^{j_1}$ and $W_u^{j_2}$ to
\[
\sum_{i_1,i_2,j_1 \neq j_2} \Exp{\left(\int_{t_0}^s B_{i_1j_1}(u)\,dW_u^{j_1}\right)^2}\Exp{\left(\int_{t_0}^s B_{i_2j_2}(u)\,dW_u^{j_2}\right)^2}.
\]
\item Finally, if $j_1=j_2$ and $k_1=k_2$ but $j_1 \neq k_1$ we define
\[
 T_3=\sum_{i_1,i_2,j_1 \neq k_1} \Exp{\int_{t_0}^s B_{i_1j_1}(u)\,dW_u^{j_1}\int_{t_0}^s B_{i_2j_1}(u)\,dW_u^{j_1}\int_{t_0}^s B_{i_1k_1}(u)\,dW_u^{k_1}\int_{t_0}^s B_{i_2k_1}(u)\,dW_u^{k_1}},
\]
which is equal, because of the independence of $W_u^{j_1}$ and $W_u^{k_1}$ to
\[
\Exp{\int_{t_0}^s B_{i_1j_1}(u)\,dW_u^{j_1}\int_{t_0}^s B_{i_2j_1}(u)\,dW_u^{j_1}}\Exp{\int_{t_0}^s B_{i_1k_1}(u)\,dW_u^{k_1}\int_{t_0}^s B_{i_2k_1}(u)\,dW_u^{k_1}},
\]
\end{enumerate}

Because of the properties of the stochastic integral, $\int_{t_0}^s B_{i_1j}(u)\,dW_u^{j}=\mathcal{N}(0,\left(\int_{t_0}^s B_{i_1j}^2(u)\,du \right)^{1/2})$ hence, because of the properties of the Gaussian integrals
\[
 \Exp{\left(\int_{t_0}^s B_{i_1j}(u)\,dW_u^{j}\right)^4}=k \left(\int_{t_0}^s B_{i_1j}^2(u)\,du \right)^2,
\]
for some positive constant $k$. This takes care of the terms of the form $T_1$. Next we have
\[
 \Exp{\left(\int_{t_0}^s B_{i_1j_1}(u)\,dW_u^{j_1}\right)^2}=\int_{t_0}^s B_{i_1j_1}^2(u)\,du,
\]
which takes care of the terms of the form $T_2$. Finally we have, because of the properties of the Itô integral
\[
\Exp{\int_{t_0}^s B_{i_1j_1}(u)\,dW_u^{j_1}\int_{t_0}^s B_{i_2j_1}(u)\,dW_u^{j_1}}=\int_{t_0}^s B_{i_1j_1}(u)B_{i_2j_1}(u)\,du,
\]
which takes care of the terms of the form $T_3$.

This shows that the term  $\Exp{\left\| \int_s^t \Phi_L(t,u) \mla(u)\,d\mW_u \right\|^4}$ in \eqref{eq:kolmo} is of the order of $(t-s)^{1+a}$ where $a \geq 1$. Therefore we have
\[
 \Exp{\|\mathcal{F}_1^{(n)}(X)_t-\mathcal{F}_1^{(n)}(X)_s\|^4} \leq C |t-s|^{1+a},\quad a \geq 1
\]
for all $s$, $t$ in $[t_0,T]$, where $C$ is a constant independent of $t,\,s$. According to Kolmogorov criterion for tightness, the sequence of processes $\left\{\mathcal{F}_1^{(n)}(X)\right\}_{n=0}^\infty$ is uniformly tight.

The proof for $\mathcal{F}_k$, $k > 1$ is similar.
\end{proof}

\subsubsection{The mean and covariance sequences are Cauchy sequences}
\noindent
Let us note $\mu^n(t)$ (respectively $C^n(t,s)$) the mean (respectively the covariance matrix) function of $X_n=\mathcal{F}_k(X_{n-1})$, $n \geq 1$. 
We have:
\begin{multline}\label{eq:mean}
\mu^n(t)= \Phi_L(t,t_0) \mu^{Z_0}+\int_{t_0}^t \Phi_L(t,u) \left(\Exp{\widetilde{\mU}^{X_n}_u}+\tilde{\mI}(u)\right) \,du=\\
\Phi_L(t,t_0) \mu^{Z_0}+\\
\int_{t_0}^t \Phi_L(t,u) \left(\left[0_P^T,\cdots,0_P^T, \left[\sum_\beta \Jbab m_\beta^{X_n}(u) \right]_{\alpha=1,\cdots,P}\right]^T+\tilde{\mI}(u)\right) \,du,
\end{multline}
where $m_\beta^{X_n}(u)$ is given by equation \eqref{eq:mean-interaction}.
Similarly we have
\begin{multline}\label{eq:cov}
 C^{n+1}(t,s)=\Phi_L(t,t_0) \Sigma^{Z_0} \Phi_L(s,t_0)^T + \int_{t_0}^{t \wedge s} \Phi_L(t,u) \mla(u) \mla(u)^T \Phi_L(s,u)^T\,du+\\
\int_{t_0}^t \int_{t_0}^s \Phi_L(t,u) {\rm Cov}\left(\widetilde{\mU}^{X_n}_u,\widetilde{\mU}^{X_n}_v\right) \Phi_L(s,v)^T\,du\,dv
\end{multline}
\noindent
Note that the $kP \times kP$ covariance matrix ${\rm Cov}\left(\widetilde{\mU}^{X_n}_u,\widetilde{\mU}^{X_n}_v\right)$ has only one nonzero $P \times P$ block:
\begin{equation}\label{eq:cov-detail}
{\rm Cov}\left(\widetilde{\mU}^{X_n}_u,\widetilde{\mU}^{X_n}_v\right)_{kk}= {\rm Cov}\left(\mU^{X_n}_u \cdot \mone,\mU^{X_n}_v \cdot \mone\right),
\end{equation}
According to definition \ref{def:effectiveInteractionProcess} we have
\[
{\rm Cov}\left(\mU^{X_n}_u \cdot \mone,\mU^{X_n}_v \cdot \mone\right)={\rm diag}\left(\sum_\beta \sigma_{\alpha \beta}^2 \Delta_\beta^{X_n}(u,v)\right),
\]
where $\Delta_\beta^{X_n}(u,v)$ is given by equation \eqref{eq:covariance-interaction} and $Dx$ is defined in proposition \ref{prop:muCexpressions}.

In order to prove our main result, that the two sequences of functions $(\mu^n)$ and $(C^n)$ are uniformly convergent,
we require the following four lemmas that we state without proofs, the proofs being found in appendixes \ref{appendix:max} to \ref{appendix:integral}.
The first lemma gives a uniform (i.e. independent of $n \geq 2$ and $\alpha=1,\cdots,kP$) strictly positive lowerbound for $C^n_{\alpha\alpha}(t,t)$. In what follows we
use the following notation: Let $C$ be a symmetric positive definite matrix, we note $\lambda_{\rm min}^C$ its smallest eigenvalue.
\begin{lemma}\label{lemma:max}
 The following uppperbounds are valid for all $n \geq 1$ and all $s,\,t\,\in [t_0,T]$.
\[
 \norm{\mu^n(t)}_\infty \leq e^{k_L (T-t_0)}\left[ \norm{\Exp{Z_0}}_\infty+(\mu+I_{\rm max})(T-t_0)  \right]\overset{\rm def}{=}\mu_{\rm max},
\]
\[
 \norm{C^n(t,s)}_\infty \leq e^{(k_L+k_{L^T})(T-t_0)}\left[ \rho(\Sigma^{Z_0})+\lambda_{\rm max}^\Gamma (T-t_0)+\sigma_{\rm max}^2(T-t_0)^2\right]\overset{\rm def}{=}\Sigma_{\rm max},
\]
where $\mu$ and $\sigma_{\rm max}$ are defined in lemma \ref{lemma:upper}, $\lambda_{\rm max}^\Gamma$ is defined in \ref{assumption:all}
\end{lemma}

\begin{lemma}\label{lemma:lwb1}
For all $t \in [t_0,T]$ all $\alpha=1,\cdots,kP$, and $n \geq 1$, we have
\[
 C^n_{\alpha\alpha}(t,t) \geq \lambda_{\rm min} \lambda_{\rm min}^{\Sigma^{Z_0}}\overset{\rm def}{=}k_0 > 0,
\]
where $\lambda_{\rm min}$ is the smallest singular value of the positive symmetric definite matrix $\Phi_L(t,t_0)\Phi_L(t,t_0)^T$ for $t \in [t_0,T]$ and $\lambda_{\rm min}^{\Sigma^{Z_0}}$ is the smallest eigenvalue of the positive symmetric definite covariance matrix $\Sigma^{Z_0}$.
\end{lemma}
The second lemma also gives a uniform lowerbound for the expression
$C_{\alpha\alpha}^n(s,s)C_{\alpha\alpha}^n(t,t)-C_{\alpha\alpha}^n(t,s)^2$ which appears in the definition of $C^{n+1}$ through equations \eqref{eq:cov-detail} and \eqref{eq:covariance-interaction}. The crucial point is that this function is $O(|t-s|)$ which is central in the proof of the third lemma \ref{lemma:integral}.
\begin{lemma}\label{lemma:lwb2}
For all $\alpha=1,\cdots,kP$ and $n \geq 1$ the quantity $C_{\alpha\alpha}^n(s,s)C_{\alpha\alpha}^n(t,t)-C_{\alpha\alpha}^n(t,s)^2$ is lowerbounded by the positive symmetric function:
\[
 \theta(s,t)\overset{\rm def}{=} |t-s| \lambda_{\rm min}^2 \lambda_{\rm min}^{\Sigma^{Z_0}}  \lambda_{\rm min}^{\Gamma},
\]
where $\lambda_{\rm min}^{\Gamma}$ is the strictly positive lower bound, introduced in \ref{assumption:all}, on the singular values of the matrix $\mla(u)$ for $u \in [t_0,T]$.
\end{lemma}
The third lemma shows that an integral that appears in the proof of the uniform convergence of the sequences of functions $(\mu^n)$ and $(C^n)$ is upperbounded by the $n$th term of a convergent series.
\begin{lemma}\label{lemma:integral}
The $2n$-dimensional integral 
 \begin{multline*}
I_n=\int_{[t_0,t \vee s]^2} \rho_1(u_1,v_1)\Bigg(\int_{[t_0,u_1 \vee v_1]^2} \cdots \Bigg(\int_{[t_0,u_{n-2} \vee v_{n-2}]^2}\rho_{n-1}(u_{n-1},v_{n-1})\\
\Bigg(\int_{[t_0,u_{n-1} \vee v_{n-1}]^2} \rho_n(u_n,v_n) du_n dv_n\Bigg)du_{n-1} dv_{n-1}\Bigg)\cdots \Bigg) du_1 dv_1,
\end{multline*}
where the functions $\rho_i(u_i,v_i)$, $i=1,\cdots,n$ are either equal to 1 or to $1/\sqrt{\theta(u_i,v_i)}$ (the function $\theta$ is defined in lemma \ref{lemma:lwb2}), is upperbounded by $k^n/(n-1)!$ for some positive constant $k$.
\end{lemma}

With these lemmas in hand we prove  proposition \ref{prop:conv-Cmu}. The proof is technical but its idea is very simple. We find upperbounds for the matrix infinite norm of $C^{n+1}(t,s)-C^n(t,s)$ and the infinite norm of $\mu^{n+1}(t)-\mu^n(t)$ by applying the mean value theorem and lemmas \ref{lemma:lwb1} and \ref{lemma:lwb2} to the these norms. These upperbounds involve integrals of the infinite norms of $C^{n}(t,s)-C^{n-1}(t,s)$ and $\mu^{n}(t)-\mu^{n-1}(t)$ and, through lemma \ref{lemma:lwb2}, one over the square root of the function $\theta$. Proceeding recursively and using lemma \ref{lemma:integral}, one easily shows that the infinite norms of $C^{n+1}-C^n$ and $\mu^{n+1}-\mu^n$ are upperbounded by the $n$th term of a convergent series from which it follows that the two sequences of functions are Cauchy sequences, hence convergent.
\begin{proposition}\label{prop:conv-Cmu}
The sequences of covariance matrix functions $C^n(t,s)$ and of mean functions $\mu^n(t)$, $s,\,t$ in $[t_0,T]$ are Cauchy sequences for the uniform norms.
\end{proposition}
\begin{proof}
 We have
\begin{multline*}
 C^{n+1}(t,s)-C^n(t,s)=\int_{t_0}^t \int_{t_0}^s \Phi_L(t,u) \Big({\rm Cov}\left(\widetilde{\mU}^{X_n}_u,\widetilde{\mU}^{X_n}_v \right)- \\
{\rm Cov}\left(\widetilde{\mU}^{X_{n-1}}_u ,\widetilde{\mU}^{X_{n-1}}_v\right)\Big)\Phi_L(s,v)^T\,du\,dv.
\end{multline*}
We take the infinite matrix norm of both sides of this equality and use the upperbounds $\norm{\Phi_L(t,u)}_\infty \leq e^{\norm{\mL}_\infty(T-t_0)}=k_L$
and $\norm{\Phi_L(t,u)^T}_\infty \leq e^{\norm{\mL^T}_\infty(T-t_0)}=k_{L^T}$ (see appendix \ref{appendix:TOop}) to obtain\footnote{The notation $\norm{\ }^v$ is introduced in appendix \ref{appendix:OpNorms}.}
\begin{align}
 \nonumber &\norm{C^{n+1}(t,s)-C^n(t,s)}_\infty \leq k_L k_{L^T} \int_{t_0}^t \int_{t_0}^s \Big\|{\rm Cov}\left(\widetilde{\mU}^{X_n}_u ,\widetilde{\mU}^{X_n}_v \right)-
{\rm Cov}\left(\widetilde{\mU}^{X_{n-1}}_u ,\widetilde{\mU}^{X_{n-1}}_v\right)\Big\|^v_\infty\,du\,dv\\
 \label{eq:upper1} &\qquad = k_L k_{L^T} \int_{t_0}^t \int_{t_0}^s \Big\|{\rm Cov}\left(\mU^{X_n}_u \cdot \mone,\mU^{X_n}_v \cdot \mone\right)-
{\rm Cov}\left(\mU^{X_{n-1}}_u \cdot \mone,\mU^{X_{n-1}}_v \cdot \mone\right)\Big\|^v_\infty\,du\,dv.
\end{align}
According to equations \eqref{eq:covariance-interaction} we are led to consider the difference $A_n-A_{n-1}$, where:
\begin{multline*}
 A_n\eqdef\\
S_{ \beta}\Big(\frac{\sqrt{C_{\beta \beta}^n(u,u)C_{\beta \beta}^n(v,v)-C_{\beta \beta}^n(u,v)^2}}{\sqrt{C_{\beta \beta}^n(u,u)}}x+
\frac{C_{\beta \beta}^n(u,v)}{\sqrt{C_{\beta \beta}^n(u,u)}}y+\mu^n_\beta(v) \Big)
S_{\beta}\left(y \sqrt{C_{\beta \beta}^n(u,u)}+ \mu^n_\beta(u)\right) \\
\eqdef
S_{ \beta}\left[\mathcal{P}_\beta^n(u,v) x+\mathcal{S}_\beta^n(u,v) y+\mu^n_\beta(v)\right]
S_{ \beta}\left[\mathcal{T}^n_\beta(u) y+\mu^n_\beta(u)\right].
\end{multline*}
We write next:
\begin{multline*}
A_n-A_{n-1}=S_{ \beta}\left[\mathcal{P}_\beta^n(u,v) x+\mathcal{S}_\beta^n(u,v) y+\mu^n_\beta(v)\right]\\
\left(S_{ \beta}\left[\mathcal{T}^n_\beta(u) y+\mu^n_\beta(u)\right]-S_{ \beta}\left[\mathcal{T}^{n-1}_\beta(u) y+\mu^{n-1}_\beta(u)\right]\right)+\\
S_{ \beta}\left[\mathcal{T}^{n-1}_\beta(u) y+\mu^{n-1}_\beta(u)\right]\\
\left(S_{ \beta}\left[\mathcal{P}_\beta^n(u,v) x+\mathcal{S}_\beta^n(u,v) y+\mu^n_\beta(v)\right]-
S_{ \beta}
\left[\mathcal{P}_\beta^{n-1}(u,v) x+\mathcal{S}_\beta^{n-1}(u,v) y+\mu^{n-1}_\beta(v) \right]\right).
\end{multline*}
The mean value theorem yields:
\begin{multline*}
 \mid A_n-A_{n-1}\mid  \leq \norm{S_{\beta}}_\infty \norm{S_{ \beta}'}_\infty\Big(
\mid x\mid  \, \mid \mathcal{P}_\beta^n(u,v)-\mathcal{P}_\beta^{n-1}(u,v)\mid +\\ \mid y\mid \,\mid \mathcal{S}_\beta^n(u,v)- \mathcal{S}_\beta^{n-1}(u,v)\mid +
\mid \mu^n_\beta(v) -\mu^{n-1}_\beta(v)\mid  +\mid y\mid \,\mid \mathcal{T}_\beta^n(u)-\mathcal{T}^{n-1}_\beta(u)\mid +\\
\mid \mu^n_\beta(u)-\mu^{n-1}_\beta(u)\mid \Big).
\end{multline*}
Using the fact that $\int_{-\infty}^\infty \mid x\mid \, Dx=\sqrt{\frac{2}{\pi}}$,  we obtain:
\begin{multline*}
 \norm{C^{n+1}(t,s)-C^n(t,s)}_\infty \leq k_L k_{L^T} k_C  \Bigg( \sqrt{\frac{2}{\pi}}\int_{t_0}^t \int_{t_0}^s \norm{\mathcal{P}^n(u,v)-\mathcal{P}^{n-1}(u,v)}_\infty dudv+\\
\sqrt{\frac{2}{\pi}}\int_{t_0}^t \int_{t_0}^s \norm{\mathcal{S}^n(u,v)-\mathcal{S}^{n-1}(u,v)}_\infty dudv+\\
(t-t_0) \int_{t_0}^s \norm{\mu^n(v) -\mu^{n-1}(v)}_\infty dv + (s-t_0) \int_{t_0}^t \norm{\mu^n(u) -\mu^{n-1}(u)}_\infty du+\\
\sqrt{\frac{2}{\pi}}(s-t_0) \int_{t_0}^t \norm{\mathcal{T}^n(u)-\mathcal{T}^{n-1}(u)}_\infty du \Bigg),
\end{multline*}
where the constants $k_L$ and $k_{L^T}$ are defined in appendix \ref{appendix:TOop} and 
\begin{equation}\label{eq:kC}
 k_C \overset{\rm def}{=}\max_\alpha \sum_\beta \sigma_{\alpha \beta}^2 \norm{S_{\beta}}_\infty \norm{S_{\beta}'}_\infty.
\end{equation}

A similar process applied to the mean values yields:
\begin{multline*}
 \norm{\mu^{n+1} (t)-\mu^n(t)}_\infty \leq k_L \mu \Big(\int_{t_0}^t \norm{\mathcal{T}^n(u)-\mathcal{T}^{n-1}(u)}_\infty du+\\
\int_{t_0}^t \norm{\mu^n(u) -\mu^{n-1}(u)}_\infty du \Big),
\end{multline*}
where $\mu$ is defined in lemma \ref{lemma:upper}.
We now use the mean value theorem and lemmas \ref{lemma:lwb1} and \ref{lemma:lwb2} 
to find upperbounds for $\norm{\mathcal{P}^n(u,v)-\mathcal{P}^{n-1}(u,v)}_\infty$,
$\norm{\mathcal{S}^n(u,v)-\mathcal{S}^{n-1}(u,v)}_\infty$ and $\norm{\mathcal{T}^n(u)-\mathcal{T}^{n-1}(u)}_\infty$. We have
\begin{multline*}
 |\mathcal{T}^n_\beta(u)-\mathcal{T}^{n-1}_\beta(u)|=\left|\sqrt{C^n_{\beta \beta}(u,u)}-\sqrt{C^{n-1}_{\beta \beta}(u,u)}\right| \leq \frac{1}{2\sqrt{k_0}} \left|C^n_{\beta \beta}(u,u)- C^{n-1}_{\beta \beta}(u,u)\right|,
\end{multline*}
where $k_0$ is defined in lemma \ref{lemma:lwb1}. Hence:
\[
 \norm{\mathcal{T}^n(u)-\mathcal{T}^{n-1}(u)}_\infty \leq \frac{1}{2\sqrt{k_0}} \norm{C^n(u,u)-C^{n-1}(u,u)}_\infty.
\]
Along the same lines we can show easily that:
\begin{multline*}
 \norm{\mathcal{S}^n(u,v)-\mathcal{S}^{n-1}(u,v)}_\infty \leq k \Big( \norm{C^n(u,v)-C^{n-1}(u,v)}_\infty+ \\
\norm{C^n(u,u)-C^{n-1}(u,u)}_\infty \Big),
\end{multline*}
and that:
\begin{multline*}
 \norm{\mathcal{P}^n(u,v)-\mathcal{P}^{n-1}(u,v)}_\infty \leq \frac{k}{\sqrt{\theta(u,v)}} \Big(\norm{C^n(u,v)-C^{n-1}(u,v)}_\infty+ \\
\norm{C^n(u,u)-C^{n-1}(u,u)}_\infty+\norm{C^n(v,v)-C^{n-1}(v,v)}_\infty
\Big),
\end{multline*}
where $\theta(u,v)$ is defined in lemma \ref{lemma:lwb2}.
Grouping terms together and using the fact that all integrated functions are positive, we write:
\begin{multline}\label{eq:upper2}
 \norm{C^{n+1}(t,s)-C^n(t,s)}_\infty \leq \\
k \Bigg( \int_{[t_0,t \vee s]^2} \frac{1}{\sqrt{\theta(u,v)}} \norm{C^n(u,v)-C^{n-1}(u,v)}_\infty\,dudv+ \\
\int_{[t_0,t \vee s]^2} \frac{1}{\sqrt{\theta(u,v)}} \norm{C^n(u,u)-C^{n-1}(u,u)}_\infty\,dudv+\\
\int_{[t_0,t \vee s]^2} \norm{C^n(u,v)-C^{n-1}(u,v)}_\infty\,dudv+ \\
\int_{[t_0,t \vee s]^2} \norm{C^n(u,u)-C^{n-1}(u,u)}_\infty\,dudv+\\
\int_{[t_0,t \vee s]^2} \norm{\mu^n(u) -\mu^{n-1}(u)}_\infty\,dudv\Bigg).
\end{multline}
Note that, because of lemma \ref{lemma:lwb1}, all integrals are well-defined. Regarding the mean functions, we write:
\begin{multline}\label{eq:upper3}
\norm{\mu^{n+1} (t)-\mu^n(t)}_\infty \leq k \Bigg( \int_{[t_0,t \vee s]^2} \norm{C^n(u,u)-C^{n-1}(u,u)}_\infty\,dudv+\\
\int_{[t_0,t \vee s]^2} \norm{\mu^n(u) -\mu^{n-1}(u)}_\infty\,dudv\Bigg).
\end{multline}

Proceeding recursively until we reach $C^0$ and $\mu^0$ we obtain an upperbound for $\norm{C^{n+1}(t,s)-C^n(t,s)}_\infty$ (respectively for $\norm{\mu^{n+1}(t)-\mu^n(t)}_\infty$) which is the sum
of less than $5^n$ terms each one being the product of $k$ raised to a power less than or equal to $n$, times  $2\mu_{\rm max}$ or $2\Sigma_{\rm max}$ (upperbounds for the norms of the mean vector and the covariance matrix defined in lemma \ref{lemma:max}), times a $2n$-dimensional integral $I_n$ given by
\begin{multline*}
\int_{[t_0,t \vee s]^2} \rho_1(u_1,v_1)\Bigg(\int_{[t_0,u_1 \vee v_1]^2} \cdots \Bigg(\int_{[t_0,u_{n-2} \vee v_{n-2}]^2}\rho_{n-1}(u_{n-1},v_{n-1})\\
\Bigg(\int_{[t_0,u_{n-1} \vee v_{n-1}]^2} \rho_n(u_n,v_n) du_n dv_n\Bigg)du_{n-1} dv_{n-1}\Bigg)\cdots \Bigg) du_1 dv_1,
\end{multline*}
where the functions $\rho_i(u_i,v_i)$, $i=1,\cdots,n$ are either equal to 1 or to $1/\sqrt{\theta(u_i,v_i)}$.
According to lemma \ref{lemma:integral}, this integral is of the order of some positive constant raised to the power $n$ divided by $(n-1)!$. Hence the sum is less than some positive constant $k$ raised to the power $n$ divided by $(n-1)!$. By taking the supremum with respect to $t$ and $s$ in $[t_0,T]$ we obtain the same result for $\norm{C^{n+1}-C^n}_\infty$ (respectively for $\norm{\mu^{n+1}-\mu^n}_\infty$). Since the series $\sum_{n \geq 1} \frac{k^n}{n!}$ is convergent, this implies that $\norm{C^{n+p}-C^n}_\infty$ (respectively $\norm{\mu^{n+p}-\mu^n}_\infty$) can be made arbitrarily small for large $n$ and $p$ and the sequence $C^n$ (respectively $\mu^n$) is a Cauchy sequence.

\end{proof}

\subsubsection{Existence and uniqueness of a solution of the mean-field equations}
It is now easy to prove our main result, that the mean-field equations \eqref{eq:IntegratedGMM} or equivalently \eqref{eq:MFEwithMatrix} are well-posed, i.e. have a unique solution.

\begin{theorem} \label{theo:ExistenceUniqueness}
 For any nondegenerate $kP$-dimensional Gaussian random variable $Z_0$, independent of the Brownian, and any initial process $X$ such that $X(t_0)=Z_0$, the map $\mathcal{F}_k$ has a unique fixed point in $\procP{kP}$ towards which the sequence $\{\mathcal{F}_k^{(n)}(X)\}_{n=1}^\infty$ of Gaussian processes converges in law.
\end{theorem}

\begin{proof}
Since $C([t_0,T], \R^{kP})$ (respectively $C([t_0,T]^2, \R^{kP \times kP})$) is a Banach space for the uniform norm, the  Cauchy sequence $\mu^n$ (respectively $C^n$) of proposition \ref{prop:conv-Cmu} converges to an element $\mu$ of $C([t_0,T], \R^{kP})$ (respectively an element $C$ of $C([t_0,T]^2, \R^{kP \times kP})$).
Therefore, according to theorem \ref{theo:convergence}, the sequence $\{\mathcal{F}_k^{(n)}(X)\}_{n=0}^\infty$ of Gaussian processes converges in law toward the Gaussian
process $Y$ with mean function $\mu$ and covariance function $C$. This process is clearly a fixed point of $\mathcal{F}_k$. 

Hence we know that there there exists at least one fixed point for the map $\mathcal{F}_k$. Assume there exist two distinct fixed points $Y_1$ and $Y_2$ of $\mathcal{F}_k$ with mean functions $\mu_i$ and covariance functions $C_i$, $i=1,2$, with the same initial condition. Since for all $n \geq 1$ we have $\mathcal{F}_k^{(n)}(Y_i)=Y_i$, $i=1,2$, the proof of proposition  \ref{prop:conv-Cmu} shows that $\norm{\mu_1^n-\mu_2^n}_\infty$ (respectively $\norm{C_1^n-C^2_n}_\infty$) is upperbounded by the product of a positive number $a_n$ (respectively $b_n$) with $\norm{\mu_1-\mu_2}_\infty$ (respectively with $\norm{C_1-C_2}_\infty$). Since $\lim_{n \to \infty} a_n=\lim_{n \to \infty} b_n=0$ and $\mu_i^n=\mu_i$, $i=1,2$ (respectively $C_i^n=C_i$, $i=1,2$), this shows that $\mu_1=\mu_2$ and $C_1=C_2$, hence the two Gaussian processes $Y_1$ and $Y_2$ are indistinguishable.

\end{proof}

\subsection*{Conclusion}
We have proved that for any non degenerate Gaussian initial condition $Z_0$ there exists a unique solution of the mean-field equations.
The proof of theorem \ref{theo:ExistenceUniqueness} is constructive, and hence provides a way for computing the solution of the mean-field equations by iterating the map $\mathcal{F}_k$ defined in \ref{def:F1Fk}, starting from any initial process $X$ satisfying $X(t_0)=Z_0$, for instance a Gaussian process such as an Ornstein-Uhlenbeck process. We build upon these facts in section \ref{sect:numerics}.
 
 Note that the existence and uniqueness is true whatever the initial time $t_0$ and the final time $T$.

\section{Existence and uniqueness of stationary solutions}\label{sect:station}
So far, we have investigated the existence and uniqueness of solutions of the mean-field equation for a given initial condition. 
We are now interested in investigating stationary solutions, which allow for some simplifications of the formalism.

A stationary solution is a solution whose probability distribution does not change under the flow of the equation. These solutions have been already investigated by several authors 
(see \cite{sompolinsky-crisanti-etal:88,brunel-hakim:99}). We propose a new framework to study and simulate these processes. Indeed we show in this section that under a certain contraction condition there exists a unique solution to the stationary problem. As in the previous section our proof is constructive and  provides a way to simulate the solutions. 

\begin{remark}
 The long-time mean-field description of a network is still a great endeavor in mathematics and statistical physics. In this section we formally take the mean-field equation we obtained and let $t_0 \to -\infty$. This way we obtain an equation which is the limit of the mean-field equation when $t_0 \to -\infty$. It means that we consider first the limit $N\to \infty$ and then $t_0\to -\infty$. These two limits do not necessarily commute and there are known examples, for instance in spin glasses, where they do not.
\end{remark}

It is clear that in order to get stationary solutions, the stochastic system has to be autonomous. More precisely, we modify assumptions \ref{assumption:all} as follows
\begin{Assump}\label{assumption:all1}\ \\
\begin{enumerate}
	\item The matrixes $\mL(t)$ and $\mla(t)$,  the input currents $\mathbf{I}(t)$ do not depend upon $t$. 
	\item The real parts of the eigenvalues of $\mL$ are negative: 
\begin{equation}\label{eq:eigenvalues}
 {\rm Re}(\lambda) < -\lambda_L \quad \lambda_L > 0
\end{equation}
for all eigenvalues $\lambda$ of $\mL$.
	\item The matrix $\mla$ has full rank.
\end{enumerate}
\end{Assump}
Under assumption a of \ref{assumption:all1}, the resolvent $\Phi_L(t,s)$ is equal to $e^{\mL(t-s)}$. Under assumption 2 we only consider first-order system since otherwise the matrix $\mL$ has eigenvalues equal to 0. 
We now prove the following proposition.
\begin{proposition}\label{prop:PhiStationary}
Under the previous assumptions we have:
\begin{enumerate}
 \item \begin{equation*}
 \begin{cases}
  \lim\limits_{t_0 \to -\infty} e^{\mL(t-t_0)} =0, \\
   \int_{-\infty}^t \norm{e^{\mL(t-s)}}\,ds = \int_0^\infty \norm{e^{\mL u}}_\infty \,du \overset{{\rm def}}{=} M_L <\infty, \\
   \int_{-\infty}^t \norm{e^{\mL^T(t-s)}}_{\infty}\,ds = \int_0^\infty \norm{e^{\mL^T u}}_\infty \,du \overset{{\rm def}}{=} M_{L^T} <\infty,
 \end{cases}
\end{equation*}
\item the process $Y_t^{t_0}=\int_{t_0}^t e^{\mL(t-s)} \mla \cdot d\mW_s$ is well-defined, Gaussian and stationary 
when $t_0 \to -\infty$.
\end{enumerate}

\end{proposition}

\begin{proof}
The first point property follows from the fact that ${\rm Re}(\lambda) < -\lambda_L$ for all eigenvalues $\lambda$ of $\mL$. This assumption also implies that there exists a norm on $\R^{P}$ such that
\[
 \norm{e^{\mL t}} \leq e^{-\lambda_L t}\ \forall t \geq 0,
\]
and hence
\begin{equation}\label{eq:expbound}
\norm{e^{\mL t}}_\infty \leq ke^{-\lambda_L t}\ \forall t \geq 0, 
\end{equation}
for some positive constant $k$. This implies the remaining two properties.

We now address the second point of the property. The stochastic integral $Y_t^{t_0} = \int_{t_0}^{t} e^{\mL(t-s)} \mla \cdot d\mW_s$ is well-defined 
$\forall t \leq T$ and is Gaussian with zero-mean. Its covariance matrix reads:
\[
 \Sigma^{Y_t^{t_0}Y_{t'}^{t_0}}=\int_{t_0}^{t \wedge t'} e^{\mL(t-s)} \mla \mla^T e^{\mL^T(t'-s)}\,ds.
\]
Let us assume for instance that $t'<t$ and perform the change of variable $u=t-s$ to obtain

\[
 \Sigma^{Y_t^{t_0}Y_{t'}^{t_0}}=\left(\int_{t-t'}^{t-t_0} e^{\mL u}\mla \mla^T  e^{\mL^T u}\,du \right) e^{\mL^T(t'-t)}.
\]

Under the previous assumptions this matrix integral is defined when $t_0 \to -\infty$ (dominated convergence theorem) and we have
\begin{equation}\label{eq:sigmaMinusInfinity}
 \lim\limits_{t_0\to -\infty} \Sigma^{Y_t^{t_0}Y_{t'}^{t_0}} \eqdef \Sigma^{Y_t^{-\infty}Y_{t'}^{-\infty}}=\left(\int_{t-t'}^{+\infty} e^{\mL u}\mla \mla^T e^{\mL^T u}\,du\right) e^{\mL^T(t'-t)} ,
\end{equation}
which is a well defined function of $t'-t$.
\end{proof}

The second point of propostion \ref{prop:PhiStationary} guarantees the existence of process
\[
 \mX_0(t)=\int_{-\infty}^t e^{\mL (t-s)} \mla \cdot d\mW_s.
\]
as the limit of the processes $Y_t^{t_0}$ when $t_0\to -\infty$. This process is a stationary distribution of the equation:
\begin{equation}\label{eq:stationary}
d\mathbf{X}_0(t) = \mL \cdot \mathbf{X}_0(t)\,dt  + \mla \cdot d\mathbf{W}_t,
\end{equation}
it is Gaussian, of mean $\Exp{\mathbf{X}_0(t)}=0$ and of covariance matrix $\Sigma^0$ is equal to $\Sigma^{Y_t^{-\infty} Y_{t}^{-\infty}}$ defined by equation \eqref{eq:sigmaMinusInfinity} and which is independent of $t$.\\

We call \textit{long term mean-field equation} (LTMFE) the implicit equation:

\begin{equation}\label{eq:LTMFE}
 \VV(t) = \int_{-\infty}^t e^{\mL(t-s)} \Big( \mathbf{U}^{\VV}_s \cdot \mone+ \mathbf{I} \Big) \, ds + \mathbf{X}_0(t)
\end{equation}

\noindent where $\mathbf{X}_0$ is the stationary process defined by equation \eqref{eq:stationary} and where $\mathbf{U}^{\VV}(t)$ is the effective interaction process introduced previously. 

We next define the long term function $\Fs: \prociP{P} \to \prociP{P}$: 

 \begin{equation*}
  \Fs(\mathbf{X})_t = \int_{-\infty}^t e^{\mL(t-s)} \Big ( \mathbf{U}^X _s\cdot \mathbf{1} + \mathbf{I}\Big) \, ds + \mathbf{X}_0(t).
 \end{equation*}

\medskip

\begin{proposition}
 The function $\Fs$ is well defined on $\prociP{P}$.
\end{proposition}

\begin{proof}
 We have already seen that the process $\mathbf{X}_0$ is well defined. The term $\int_{-\infty}^t e^{\mL(t-s)} \mathbf{I} \, ds=\left(\int_{-\infty}^t e^{\mL(t-s)}\,ds\right)\,\mathbf{I}$ is also well defined because of the assumptions on $\mL$.
 
 Let $X$ be a given process in $\prociP{P}$. To prove the proposition we just have to ensure that the Gaussian process $\int_{-\infty}^t e^{\mL(t-s)} \mathbf{U}^X_s\cdot \mone\, ds$ is well defined. This results from the contraction assumption on $\mL$ and the fact that the functions $S_{\beta}$ are bounded. We decompose this process into a ``long memory'' term $\int_{-\infty}^0 e^{\mL(t-s)} \mathbf{U}^X_s\cdot \mone \,ds$ and the interaction term from time $t=0$, namely $\int_0^t e^{\mL(t-s)} \mathbf{U}^X_s\cdot \mone \, ds$. This latter term is clearly well defined. We show that the memory term is also well defined as a Gaussian random variable.\\

We write this term  $e^{\mL t}\int_{-\infty}^0 e^{-\mL s} \mathbf{U}^X_s \cdot \mone \, ds$ and consider the second factor. This random variable is Gaussian, its mean reads $\int_0^\infty e^{\mL s} \mu^{\mU^X_{-s}} \cdot \mone \,ds$ where 
\[
 \mu^{\mU^X_{-s}}=
                                             \left(\sum_{\beta=1}^P \Jbab \Exp{S_{\beta}(X_{\beta}(-s))} +I_{\alpha} \right)_{\alpha=1\ldots P}
\]
The integral defining the mean is well-defined because of \eqref{eq:expbound} and the fact that the functions $S_{\beta}$ are bounded. A similar reasoning shows that the corresponding covariance matrix
is well-defined. Hence the Gaussian process $\int_{-\infty}^0  e^{-\mL s} \mathbf{U}^X _s\cdot \mone ds$ is well defined, and hence for any process $X \in \prociP{P}$, the process $\Fs(X)$ is well defined.
\end{proof}
\noindent
We can now prove the following proposition.
\begin{proposition}\label{prop:mean-cov}
The mean vectors and the covariance matrices of the processes in the image of $\Fs$ are bounded. 
\end{proposition}
\begin{proof}
Indeed, since $\Exp{X_0(t)}=0$,  we have:
\[
 \norm{ \Exp{\Fs(\mathbf{X})_t} }_{\infty} = 
 \norm{ \int_{-\infty}^t e^{\mL(t-s)} \mu^{\mU^X_{s}} ds }_{\infty} 
  \leq M_L (\mu+\norm{I}_\infty) \overset{{\rm def}}{=} \mu_{LT}.
\]
In a similar fashion the covariance matrices of the processes in the image of $\Fs$ are bounded. Indeed we have:
\begin{multline*}
 \Exp{\Fs(\mX)_t \Fs(\mX)_t^T}=\Sigma^0+\\
\int_{-\infty}^t \int_{-\infty}^t e^{\mL(t-s_1)} {\rm diag}\left(\sum_\beta \sigma_{\alpha \beta}^2 \Exp{S_{\beta}(X_\beta(s_1))S_{\beta}(X_\beta(s_2))}\right)e^{\mL^T(t-s_2)}\,ds_1\,ds_2,
\end{multline*}
resulting in
\[
 \norm{\Exp{\Fs(\mX)_t \Fs(\mX)_t^T}}_\infty \leq \norm{\Sigma^0}_\infty+k^2\left(\frac{\sigma_{\rm max}}{\lambda_L}\right)^2 \overset{{\rm def}}{=} \Sigma_{LT}.
\]

\end{proof}

\begin{lemma}\label{lem:stationaryStab}
 The set of stationary processes is invariant by $\Fs$.
\end{lemma}

\begin{proof}
 Since the processes in the image of $\Fs$ are Gaussian processes, one just needs to check that the mean of the process is constant in time and that its covariance matrix  $C(s,t)$ only depends on $t-s$. 

 Let $Z$ be a stationary process and $Y=\Fs(Z)$. We denote by $\mu^Z_{\alpha}$ the mean of the process $Z_{\alpha}(t)$ and by $C^Z_{\alpha}(t-s)$ its covariance function. The mean of the process $U^{Z}_{\alpha\beta}$ reads: 
\[
m_{\alpha,\beta}^Z(t) = \Exp{S_{\beta}(Z_\beta(t))}=\frac 1 {\sqrt{2\pi C^Z_{\beta}(0)}} \int_{\R} S_{\beta}(x) e^{\frac{(x-\mu^Z_\beta)^2}{2 C^Z_{\beta}(0)}}dx
\]
\noindent and hence does not depends on time. We note $\mu^Z$ the mean vector of the stationary process $\mU^Z \cdot \mone$.

 Similarly, its covariance function reads:

\begin{multline*}
\Delta_{\alpha\beta}^Z(t,s)=\Exp{S_{\beta}(Z_\beta(t))S_{\beta}(Z_\beta(s))}=\\
 \int_{\R^2} S_{\beta}(x)S_{\beta}(y) 
\exp\left (-\displaystyle\frac{1}{2}\dbinom{x-\mu^Z_{\beta}}{y-\mu^Z_{\beta}}^T \left ( \begin{array}{cc} C^Z_\beta(0) & C^Z_\beta(t-s)\\
 C^Z_\beta(t-s) & C^Z_\beta(0) \end{array} \right)^{-1} \dbinom{x-\mu^Z_{\beta}}{y-\mu^Z_{\beta}}  \right)\,dx\,dy
\end{multline*}
which is clearly a function, noted $\Delta^Z_{\alpha\beta}(t-s)$, of $t-s$. Hence $\mU^Z \cdot \mone$ is stationary and we denote by $C^{U^Z}(t-s)$ its covariance function. \\
 
It follows that the mean of $Y_t$ reads: 
\begin{align*}
 \mu^Y(t) &= \Exp{\Fs(Z)_t} \\
 & = \Exp{X_0(t)} + \Exp{\int_{-\infty}^t e^{\mL(t-s)} \left( \mI + \mU_s^Z \cdot \mone\right) \, ds} \\
 & = \int_{-\infty}^t e^{\mL(t-s)} \left( \mI + \Exp{\mU_s^Z \cdot \mone} \right) \, ds\\ 
 & = \left(\int_{-\infty}^0 e^{\mL u} \, du\right) \left( \mI + \mu^Z\right) 
\end{align*}
 Since we proved that $\Exp{\mU_s^Z \cdot \mone}=\mu^Z$ was not a function of $s$. 

Similarly, we compute the covariance function and check that it can be written as a function of $(t-s)$. Indeed, it reads:
\begin{align*}
 C^Y(t,s) &= \int_{-\infty}^t\int_{-\infty}^s e^{\mL(t-u)} {\rm Cov}(\mU^Z_u \cdot \mone, \mU^Z_v \cdot \mone) e^{\mL^T(s-v)} \, du\,dv + {\rm Cov}(X_0(t),X_0(s)) \\
 & = \int_{-\infty}^0\int_{-\infty}^0 e^{\mL u} C^{U^Z}(t-s + (u-v) ) e^{\mL^T v} \, du\,dv + {\rm Cov}(X_0(t),X_0(s))
\end{align*}

\noindent since the process $X_0$ is stationary. $C^Y(t,s)$ is clearly a function of $t-s$.
Hence $Y$ is a stationary process, and the proposition is proved. 

\end{proof}

\begin{theorem}\label{theo:tightstationary}
The sequence of processes $\{\Fs^{(n)}(X)\}_{n=0}^\infty$ is uniformly tight.
\end{theorem}

\begin{proof}
 The proof is essentially the same as the proof of theorem \ref{theo:tight}, since we can write 
 \[\Fs (X)_t = e^{\mL t} \Fs(X)_0 + \int_0^t e^{\mL(t-s)} (\mU^{X}_s \cdot \mone+ \mathbf{I} )\, ds + \int_0^t e^{\mL(t-u)} \mla d\mW_s\]
 $\Fs(X)_t$ appears as the sum of the random variable $\Fs(X)_0$ and the  Gaussian process defined by $\int_0^t e^{\mL(t-s)} (\mU^{X}_s \cdot \mone+ \mathbf{I} )\, ds+\int_0^t e^{\mL(t-u)} \mla d\mW_s$ which is equal to $\mathcal{F}_k(X)_t$ defined in section \ref{sect:existUniq} for $t_0=0$. Therefore $\Fs^{(n)}(X)_t=\mathcal{F}_k^{(n)}(X)_t$ for $t>0$. We have proved the uniform tightness of the sequence of processes $\{\mathcal{F}_k^{(n)}(X)\}_{n=0}^\infty$ in theorem \ref{theo:tight}.
 Hence, according to Kolmogorov's criterion for tightness, we just have to prove that the sequence of Gaussian random variables:
 \[
\Fs^{(n)}(X)_0=\left\{\int_{-\infty}^0 \Phi_L(-u) (\mathbf{U}^{\Fs^{(n)}(X)}_u \cdot \mone + \mathbf{I} ) du + \mX_0(0)\right\}_{n \geq 0}
\]
is uniformly tight. Since it is a sequence of Gaussian random variables, it is sufficient to prove that their means are bounded and their covariance matrices upperbounded to obtain that for any $\varepsilon > 0$ there exists a compact $K_{\varepsilon}$ such that for any $n\in \N$, we have $\Pro(\Fs^{(n)}(X)_0 \in K_{\varepsilon}) \geq 1-\varepsilon$. This is a consequence of proposition \ref{prop:mean-cov} for the first random variable and of the definition
of $\mX_0$ for the second. By Kolmogorov's criterion the sequence of processes $\{\Fs^{(n)}(X)\}_{n=0}^\infty$ is uniformly tight

\end{proof}

In order to apply theorem \ref{theo:convergence} we need to prove that the sequences of covariance and mean functions are convergent. Unlike the case of $t_0$ finite, this is not always true. Indeed, to ensure existence and uniqueness of solutions in the stationary case, the parameters of the system have to satisfy a contraction condition, and proposition \ref{prop:conv-Cmu} extends as follows.

\begin{proposition}\label{prop:conv-Cmu-stat}
 If $\lambda_L$ defined in \eqref{eq:eigenvalues} satisfies the conditions \eqref{eq:cauchy} defined in the proof, depending upon $k_C$ (defined in \eqref{eq:kC}), $k_0$, $\mu_{LT}$ and $\Sigma_{LT}$ (defined in proposition \ref{prop:mean-cov})then the sequences of covariance matrix functions $C^n(t,s)$ and of mean functions $\mu^n(t)$, $s,\,t$ in $[t_0,T]$ are Cauchy sequences for the uniform norms.
\end{proposition}
\begin{proof}

 The proof follows that of proposition \ref{prop:conv-Cmu} with a few modifications that we indicate. In establishing the equation corresponding to \eqref{eq:upper1} we use the fact that $\norm{\Phi_L(t,u)}_\infty \leq k e^{-\lambda_L (t-u)}$ for some positive constant $k$ and all $u$, $t$, $u \leq t$. We therefore have:
\begin{multline*}
 \norm{C^{n+1}(t,s)-C^n(t,s)}_\infty \leq \\
k^2 e^{-\lambda_L(t+s)} \int_{-\infty}^t \int_{-\infty}^s e^{\lambda_L(u+v)}\Big\|{\rm Cov}\left(\mU^{X_n}_u ,\mU^{X_n}_v \right)-
{\rm Cov}\left(\mU^{X_{n-1}}_u ,\mU^{X_{n-1}}_v\right)\Big\|^v_\infty\,du\,dv
\end{multline*}
The rest of the proof proceeds the same way as in proposition \ref{prop:conv-Cmu}. Equations \eqref{eq:upper2} and \eqref{eq:upper3} become:
\begin{multline*}
 \norm{C^{n+1}(t,s)-C^n(t,s)}_\infty \leq \\
K e^{-\lambda_L(t + s)}\Bigg( \int_{[-\infty,t \vee s]^2} \frac{e^{\lambda_L(u+v)}}{\sqrt{f(u,v)}} \norm{C^n(u,v)-C^{n-1}(u,v)}_\infty\,dudv+ \\
\int_{[-\infty,t \vee s]^2} \frac{e^{\lambda_L(u+v)}}{\sqrt{f(u,v)}} \norm{C^n(u,u)-C^{n-1}(u,u)}_\infty\,dudv+\\
\int_{[-\infty,t \vee s]^2} e^{\lambda_L(u+v)} \norm{C^n(u,v)-C^{n-1}(u,v)}_\infty\,dudv+ \\
\int_{[-\infty,t \vee s]^2} e^{\lambda_L(u+v)} \norm{C^n(u,u)-C^{n-1}(u,u)}_\infty\,dudv+\\
\int_{[-\infty,t \vee s]^2} e^{\lambda_L(u+v)} \norm{\mu^n(u) -\mu^{n-1}(u)}_\infty\,dudv\Bigg),
\end{multline*}
and
\begin{multline*}
 \norm{\mu^{n+1} (t)-\mu^n(t)}_\infty \leq K e^{-\lambda_L(t + s)} \Bigg( \int_{[-\infty,t \vee s]^2} e^{\lambda_L(u+v)} \norm{C^n(u,u)-C^{n-1}(u,u)}_\infty\,dudv+\\
\int_{[-\infty,t \vee s]^2} e^{\lambda_L(u+v)} \norm{\mu^n(u) -\mu^{n-1}(u)}_\infty\,dudv\Bigg),
\end{multline*}
for some positive constant $K$, function of $k$, $k_C$ (defined in \eqref{eq:kC}), and $k_0$.

Proceeding recursively until we reach $C^0$ and $\mu^0$ we obtain an upperbound for $\norm{C^{n+1}(t,s)-C^n(t,s)}_\infty$ (respectively for $\norm{\mu^{n+1}(t)-\mu^n(t)}_\infty$) which is the sum of less than $5^n$ terms each one being the product of $K^n$, times  $2\mu_{LT}$ or $2\Sigma_{LT}$, times a $2n$-dimensional integral $I_n$ given by:
\begin{multline*}
\int_{[-\infty,t \vee s]^2} \rho_1(u_1,v_1)\Bigg(\int_{[-\infty,u_1 \vee v_1]^2} \cdots \Bigg(\int_{[-\infty,u_{n-2} \vee v_{n-2}]^2}\rho_{n-1}(u_{n-1},v_{n-1})\\
\Bigg(\int_{[-\infty,u_{n-1} \vee v_{n-1}]^2} e^{\lambda_L(u_n+v_n)} \rho_n(u_n,v_n) du_n dv_n\Bigg)du_{n-1} dv_{n-1}\Bigg)\cdots \Bigg) du_1 dv_1,
\end{multline*}
where the functions $\rho_i(u_i,v_i)$, $i=1,\cdots,n$ are either equal to 1 or to $1/\sqrt{\theta(u_i,v_i)}$.

It can be shown by straightforward calculation that each sub-integral contributes at most either 
\[
 \frac{K_0}{\lambda_L^2} \quad \text{if} \quad \rho_i=1 \quad \text{or} \quad \sqrt{\frac{\pi}{2}}\, \frac{K_0}{\lambda_L^{3/2}},
\]
in the other case. Hence we obtain factors of the type
\[
 K_0^n \left(\frac{1}{\lambda_L^2}\right)^p\,\left( \sqrt{\frac{\pi}{2}}\, \frac{1}{\lambda_L^{3/2}}\right)^{n-p}= \left( \sqrt{\frac{\pi}{2}}\right)^{n-p}\,\left(\frac{1}{\lambda_L}\right)^{(3n+p)/2} K_0^n,
\]
where $0 \leq p \leq n$.
If $\lambda_L < 1$, $(\lambda_L)^{(3n+p)/2} \geq \lambda_L^{2n}$ and else $(\lambda_L)^{(3n+p)/2} \geq \lambda_L^{3n/2}$. Since $\left(\sqrt{\frac{\pi}{2}}\right)^{n-p} \leq \left(\sqrt{\frac{\pi}{2}}\right)^{n}$ we obtain the two conditions
\begin{equation}\label{eq:cauchy}
 1 > \lambda_L^2 \geq 5 \sqrt{\frac{\pi}{2}} K K_0 \quad \text{or} \quad \left\{ \lambda_L^{3/2} > 5 \sqrt{\frac{\pi}{2}} K K_0 \quad \text{and} \quad \lambda_L \geq 1 \right\}
\end{equation}

\end{proof}

Putting all these results together we obtain the following theorem of existence and uniqueness of solutions for the long term mean-field equations:

\begin{theorem}\label{theo:StationaryExistUniq}
 Under the contraction conditions \eqref{eq:cauchy}, the function $\Fs$ has a unique solution in $\prociP{P}$ which is stationary, and for any  process $X$, the sequence $\{\Fs^{(n)}(X)\}_{n=0}^\infty$ of  Gaussian processes converges in law toward the unique fixed point of the function $\Fs$.
\end{theorem}

\begin{proof}
 The proof is essentially similar to the one of theorem \ref{theo:ExistenceUniqueness}. Indeed, the mean and the covariance matrixes converge since they are Cauchy sequences in the complete space of continuous functions equipped with the uniform norm. Using theorem \ref{theo:convergence}, we obtain that the sequence converges to a process $Y$ which is necessarily a fixed point of $\Fs$. Hence we have existence of a fixed point for $\Fs$. The uniqueness comes from the results obtained in the proof of proposition \ref{prop:conv-Cmu-stat}. 
 The limiting process is necessarily stationary. Indeed, let $X$ be a stationary process. Then for any $n\in \N$, the process $\Fs^{(n)}(X)$ will be stationary by the virtue of lemma \ref{lem:stationaryStab}, and hence so will be the limiting process which is the only fixed point of $\Fs$. 
 
\end{proof}

Hence in the stationary case, the existence and uniqueness of a solution is not always ensured. For instance if the leaks are too small (i.e. when the time constants of the decay of the membrane potentials are too long) then the sequence can diverge or have multiple fixed points.

\section{Numerical experiments}\label{sect:numerics}
\subsection{Simulation algorithm}\label{Snum}

Beyond the mathematical results, the framework that we introduced in the previous sections gives us a  
strategy to compute numerically the solutions of the dynamic mean-field equations. 
Indeed, we proved in section \ref{sect:existUniq} that 
under very moderate assumptions on the covariance matrix of the noise,
the iterations of the map $\mathcal{F}_k$ starting from any initial condition converge
to the solution of the mean-field equations. 

This convergence result gives us a direct way to compute numerically the solution of the mean-field equations. Since we are dealing with Gaussian processes, 
determining the law of the iterates of the map $\mathcal{F}_k$ amounts to computing its mean and covariance functions. In this section we describe our numerical algorithm in the case of the Simple Model of section \ref{ssect:SimpleModel}.

\sssu{Computing $\mathcal{F}_k$.}
Let $X$ be a $P$-dimensional Gaussian process of mean 
$\mu^X = (\muax(t))_{\alpha=1\ldots P}$ and 
covariance $C^X = (C^X_{\alpha\beta}(s,t))_{\alpha, \beta \in \{1\ldots P\}}$. We fix a time interval $[t_0=0,T]$ and denote by $Y$ the image of the process $X$ under $\mathcal{F}_1$. In the case of the simple model, the covariance of $Y$ is diagonal. Hence in this case the expressions we obtain in section \ref{sect:existUniq} simply read:

\begin{equation*}
\baR{lll} \muay(t) &=\muax(0)e^{-t/\ta}+\int_{0}^t e^{-(t-s)/\ta}(\sum_{\beta=1}^P \Jbab \Exp{S_{ \beta}(X_{\beta}(s))}+\Ia(s))ds \\
 & = \muax(0)e^{-t/\ta}+\int_{0}^t e^{-(t-s)/\ta} \Ia(s)ds  \\
 & \qquad + \sum_{\beta=1}^P \Jbab \int_{0}^t  e^{-(t-s)/\ta}
\int_{-\infty}^{+\infty} S_{\beta} \left ( x\sqrt{\vbx(s)} + \mub^X(s) \right) Dx ds.
\eaR
\end{equation*}
where we denoted $\vax(s)$ the standard deviation of $X_\alpha$ at time $s$, instead of $\Cax(s,s)$. Thus, knowing $\vax(s), s \in [0,t]$ we can compute $\muay(t)$ using a standard discretization scheme of the integral, with a small time step
compared with $\ta$ and the characteristic time of variation of the input current $\Ia$. Alternatively, we can use the fact that $\muay$ satisfies the differential equation:

\begin{equation*}
 \frac{d\muay}{dt} =-\frac{\muay}{\ta}  
+ \sum_{\beta=1}^P \Jbab  \int_{-\infty}^{+\infty} S_{\beta} \left ( x\sqrt{\vbx(t)} + \mub^X(t) \right) Dx
+ \Ia(t),
\end{equation*}
and compute the solution using a Runge-Kutta algorithm (which is faster and more accurate).
Note that, when all the standard deviations of the process $X$ are null for all time $t\in [0,T]$, we obtain a standard dynamical system. Nevertheless, in the general case, $\vbx(t)>0$ for some $\beta$'s, and the dynamical evolution of $\muay$ depends on the Gaussian fluctuations of the field $X$. These fluctuations must be computed via the complete equation of the covariance diagonal coefficient $\Cay(t,s)$, which reads:
\begin{multline*}
\Cay(t,s)=e^{-(t+s)/\ta}\Big[\vax(0)+ \frac{\ta\sa^2}{2}\left(e^{\frac{2s}{\ta}}-1\right) \\ 
+ \sum_{\beta=1}^P \Jdab \int_{0}^t\int_{0}^s e^{(u+v)/\ta}\Deabx(u,v)dudv\Big],
\end{multline*}
where:
\begin{multline*}
\Deabx(u,v)= \int_{\bbbr^2} S_{\beta}\left( x\frac{\sqrt{\vbx(u)\vbx(v) - \Cbx(u,v)^2}}{\sqrt{\vbx(v)}}
+y\frac{\Cbx(u,v)}{\sqrt{\vbx(v)}} +\mu^X_\beta (u)\right) \\
\times S_{ \beta}\left(y \sqrt{\vbx(v)}+\mu^X_\beta (v)\right)\,Dx\,Dy.
\end{multline*}

Unless if we assume the stationarity of the process (see e.g. section \ref{SSomp}), this equation cannot be written as an ordinary differential equation. We clearly observe here the non-Markovian nature of the problem: $\Cax(t,s)$ depends on the whole past of the process until time $t\vee s$.\\

This covariance can be split into the sum of two terms: the external noise contribution $\Ca^{OU}(t,s)=e^{-(t+s)/\ta}\left[\vax(0)+ \frac{\ta\sa^2}{2}\left(e^{\frac{2s}{\ta}}-1\right)\right]$ and the interaction between the neurons. The external noise contribution is a simple function and can be computed directly. To compute the interactions contribution to the standard deviation
we have to compute  the symmetric two-variables function:

\[ \Habx(t,s)=e^{-(t+s)/\ta} \int_0^t\int_0^s e^{(u+v)/\ta}\Deabx(u,v)dudv,\]
from which one obtains the standard deviation using the formula

\begin{equation*}
\Cay(t,s)=\Ca^{OU}(t,s)+\sum_{\beta=1}^P \Jdab \Habx(t,s).
\end{equation*}
To compute the function $\Habx(t,s)$, we start from $t=0$ and $s=0$, where $\Habx(0,0)=0$. We only compute $\Habx(t,s)$ for $t>s$ because of the symmetry. It is straightforward to see that:
\begin{equation*}
\Habx(t+dt,s)=\Habx(t,s)\left[1-\frac{dt}{\ta} \right]+\Dabx(t,s)dt + o(dt),
\end{equation*}
with
\begin{equation*}
\Dabx(t,s)= e^{-s/\ta}\int_0^s e^{v/\ta}\Deabx(t,v)dv.
\end{equation*}
Hence computing $\Habx(t+dt,s)$ knowing $\Habx(t,s)$ amounts to computing $\Dab(t,s)$. Fix $t\geq 0$. We have $\Dab(t,0)=0$ and 
\begin{equation*}
\Dabx(t,s+ds)=\Dabx(t,s)(1-\frac{ds}{\ta})+\Deabx(t,s)ds + o(ds).
\end{equation*}
This algorithm enables us to compute $\Habx(t,s)$ for $t>s$. We deduce $\Habx(t,s)$ for $t<s$ using the symmetry of this function. Finally, to get the values of $\Habx(t,s)$ for $t=s$, we use the symmetry property of this function and get:
\begin{equation*}
\Habx(t+dt,t+dt)=\Habx(t,t)\left[1-\frac{2dt}{\ta} \right]+2\Dabx(t,t)dt + o(dt).
\end{equation*}

These numerical schemes provide an efficient way for computing the mean and the covariance functions of the Gaussian process $\mathcal{F}_1(X)$ (hence its probability distribution) knowing the law of the Gaussian process $X$. 
The algorithm used to compute the solution of the mean-field equations for the general models GM1 and GM$k$ is a straightforward generalization. 

\sssu{Analysis of the algorithm}

\paragraph{Convergence rate}
As proved in theorem \ref{theo:ExistenceUniqueness}, given $Z_0$ a nondegenerate $kP$-dimensional Gaussian random variable and $X$ a Gaussian process such that $X(0)=Z_0$, the sequences of means and covariance functions computed theoretically converge uniformly towards those of the unique fixed point of the map $\mathcal{F}_k$. It is clear that our algorithm converges uniformly towards the real function it emulates. Hence for a finite $N$, the algorithm will converge uniformly towards the mean and covariance matrix of the process $\mathcal{F}_k^N(X)$. 

Denote by $X_f$ the fixed point of $\mathcal{F}_k$ in $\procP{kP}$, of mean $\mu^{X_f}(t)$ and covariance matrix $C^{X_f}(t,s)$, and by $\widehat{\mathcal{F}_k^N}(X)$ the numerical approximation of $\mathcal{F}_k^N(X)$ computed using the algorithm previously described, whose mean is noted $\mu^{\widehat{\mathcal{F}_k^N}(X)}(t)$ and whose covariance matrix is noted $C^{\widehat{\mathcal{F}_k^N}(X)}(t,s)$. The uniform error between the simulated mean after $N$ iterations with a time step $dt$ and the fixed point's mean and covariance is the sum of the numerical error of the algorithm and the distance between the simulated process and the fixed point, is controlled by:
\begin{equation}\label{eq:ConvergenceAlgorithmCmu}
  \|\mu^{\widehat{\mathcal{F}_k^N}(X)} - \mu^{X_f} \|_{\infty} + \|C^{\widehat{\mathcal{F}_k^N}(X)} - C^{X_f} \|_{\infty} = O(\;(N + T)\,dt + R_N(k_{\max})\;)
\end{equation}
where $k_{\max}=\max(k,\tilde{k})$ and $k$ and $\tilde{k}$) are the constants that appear in the proof of proposition \ref{prop:conv-Cmu} for the mean and covariance functions, and $R_N(x)$ is the exponential remainder, i.e. $R_N(x) = \sum_{n=N}^{\infty} x^n/n!$.

Indeed, we have:
\begin{equation}\label{eq:DistanceTriangIneg}
  \|\mu^{\widehat{\mathcal{F}_k^N}(X)} - \mu^{X_f} \|_{\infty} \leq \|\mu^{\widehat{\mathcal{F}_k^N}(X)} - \mu^{\mathcal{F}_k^N(X)}\|_{\infty} + \|\mu^{\mathcal{F}_k^N(X)} - \mu^{X_f}\|_{\infty}
\end{equation}

The discretization algorithm used converges in $O(dt)$. Let us denote by $C_1$ the convergence constant, which depends on the sharpness of the function we approximate, which can be uniformly controlled over the iterations. Iterating the numerical algorithm has the effect of propagating the errors. Using these simple remarks we can bound the first term of the righthand side of \eqref{eq:DistanceTriangIneg}, i.e. the approximation error at the $N$th iteration:
\[\|\mu^{\widehat{\mathcal{F}_k^N}(X)} - \mu^{\mathcal{F}_k^N(X)}\|_{\infty} \leq C_1\,N\, dt\]

Because the sequence of means is a Cauchy sequence, we can also bound the second term of the righthand side of \eqref{eq:DistanceTriangIneg}: 
\begin{align*}
 \|\mu^{\mathcal{F}_k^N(X)} - \mu^{X_f}\|_{\infty} &\leq \sum_{n=N}^{\infty} \|\mu^{\mathcal{F}_k^{n+1}(X)} - \mu^{\mathcal{F}_k^{n}(X)}\|_{\infty} \\
 & \leq \sum_{n=N}^{\infty} \frac{k^n}{n!}=:R_N(k)
\end{align*}
for some positive constant $k$ introduced in the proof of proposition \ref{prop:conv-Cmu}. The remainders sequence $(R_n(k))_{n\geq 0}$ converges fast towards $0$ (an estimation of its convergence can be obtained using the fact that  $\limsup_{k \to \infty} (1/k!)^{1/k} = 0$ by Stirling's formula). 

Hence we have: 

\begin{equation}\label{eq:ConvergenceMuAlgo}
  \|\mu^{\widehat{\mathcal{F}_k^N}(X)} - \mu^{X_f} \|_{\infty} \leq C_1\,N\, dt + R_N(k)
\end{equation}

For the covariance, the principle of the approximation is exactly the same:

\begin{equation*}
  \|C^{\widehat{\mathcal{F}_k^N}(X)} - C^{X_f} \|_{\infty} \leq \|C^{\widehat{\mathcal{F}_k^N}(X)} - C^{\mathcal{F}_k^N(X)}\|_{\infty} + \|C^{\mathcal{F}_k^N(X)} - C^{X_f}\|_{\infty}
\end{equation*}

The second term of the righthand side can be controlled using the same evaluation by $R_N(\tilde{k})$ where $\tilde{k}$ is the constant introduced in the proof of proposition \ref{prop:conv-Cmu}, and the first term is controlled by the rate of convergence of the approximation of the double integral, which is bounded by $C_2(N+T)\, dt$ where $C_2$ depends on the parameters of the system and the discretization algorithm used. 

Hence we have: 

\begin{equation}\label{eq:ConvergenceCAlgo}
  \|C^{\widehat{\mathcal{F}_k^N}(X)} - C^{X_f} \|_{\infty} \leq C_2\,(N+T-t_0)\, dt+ R_N(\tilde{k})
\end{equation}

The expressions \eqref{eq:ConvergenceMuAlgo} and \eqref{eq:ConvergenceCAlgo} are the sum of two terms, one of which is increasing with $N$ and $T$ and decreasing with $dt$ and the other one decreasing in $N$. If we want to obtain an estimation with an error bounded by some $\varepsilon>0$, we can for instance fix $N$ such that $\max(R_N({k}), R_N(\tilde{k}))<\varepsilon/2$ and then fix the time step $dt$ smaller than $\min(\;\varepsilon/(2C_1 N), \varepsilon/(2 C_2 (N+T-t_0))\;)$. 

\paragraph{Complexity}

The complexity of the algorithm depends on the complexity of the computations of the integrals. The algorithm described hence has the complexity $O(N (\frac{T}{dt})^2)$.

\subsection{The importance of the covariance: Simple Model, one population.}\label{SSomp}

As a first example and a benchmark for our numerical scheme we revisit the work of Sompolinsky and coworkers \cite{sompolinsky-crisanti-etal:88}. 
These authors studied the case of the simple model  with one population ($P=1$), with 
the centered sigmoidal function $S(x)=\tanh(gx)$, centered connectivity weights $\bar{J}=0$ of standard deviation $\sigma=1$ and no input ($I=0,\Lambda=0$).
Note therefore that there is no ``noise'' in the system, which therefore does not match the non degeneracy conditions
of proposition \ref{lemma:lwb2} and of theorem \ref{theo:ExistenceUniqueness} .
This issue is discussed below.
In this case, the mean equals $0$ for all $t$. Nevertheless, the Gaussian process is non trivial as
revealed by the study of the covariance $C(t,s)$.

\sssu{Stationary solutions} \label{SSompStat}

Assuming that the solution of the mean-field equation is a stationary solution with $C(t,s) \equiv C(t-s)=C(\tau)$,
Sompolinsky and his collaborators found that the covariance  obeyed a second order differential equation :

\begin{equation}\label{EDOSomp}
\frac{d^2 C}{d\tau^2}=-\frac{\partial V_{q}}{\partial C}.
\end{equation} 

\nid This form corresponds to the motion of a particle in a potential well and it is easy to draw the phase
portrait of the corresponding dynamical system. However, there is a difficulty. The potential $V_q$ depends on 
a parameter $q$ which is in fact precisely the covariance at $\tau=0$ ($q=C(0)$). In the stationary case, this covariance depends on the whole solution, and hence cannot be really considered as a parameter of the system. This is one of the main difficulties in this approach: mean-field equations in the stationary regime are self-consistent. 

Nevertheless, the study of the shape of $V_q$, considering $q$ as a free parameter gives us some informations. Indeed, $V_q$ has the following Taylor expansion ($V_q$ is even because $S$ is odd):

\begin{equation*}
V_q(C)=\frac{\lambda}{2}C^2+\frac{\gamma}{4}C^4 + O(C^6)
\end{equation*}

\nid where $\lambda=(1-g^2J^2 \langle S'\rangle^2_{q})$ and $\gamma=\frac{1}{6}J^2g^6 \langle S^{(3)}\rangle^2_{q})$,
$\langle \phi\rangle_{q}$ being the average value of $\phi$ under the Gaussian distribution with
mean zero and variance $q=C(0)$.

If $\lambda>0$, i.e. when $g^2J^2 \langle S'\rangle^2_{q} < 1$, then the
dynamical system (\ref{EDOSomp}) has a unique solution $C(t)=0, \forall t \geq 0$.
This corresponds to a stable fixed point (i.e. a deterministic trajectory, $\mu=0$ with no fluctuations) for
the neural network dynamics. On the other hand, if   $g^2J^2 \langle S'\rangle^2_{q} \geq 1$ there is a homoclinic
trajectory in (\ref{EDOSomp}) connecting the point $q=C^\ast>0$ where
$V_q$ vanishes to the point $C=0$. 
This solution is interpreted by the authors as a chaotic solution in the neural network.
A stability analysis shows that this is the only stable\footnote{More precisely, this is the only
minimum for the large deviation functional.} stationary solution \cite{sompolinsky-crisanti-etal:88}. \\

The equation for the homoclinic solution is easily found using energy conservation
and the fact that $V_q(q)=0$ and $\frac{dV_q}{dC}(q)=0$. One finds:

$$u=\frac{dC}{dx}=-\sqrt{-V_q(C)}.$$

At the fourth order in the Taylor expansion of $V_q$ this gives

\begin{equation*}
C(\tau)=\frac{\sqrt{\frac{-2\lambda}{\gamma}}}{\cosh(\sqrt{-\frac{\lambda}{2}}\tau)}.
\end{equation*} 

Though $\lambda$ depends on $q$ it can be used as a free parameter for interpolating  the curve
of $C(\tau)$ obtained from numerical data.
 
\sssu{Numerical experiments}

This case is a good benchmark for our numerical procedure since we know analytically the solutions we are searching for. We expect to find two regimes. In one case the correlation function is identically zero in the stationary regime, for 
sufficiently small $g$ values or for a sufficiently small $q$ (trivial case). The other case corresponds to a regime where $C(\tau)>0$ and $C(\tau) \to 0$ has $\tau \to +\infty$  (``chaotic'' case). This regime requires that $g$ be sufficiently large and that  $q$
be large too. We took $\ta=0.25, \sigma_{\alpha\alpha}=1$. For these values, the change in dynamics predicted by Sompolinsky and collaborators is $g_c=4$.

\begin{figure}
\centerline{\includegraphics[width=.7\textwidth]{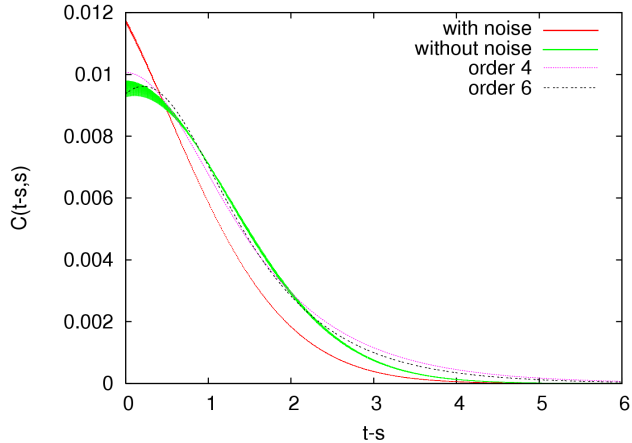}}
  \caption{Numerical solution of the mean-field equation after 14 iterations in the chaotic case ($g=5$). We clearly see the numerical instabilities in the no-noise case, which do not exist in the low-noise case.}
  \label{fig:NoNoiseVsSmallNoise}
\end{figure}

In  sections \ref{sect:existUniq} and \ref{sect:station} we have introduced the assumption  of non-degeneracy of the noise, in order to ensure that the mean-field process was non degenerate.
However, in the present example, there is no external noise in the evolution, so we can observe the effects of relaxing this hypothesis in a situation
where the results of proposition \ref{lemma:lwb2} and of theorem \ref{theo:ExistenceUniqueness} 
cannot be applied. First,  we observed numerically that, without external noise, the process could become degenerate (namely some eigenvalues of the covariance matrix $C_\alpha(t,s)$ become very small and even vanish.). This has also an incidence on the convergence of the method
which  presents numerical instabilities, though the iterations leads to a curve which is well
fitted by the theoretical results of  Sompolinsky et al. (see Fig. \ref{fig:NoNoiseVsSmallNoise}) . 
The instability essentially disappears if one adds a small noise. 
 But, note that in this case, the
solution does not match with Sompolinsky et al. theoretical calculation (see Fig. \ref{fig:NoNoiseVsSmallNoise}).\\

Modulo this remark, we have first considered 
the trivial case corresponding to small $g$ values. We took $g=0.5$ and $T=5$. 
We choose as initial process the stationary Ornstein-Uhlenbeck process corresponding to the uncoupled system with $\Lambda=0.1$. We drew $\mua(0)$ randomly from the uniform distribution in $[-1,1]$ and $\va(0)$ randomly from the uniform distribution in $[0,1]$. 

Starting from this initial stationary process, we iterated the function $\mathcal{F}_1$.
Then, during the iterations, we set $\sa=0$  in order to match the conditions imposed by Sompolinsky and coworkers. We observe that the method converges towards the expected solution: the mean function converges to zero, while the variance $v(t)$ decreases exponentially fast in time towards a constant value corresponding to the stationary regime. This asymptotic value decreases between two consecutive iterations, which is consistent with the theoretical expectation that $v(t)=0$ in the stationary regime of the trivial case. Finally, we observe that the covariance $C(t-s,s)$ stabilizes to a curve that does not depend on $s$ and the stationary value (large $t-s$) converges to zero. 

We applied the same procedure for $g=5$ corresponding to
the ``chaotic'' regime. The behavior was the same for
$\mu(t)$ but was quite different for the covariance function $C(t,s)$. Indeed,
while in the first case the stationary value of $v(t)$ tends to zero
with the number of iterations, in the chaotic case it stabilizes
to a finite value. In the same way, the covariance $C(t-s,s)$ 
stabilizes to a curve that does not depend on $s$. The shape
of this curve can be extrapolated thanks to Sompolinsky et al.
results. We observe a very good agreement with the theoretical predictions
with a fit $f_4(x)=\frac{a}{\cosh(b(x-\delta))}$, corresponding to the fourth expansion of $V_q$.
Using a $6$-th order expansion of $V_q(x)=\frac{a}{2}x^2+\frac{b}{4}x^4+\frac{c}{6}x^2$ 
gives a fit $f_6(x)=\frac{\rho}{\cosh(\lambda(x-\delta))}\frac{1}{\sqrt{1+K^2-\frac{1}{\cosh^2(\lambda(x-\delta))}}}$,
where $\rho,K,\lambda$ are explicit functions of $a,b,c$,
 we obtain a slightly better approximation.

 \ssu{Mean-field equations for two populations with a negative feedback loop.}\label{Sdeuxpop}

Let us now present a case  where the fluctuations of the Gaussian field 
act on the dynamics of $\mua(t)$ in a non trivial way, with a behavior strongly departing
from the naive mean-field picture.
We consider  two interacting populations where the connectivity weights are Gaussian random variables $J_{\alpha\beta} \equiv \mathcal{N}(\Jb_{\alpha\beta}, \sigma_{\alpha\beta} =1)$  for $(\alpha,\beta) \in \{1,2\}^2$.
We set $S_{\beta}(x)=\tanh(gx)$ and $\Ia=0,\sa=0, \alpha=1,2$. 

\sssu{Theoretical framework.}
 
 The dynamic mean-field equation for $\mua(t)$ is given, in differential form, by:
 
 \begin{equation*}
 \frac{d\mua}{dt}=-\frac{\mua}{\ta}+\sum_{\beta=1}^2 \Jbab \int_{-\infty}^\infty S\left(\sqrt{\vb(t)}x+\mub(t)\right) Dx,\ \alpha=1,2.
 \end{equation*}

Let us denote by $G_\alpha(\mu,v(t))$ the function in the righthand side of the equality. 
 Since $S$ is odd, $\int_{-\infty}^\infty S(\sqrt{\vb(t)}x) Dx =0$. Therefore, we have $G_\alpha(0,v(t))=0$ whatever $v(t)$, and hence the point $\mu_1=0, \mu_2=0$ is always a fixed point of this equation. 
 
 Let us study the stability of this fixed point. To this purpose, we compute the partial derivatives of $G_{\alpha}(\mu,v(t))$ with respect to $\mub$ for $(\alpha, \beta)\in\{1,2\}^2$. We have:
  
 \[\frac{\partial G_\alpha}{\partial \mub}(\mu,v(t))= -\frac{\delta_{\alpha\beta}}{\ta}+g \Jbab \int_{-\infty}^\infty \left(1-\tanh^2\left(\sqrt{\vb(t)}x+\mub(t)\right)\right) Dx,\] 
 and hence at the point $\mu_1=0, \mu_2=0$, these derivatives read:
  \[\frac{\partial G_\alpha}{\partial \mub}(0,v(t))=-\frac{\delta_{\alpha\beta}}{\ta}+g \Jbab h(\vb(t)),\] 
  where $h(\vb(t))=1-\int_{-\infty}^\infty \tanh^2(\sqrt{\vb(t)}x) Dx$.

In the case $\va(0)=0,J=0,\sa=0$, implying $\va(t)=0, t \geq 0$, the equation for
 $\mua$ reduces to:
 
 \begin{equation*}
 \frac{d\mua}{dt}=-\frac{\mua}{\ta}+\sum_{\beta=1}^2 \Jbab S(\mub(t))
 \end{equation*}
 
 \nid which is the standard Amari-Cohen-Grossberg-Hopfield  system.
This corresponds to the naive mean-field approach where Gaussian fluctuations are neglected. 
In this case the stability of the fixed point $\mu=0$ is given by the sign of the largest eigenvalue of the
Jacobian matrix of the system that reads:

\[\left (\begin{array}{cc} -\frac{1}{\tau_1} & 0 \\ 0 & -\frac{1}{\tau_2} \end{array}\right)+ g\left(\baR{ccc} \Jb_{11}&\Jb_{12}\\ \Jb_{21}&\Jb_{22} \eaR\right).\]

For the sake of simplicity we assume that the two time constants $\ta$ are equal and we denote this value $\tau$.
The eigenvalues  are in this case $-\frac{1}{\tau}+g\lambda$,
where $\lambda$ are the eigenvalues of $\bar{\cJ}$
and have the form:
\[\lambda_{1,2}=\frac{\Jb_{11}+\Jb_{22} \pm \sqrt{(\Jb_{11}-\Jb_{22})^2+4 \Jb_{12}\Jb_{21}}}{2}.\]
 
 Hence, they are complex whenever $\Jb_{12}\Jb_{21}< -(\Jb_{11}-\Jb_{22})^2/4$, corresponding
 to a negative feedback loop between population 1 and 2. Moreover,
 they have a real part only if $\Jb_{11} + \Jb_{22}$ is non zero (self interaction).

 This opens up the possibility to have an instability of the fixed point ($\mu=0$) leading to a
regime  where the average value of the membrane potential
 oscillates. This occurs if
   $\Jb_{11}+\Jb_{22}>0$  and if $g$ is larger than:
 \begin{equation*}
 g_c=\frac{2}{\tau(\Jb_{11}+\Jb_{22})}.
 \end{equation*}
The corresponding bifurcation is a Hopf bifurcation.

The situation is different if one takes into account the fluctuations of the Gaussian
field. Indeed, in this case the stability of the fixed point $\mu=0$ depends on $v(t)$.
More precisely, the real and imaginary part of the eigenvalues of $DG(0,v(t))$ depend on $v(t)$.
Therefore, the variations of $v(t)$ act on the stability and oscillations period of $v(t)$.
Though the evolution of $\mu(t),v(t)$ are coupled we cannot consider this evolution as
a coupled dynamical system, since $v(t)=C(t,t)$ is determined by the mean-field equation
for $C(t,s)$ which cannot be written as an ordinary differential equation. Note that we cannot
assume stationarity here, as in the previous case, since $\mu(t)$ depends on time for sufficiently large $g$.
This  opens up the possibility of having  complex dynamical regimes when $g$ is large.

\sssu{Numerical experiments}

\begin{figure}[htb]
\centerline{\includegraphics[width=.7\textwidth]{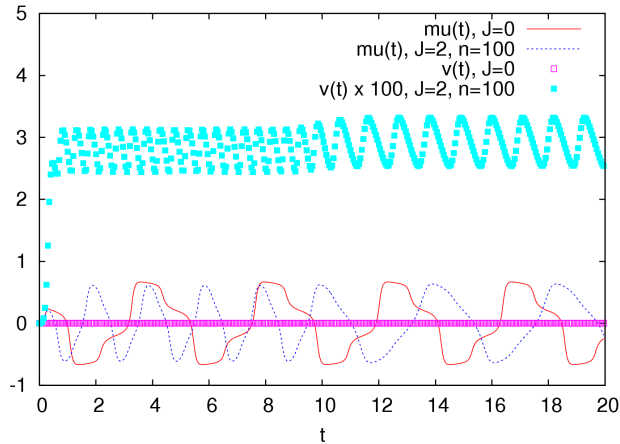}}
\caption{Evolution of the mean $\mu_1(t)$ and variance $v_1(t)$ for the mean-field of population
$1$, for $J=0$ and $J=2$, over a time window $[0,20]$. $n$ is the number of iterations
of $\mathcal{F}_1$ defined in section \ref{sect:existUniq}. This corresponds to a number of iterations for which the method
has essentially converged (up to some precision). Note that $v_1(t)$ has been magnified by
a factor of $100$. Though Gaussian fluctuations are small, they have
a strong influence on $\mu_1(t)$.}
\label{F2popmu}
\end{figure}

 We have considered the case $\Jb_{11}=\Jb_{22}=5$,$\tau=0.1$ giving a Hopf bifurcation for $g_c=2$ when $J=0$
(fig. \ref{F2popmu}). The trajectory of $\mu_1(t)$ and $v_1(t)$ is represented in Figure \ref{F2popmu} in the case $g=3$.
When $J=0$, $\mu_1(t)$ presents regular oscillations (with non linear effects since $g=3$ is larger than the critical
value for the Hopf bifurcation, $g_c=2$). In this case, the solution $v_1(t)=0$ is stable
as seen on the figure. When $J \neq 0$ the Gaussian field has (small) fluctuations which nevertheless
strongly interact with the dynamics of $\mu_1(t)$, leading to a regime where $\mu_1(t)$ and
$v_1(t)$ oscillate periodically

\section{Discussion}\label{section:discussion}
The problem of bridging scales is overwhelming in general when studying complex systems and in particular in neuroscience. After many others we looked at this difficult problem from the theoretical and numerical viewpoints, hoping to get closer to its solution from relatively simple and physically/biologically plausible first principles and assumptions. One of our motivations is to better understand such phenomenological neural mass models as that of Jansen and Rit \cite{jansen-rit:95}.

We consider several populations of neurons and start from a microscopic, i.e. individual, description of the dynamics of the membrane potential of each neuron that contains four terms. 

The first one controls the intrinsic dynamics of the neuron. It is linear in this article but this assumption is not essential and could probably be safely removed if needed. 

The second term is a stochastic input current, correlated or uncorrelated. The corresponding noise model is very rich, depending on the degree $k$ of smoothness of the g-shapes. It features integrated Brownian noise up to order $k-1$. 

The third term is a deterministic input current, and the fourth one describes the interaction between the neurons through random connectivity coefficients that weigh the contributions of other neurons through a set of functions that are applied to their membranes potentials. The only hypothesis on these functions is that they are smooth and bounded, as well as their first order derivative. The obvious choice of sigmoids is motivated by standard rate models ideas. Another appealing choice is a smooth approximation to a Dirac delta function thereby opening a window on the world of spiking neurons. Thus, the model presented in this paper is {\em more general} than the instantaneous rate model that is underlying Ermentrout's voltage-based model \cite{ermentrout:98} even though we have not explored this avenue.

We then derive the mean-field equations and  provide a constructive and new proof, under some mild assumptions, of the existence and uniqueness of a solution of these equations over finite and infinite time intervals.  The key idea is to look at this mean-field description as a \emph{global problem} on the probability distribution of the membranes potentials, unlike previous studies. Our proof provides an efficient way of computing this solution and our numerical experiments show a good agreement with previous studies. It is interesting to note that a sufficient condition for the convergence of our algorithm is related to the previously mentioned noise model. We prove that if the noise matrix $\mla$ is full rank, with bounded eigenvalues, then the algorithm is in general convergent. An important fact to note is that the solutions of the mean-field equations that we construct are fundamentally {\em non-Markovian}  eliminating the need for such approximations as the introduction of the $q$ parameter summarizing the whole history of the non-Markovian process, see below.

In the case where the nonlinearities are chosen to be sigmoidal our results shed a new light on existing neural mass models. Indeed, as shown in section \ref{subsubsection:mean-field}, these appear as approximations of the mean-field equations where the intricate but fundamental coupling between the time variations of the mean membrane potentials and their fluctuations, as represented by the covariance functions, is neglected. \\

An alternative approach has been recently proposed by Chizhov and collaborators\footnote{We 
thank one of the referees for pointing out these references to us.} 
\cite{chizhov-graham:07, chizhov-rodrigues-etal:07}. The approach of these authors
consists in reducing the large number, $N$, of degrees of freedom of the neural assembly by
constructing a probability density $\rho$ on the phase space of neurons states in the 
limit $N \to \infty$. This is a non rigorous approach where the evolution equations
for $\rho$ are heuristically derived. Especially, it is assumed that $\rho$
depends on two parameters only: the current time $t$ and the time elapsed 
since the last spike $t^\ast$. Under these assumptions the initial phase space
of neurons states is mapped to a two dimensional space $t,t^\ast$, while $\rho(t,t^\ast)dt$
characterizes the fraction of neurons which have fired in the time interval $[t-t^\ast,t-t^\ast+dt]$.
Therefore, this approach intrinsically holds for integrate and fire neurons models where the
neuron's membrane potential history is summarized by the last spike time, when it is reset
to a constant value. As noticed by these authors, this allows to circumvent
the main problem in mean-field approaches for firing rate models, that we also
discuss in the present paper: When using mean-field theory to characterize
stationary regimes, one needs to introduce ad hoc parameters (see e.g. the parameter
$q$ introduced in section \ref{SSompStat}) summarizing the whole history of the non-Markovian 
process. Introducing a ``history cut-off'' while reseting the membrane potential
to a constant value indeed removes this difficulty. Therefore, it might be interesting
to compare our approach in the case of integrate-and-fire models (see above remark on the choice of the nonlinearity), to the approach of Chizov and collaborators.
This could provide some rigorous basis for their analysis and allow to elucidate
the role of field fluctuations which does not appear explicitely in the
probability density approach.

\section{Conclusion and further work}
On more general grounds, our goal is now to extend the present
work in several directions.
\paragraph{Bifurcations analysis of the dynamic mean-field equations.}
From the present analysis, and as shown in the simple examples of section
\ref{sect:numerics}, the mesoscopic dynamics of the average membrane potential
of a neurons population can be really different from the classical
phenomenological equations à la Jansen-Rit if one includes
the non-Markovian fluctuations of the interaction fields, which summarize
the cumulative effects of the nonlinear interactions of a given neuron 
with the bulk of other neurons. Jansen-Rit equations
are commonly used in the neuroscience community either to anticipate the dynamics
of local field potential in relation with imaging (Optical Imaging, MEG-EEG),
or to understand neurophysiological disorders such as epilepsy. Bifurcations
analysis of these equations reveal dynamical regimes that can be related
to experiments  \cite{grimbert-faugeras:06}. They can be generalised using
more accurate neural models \cite{wendling-hernandez-etal:05}. Is there any need to generalize
these equations, that we claim to be incomplete, while people commonly use them
with some satisfaction? Are there new phenomena, \textit{experimentally accessible},
that can be exhibited by the generalised mean-field equations and that do
not appear in the naive ones? These are obviously important questions that we intend
to address in the near future.   On mathematical grounds, the goal
is to make a bifurcation analysis of the map ${\mathcal F}$ on the space of trajectories,
introduced in the present paper. Do any new salient dynamical regimes appear?
If such regimes exist, the goal will be, on experimental grounds,
to interact with experimentalists in order to see in which conditions
such a regime can be exhibited, and what are its implications 
on cortical columns dynamics or function.
\paragraph{Investigations of non stationary regimes.} As discussed in this paper, and as is well-known
in the physicists' community (especially spin-glasses community), the dynamic mean-field approach
raises serious difficulties as far as one is trying to describe \textit{stationary dynamics}.
On technical grounds, this relies on the non-commutativity of the two limits $N \to \infty$ and
$t \to \infty$ already discussed in \cite{sompolinsky-zippelius:82}. As a result,
one is led to introduce ad-hoc phenomenological parameters,  depending
on initial conditions, that can be determined in statistical physics models 
where the distribution of equilibria is known (Gibbs distribution), using sophisticated
techniques such as the replica ``trick'' \cite{houghton-jain-etal:83}. For spin-glasses it is only
in the high temperature regime that  a simple solution to this problem is known.
This restriction also appears in the present paper, where the existence and uniqueness
of a stationary solution is proved only for low values of the gain parameter $g$
(which plays a role similar to the inverse temperature).  However, we are not
so much interested in stationary dynamics, since brain processes are ultimately
non stationary. Our approach, valid for any finite time $T$, opens up
the possibility to characterize mean-field equations in transient regimes, with 
an analysis strategy that can moreover be easily implemented. To the best of
our knowledge, this type of techniques has never been used in the statistical physics
community, where iterations on space trajectories are not in the standard toolbox.
Therefore, our work could allow the (numerical) investigation of cortical
columns submitted to non stationary inputs, with strong implications on  neuroscience. 
\paragraph{Extension to a larger class of models.}  A very challenging question is the application
of this theory to spiking neurons models. We have briefly mentioned in section \ref{section:discussion} that this may be possible
through the use of non-sigmoidal functions in the interaction terms. This idea could be applied to the analysis
of Integrate and Fire models with conductance based synapses, which constitute good
models of spiking neurons. As discussed at the end of section \ref{section:discussion},
the analysis of the mean-field equations could be simplified by the fact that memory
is reset after a neuron fires. There is however a need to characterize
parameter space regions where neurons can take an arbitrary large time to fire
for the first time \cite{cessac:08,cessac-vieville:08}. This is the main
obstruction in the application of our theory to this type of models. 

\section*{Conflict of Interest Statement}
The authors declare that the research was conducted in the absence of any commercial or financial relationships that could be construed as a potential conflict of interest.

\section*{Aknowledgement}
This research was partly supported by funding of the
European Union under the grant no. 15879 (FACETS) and
the Fondation d'Entreprise EADS. It was also supported
by the MACACC ARC INRIA and the Doeblin CNRS Federation. 

\appendix

\section{Identification of the mean-field equations}
\label{appendix:MFECalculus}
Ben-Arous and Guionnet studied from a mathematical point of view the problem of finding a mean-field description of large networks of spin glasses. They obtained using different methods of stochastic analysis a weak limit of the law of a given spin and proved their independence. 

Our equations do not directly fit in their study: indeed, the spin intrinsic dynamics is nonlinear while the interaction is linear, and everything in done in dimension one. Nevertheless, their proof extends to our case which is somehow more simple. For instance in the case of the Simple Model with one population, we can readily adapt their proof in our case. More precisely, let $P=1$, the equation of the network reads:

\[\tau dV^j_t = (-V^j_t + \sum_{i=1}^N J_{ij} S(V^i_t) )\, dt + \sigma dW^j_t\]

In this case, we define for $X \in \proc$  the effective interaction term $(U^{X}_t)$ which is the effective interaction process defined in \ref{def:effectiveInteractionProcess}, i.e. the Gaussian process of mean $\Jbab \Exp{S(X_t)}$ and of covariance: 
$ \Cov\Big(U^{X}_t,\,U^{X}_s\Big)=: \Jdab \Exp{S(X_t)S(X_s)}$.

Let us note $\mathcal{P}$ the law of the membrane potential when there is no interaction (it is an Ornstein-Ulhenbeck process), and the empirical measure $\hat{V}^N = \frac 1 N \sum_{i=1}^N \delta_{V^i}$. We can prove that under the probability distribution averaged over the connectivities, see below, the empirical measure satisfies a large deviation principle with good rate function $H$ defined as in \cite{guionnet:97}. Using this large deviation result, we can prove annealed and quenched tightness of the empirical measure, and finally its convergence towards the unique process where the good rate function $H$ achieves its unique minimum, which is defined by the property of having a density with respect to $\mathcal{P}$ and whose density satisfies the implicit equation:

\begin{equation}\label{eq:BenArousImplicit}
Q \ll \mathcal{P} \qquad \frac{dQ}{d\mathcal{P}} = \mathcal{E} \left [ \exp \left \{ \int_0^T U^Q_t dW_t - \frac 1 2 \int_0^T (U^Q_t)^2 \, dt \right \}\right ] 
\end{equation}
\noindent where $\mathcal E$ denotes the expectation over the effective interaction process $U^Q$. 

We can also prove following the steps of Ben-Arous and Guionnet in \cite{ben-arous-guionnet:97} that there exists a unique solution to this equation, and that this solution satisfies the nonlinear nonmarkovian stochastic differential equation:

\begin{equation}\label{eq:BenArousSDE}
 \begin{cases}
  \tau dV_t = -V_t \, dt + dB_t  \\
  dB_t = dW_t + \int_0^t dB_s \mathcal{E} \left [ U^Q_s U^Q_t \frac{\exp\{-\frac 1 2 \int_0^t (U^Q_u)^2 du\}}{\mathcal{E} [\exp\{-\frac 1 2 \int_0^t (U^Q_u)^2 du \}]}\right] \\
  \textrm{ Law of } (V) = Q, \;\; \textrm{law of }(V_0) = Z_0
 \end{cases}
\end{equation}
which can also be written as our mean-field equation, averaged on the connectivities (see \cite{ben-arous-guionnet:95}). More precisely, let $L^V$ be the law of the solution of the equation:
\[
\begin{cases}
 \tau dV_t = -V_t dt + dW_t + U^V_t  dt \\
 \textrm{Law of } V_0 = Z_0
\end{cases},
\]
which is exactly equation \eqref{eq:V_MFE}. They prove that $V$ satisfies the nonlinear equation:

\[V \eqlaw \mathcal{E}(L^V)\]
This result is probably extendable to the multi-population case using the multidimensional Girsanov's theorem, but the corresponding
mathematical developments are out of the scope of this paper.

\section{The resolvent}
\label{appendix:TOop}
In this appendix we introduce and give some useful properties of the resolvent $\Phi_L$ of a homogeneous differential equation
\begin{equation}\label{eq:ode}
 \der{x}{t}=\mL(t) x(t) \quad x(t_0)=x_0 \in \R^P,
\end{equation}
where $\mL : [t_0,T] \to \mathcal{M}_{P \times P}$ (or $(-\infty,T] \to \mathcal{M}_{P \times P}$) is $C^0$.

\begin{definition}
 The resolvent of \eqref{eq:ode} is defined as the unique solution of the linear equation:
 \begin{equation}\label{eq:resolv}
 \begin{cases}
  \der{\Phi_L(t,t_0)}{t} &= \mL(t) \Phi_L(t,t_0)\\
  \Phi_L(t_0,t_0) &= {\rm Id}_{P}
 \end{cases}
 \end{equation}
 \noindent where ${\rm Id}_P$ is the $P \times P$ identity matrix. 
\end{definition}

\renewcommand{\theenumi}{(\roman{enumi})}
\begin{proposition}\label{prop:ETO}
 The resolvent satisfies the following properties:
 \begin{enumerate}
   \item $\Phi_L(t + s,t_0) =  \Phi_L(t+s,t)\cdot \Phi_L(t,t_0)$
   \item $\Phi_L(t,t_0)$ is invertible of inverse $\Phi_L(t_0,t)$ which satisfies:
\begin{equation}\label{eq:resolvinv}
 \begin{cases}
  \der{\Phi_L(t_0,t)}{t} &=  - \Phi_L(t_0,t)\mL(t)\\
  \Phi_L(t_0,t_0) &= {\rm Id}_{P\times P}
 \end{cases}
\end{equation}
   \item Let $\norm{\ }$ be a norm on $\mathcal{M}_{P \times P}$ and assume that $\norm{\mL(t)} \leq k_L$  on $[t_0,T]$. Then we have:
\begin{equation}\label{eq:upresol}
\norm{\Phi_L(t,t_0)} \leq e^{k_L \vert t-t_0 \vert} \quad \forall t \in [t_0,T]
\end{equation}
Similarly, if $\norm{\mL^T(t)} \leq k_{L^T}$  on $[t_0,T]$ we have:
\begin{equation}\label{eq:upresolt}
\norm{\Phi_L^T(t,t_0)} \leq e^{k_{L^T} \vert t-t_0 \vert} \quad \forall t \in [t_0,T]
\end{equation}
   \item We have
\[
 {\rm det}\Phi_L(t,t_0)=\exp \int_{t_0}^t {\rm Tr} \mL(s)\,ds
\]
 \end{enumerate}
\end{proposition}

\begin{proof}
 The properties (i) and (ii) are directly linked with the property of group of the flow of a reversible ODE.
 (iii) is an application of Gronwald's lemma. (iv) is obtained by a first order Taylor series expansion.
\end{proof}

\begin{theorem}[Solution of an inhomogeneous linear SDE]
 The solution of the inhomogeneous linear Stochastic Differential Equation:
 \begin{equation}\label{eq:AppendixResolventEquation}
   \begin{cases}
    dX_t &= (\mL(t) X(t) + \mI(t))\,dt + \mla(s) d\mW_s  \\
    X_{t_0} & = X_0
   \end{cases}
 \end{equation}

 can be written using the resolvent:
 \begin{equation}\label{eq:AppendixResolventSolution} 
    X_t = \Phi_L(t,t_0) X_0 + \int_{t_0}^t \Phi_L(t,s) \mI(s)\, ds + \int_{t_0}^t \Phi_L(s,t) \mla(s)d\mW_s
 \end{equation}
\end{theorem}

\begin{proof}
 Pathwise (strong) uniqueness of solution directly comes from the results on the SDE with Lipschitz coefficients (see e.g. \cite[Theorem 2.5 of Chapter 5]{karatzas-shreve:91}). It is clear that $X_{t_0} = X_0$. We use It\^o's formula for the product of two stochastic processes to prove that the process \eqref{eq:AppendixResolventSolution} is solution of equation \eqref{eq:AppendixResolventEquation}:
 
 \begin{align*}
  dX_t &= \Big (\mL(t) \Phi_L(t,t_0) X_0 + \Phi_L(t,t) \mI(t) + \int_{t_0}^t \mL(t) \Phi_L(t,s) \mI(s)\, ds \Big )\, dt \\
  &  + \Phi_L(t,t) \mla(t) d\mW_t + \int_{t_0}^t \mL(t) \Phi_L(s,t) \mla(s)d\mW_s \, dt \\
  & = \Big ( \mL(t) \Big[ \Phi_L(t,t_0) X_0 + \int_{t_0}^t \Phi_L(s,t) I(s)\, ds + \int_{t_0}^t \Phi_L(s,t) \mla(s)d\mW_s \Big] + \mI(t) \Big) \, dt  \\
  & + \mla(t) d\mW_t \\
  & = (\mL(t) X(t) + \mI(t))\, dt + \mla(t) d\mW_t 
 \end{align*}
 Hence the theorem is proved.
\end{proof}

\section{Matrix norms}
\label{appendix:OpNorms}
In this section we recall some definitions on matrix and vector norms. Let $\mathcal{M}_{n \times n}$ be the set of $n\times n$ real matrices. It is a vector space of dimension $n^2$ and the usual $L^p$ norms $1 \leq p \leq \infty$ can be defined. Given $\mL \in \mathcal{M}_{n \times n}$, we note $\norm{\mL}_p^v$ the corresponding norm. Given a vector norm, noted $\norm{\ }$, on $\R^n$ the induced norm, noted $\norm{\ }$, on $\mathcal{M}_{n \times n}$ is defined as 
\[
\norm{\mL}= \sup_{x \in \R^n,\, \norm{x} \leq 1} \frac{\norm{\mL x}}{\norm{x}}
\]
Since $\mathcal{M}_{n \times n}$ is finite dimensional all norms are equivalent.
In this article we use the following norms
\begin{enumerate}
    \item $\norm{\mL}_\infty = \max_i \sum_{j=1}^n \vert L_{ij} \vert$.
    \item $\norm{\mL}_\infty^v = \max_{i,\,j} \vert L_{ij} \vert$
    \item $\norm{\mL}_2= \sup_{x \in \R^n,\, \norm{x}_2 \leq 1} \frac{\norm{\mL x}_2}{\norm{x}_2}$. This so-called {\em spectral} norm is equal to the square root of the largest singular value of $\mL$ which is the largest eigenvalue of the positive matrix $\mL^T \mL$. If $\mL$ is positive definite this is its largest eigenvalue which is also called its spectral radius, noted $\rho(\mL)$.
\end{enumerate}

\section{Important Constants}
\label{appendix:quant}
Table \ref{table:summ} summarizes some notations which are introduced in the article and used in several places.

\begin{table}[!h]
 \begin{center}
  \begin{tabular}{|c|c|c|}
   \hline
   Constant & Expression & Defined in\\
   \hline
   $\mu$ & $\max_\alpha \sum_\beta |\Jbab|\, \|S_{\beta}\|_\infty$ & lemma \ref{lemma:upper} \\
    &  & equation \eqref{eq:mu}\\
   \hline
   $\sigma_{\rm{max}}^2$ & $\max_\alpha \sum_\beta \Jdab \, \|S_{\beta}\|_\infty^2$ & lemma \ref{lemma:upper} \\
   \hline
   $\sigma_{\rm{min}}$ & $\min_{\alpha,\beta} \sigma_{\alpha\beta}^2$ & lemma \ref{lemma:upper} \\
   \hline
   $\mu_{\rm max}$ & $e^{k_L (T-t_0)}\left[ \norm{\Exp{Z_0}}_\infty+(\mu+I_{\rm max})(T-t_0)  \right]$ & lemma \ref{lemma:max} \\
   \hline
    $\Sigma_{\rm max}$ & $e^{(k_L+k_{L^T})(T-t_0)}\left[ \rho(\Sigma^{Z_0})+\lambda_{\rm max}^\Gamma (T-t_0)+\sigma_{\rm max}^2(T-t_0)^2\right]$ & lemma \ref{lemma:max} \\
   \hline
   $k_0$ & $\lambda_{\rm min} \lambda_{\rm min}^{\Sigma^{Z_0}}$ & lemma \ref{lemma:lwb1} \\
   \hline
   $K$ & $\lambda_{\rm min} \sqrt{\lambda_{\rm min}^{\Sigma^{Z_0}} \lambda_{\rm min}^\Gamma (T -t_0)}$ & proof of lemma \ref{lemma:integral} \\
   \hline
   $k_C$ & $\max_\alpha \sum_\beta \sigma_{\alpha \beta}^2 \norm{S_{\beta}}_\infty \norm{S_{\beta}'}_\infty$  &proposition \ref{prop:conv-Cmu} \\
   & & equation \eqref{eq:kC}\\
   \hline
   $\lambda_L$ & & equation \eqref{eq:eigenvalues}\\
   \hline
  \end{tabular}
  \caption{Some important quantities defined in the article.}
  \label{table:summ}
 \end{center}
\end{table}

\section{Proof of lemma \ref{lemma:max}}\label{appendix:max}
\begin{lemma}
 The following uppperbounds are valid for all $n \geq 1$ and all $s,\,t\,\in [t_0,T]$.
\[
 \norm{\mu^n(t)}_\infty \leq e^{k_L (T-t_0)}\left[ \norm{\Exp{Z_0}}_\infty+(\mu+I_{\rm max})(T-t_0)  \right]\overset{\rm def}{=}\mu_{\rm max},
\]
\[
 \norm{C^n(t,s)}_\infty \leq e^{(k_L+k_{L^T})(T-t_0)}\left[ \rho(\Sigma^{Z_0})+\lambda_{\rm max}^\Gamma (T-t_0)+\sigma_{\rm max}^2(T-t_0)^2\right]\overset{\rm def}{=}\Sigma_{\rm max},
\]
where $\mu$ and $\sigma_{\rm max}$ are defined in lemma \ref{lemma:upper}, $\lambda_{\rm max}^\Gamma$ is defined in \ref{assumption:all}
\end{lemma}
\begin{proof}
The first inequality follows  from taking the infinite norm of both sides of equation \eqref{eq:mean} and using assumption 1. in  \ref{assumption:all} and equation \eqref{eq:upresol}, lemma \ref{lemma:upper}, and assumption 3. in \ref{assumption:all}.

The second inequality follows from taking the infinite norm of both sides of equation \eqref{eq:cov} and using assumption 1. in  \ref{assumption:all} and equations \eqref{eq:upresol} and \eqref{eq:upresolt}, lemma \ref{lemma:upper}, and assumption 2. in \ref{assumption:all}.
\end{proof}

\section{Proof of lemma \ref{lemma:lwb1}}\label{appendix:lwb1}
\begin{lemma}
For all $t \in [t_0,T]$ all $\alpha=1,\cdots,kP$, and $n \geq 1$, we have
\[
 C^n_{\alpha\alpha}(t,t) \geq \lambda_{\rm min} \lambda_{\rm min}^{\Sigma^{Z_0}}\overset{\rm def}{=}k_0 > 0,
\]
where $\lambda_{\rm min}$ is the smallest singular value of the symmetric positive definite matrix $\Phi_L(t,t_0)\Phi_L(t,t_0)^T$ for $t \in [t_0,T]$ and $\lambda_{\rm min}^{\Sigma^{Z_0}}$ is the smallest eigenvalue of the symmetric positive  definite covariance matrix $\Sigma^{Z_0}$.
\end{lemma}
\begin{proof}
 $C^n_{\alpha\alpha}(t,t)$ is larger than $(\Phi_L(t,t_0) \Sigma^{Z_0} \Phi_L(t,t_0)^T)_{\alpha\alpha}$ which is larger than the smallest eigenvalue of the matrix $\Phi_L(t,t_0) \Sigma^{Z_0} \Phi_L(t,t_0)^T$. This smallest eigenvalue  is equal to
\begin{multline*}
  \min_{\|x\| \leq 1} \frac{x^T \Phi_L(t,t_0) \Sigma^{Z_0} \Phi_L(t,t_0)^T x}{x^Tx}=\\
\min_{\|x\| \leq 1} \left(\frac{x^T \Phi_L(t,t_0) \Sigma^{Z_0} \Phi_L(t,t_0)^T x}{x^T \Phi_L(t,t_0)\Phi_L(t,t_0)^T x}\,\frac{x^T \Phi_L(t,t_0)\Phi_L(t,t_0)^T x}{x^Tx}\right) \geq \\
\min_{\|x\| \leq 1} \frac{x^T \Phi_L(t,t_0) \Sigma^{Z_0} \Phi_L(t,t_0)^T x}{x^T \Phi_L(t,t_0)\Phi_L(t,t_0)^T x}\,\min_{\|x\| \leq 1} \frac{x^T \Phi_L(t,t_0)\Phi_L(t,t_0)^T x}{x^Tx}.
\end{multline*}

In the last expression the first term is larger than the smallest eigenvalue $\lambda_{\rm min}^{\Sigma^{Z_0}}$ of the matrix $\Sigma^{Z_0}$ which is positive definite since we have assumed the Gaussian random variable $Z_0$ nondegenerate. The second term is equal to the smallest singular value $\lambda_{\rm min}$ of the matrix $\Phi_L(t,t_0)$ which is also strictly positive since $\Phi_L(t,t_0)$ is invertible for all $t \in [t_0,T]$, see appendix \ref{appendix:TOop}.

\end{proof}

\section{Proof of lemma \ref{lemma:lwb2}}\label{appendix:lwb2}
\begin{lemma}
For all $\alpha=1,\cdots,kP$ and $n \geq 1$ the quantity $C_{\alpha\alpha}^n(s,s)C_{\alpha\alpha}^n(t,t)-C_{\alpha\alpha}^n(t,s)^2$ is lowerbounded by the positive symmetric function:
\[
 \theta(s,t)\overset{\rm def}{=} |t-s| \lambda_{\rm min}^2 \lambda_{\rm min}^{\Sigma^{Z_0}}  \lambda_{\rm min}^{\Gamma},
\]
where $\lambda_{\rm min}^{\Gamma}$ is the strictly positive lower bound, introduced in \ref{assumption:all}, on the singular values of the matrix $\mla(u)$ for $u \in [t_0,T]$.
\end{lemma}
\begin{proof}
We use equation \eqref{eq:cov} which we rewrite as follows, using the group property of the resolvent $\Phi_L$~:
\begin{multline*}
 C^{n+1}(t,s)=\Phi_L(t,t_0) \Bigg( \Sigma^{Z_0}  + \int_{t_0}^{t \wedge s} \Phi_L(t_0,u) \mla(u) \mla(u)^T \Phi_L(t_0,u)^T\,du+\\
\int_{t_0}^t \int_{t_0}^s \Phi_L(t_0,u) {\rm Cov}\left(\widetilde{\mU}^{X_n}_u,\widetilde{\mU}^{X_n}_v\right) \Phi_L(t_0,v)^T\,du\,dv \Bigg)\Phi_L(s,t_0)^T.
\end{multline*}
We now assume $s < t$ and introduce the following notations:
\[
\begin{array}{rcl}
 A(s)&=&\Sigma^{Z_0}+\int_{t_0}^{s} \Phi_L(t_0,u) \mla(u) \mla(u)^T \Phi_L(t_0,u)^T\,du\\
 B(s,t)&=&\int_{s}^{t} \Phi_L(t_0,u) \mla(u) \mla(u)^T \Phi_L(t_0,u)^T\,du\\
a(t,s)&=&\int_{t_0}^t \int_{t_0}^s \Phi_L(t_0,u) {\rm Cov}\left(\widetilde{\mU}^{X_n}_u,\widetilde{\mU}^{X_n}_v\right) \Phi_L(t_0,v)^T\,du\,dv
\end{array}
\]
Let $e_\alpha$, $\alpha=1,\cdots,kP$, be the unit vector of the canonical basis whose coordinates are all equal to 0 except the $\alpha$th one which is equal to 1. We note $E_\alpha (t)$ the vector $\Phi_L(t,t_0)^T e_\alpha$. We have, dropping the index $n$ for simplicity:
\[
\begin{array}{lcl}
C_{\alpha \alpha}(t,s)&=&E_\alpha (t)^T \left(A(s)+a(t,s)\right) E_\alpha(s)\\
C_{\alpha \alpha}(s,s)&=&E_\alpha (s)^T (A(s)+a(s,s)) E_\alpha(s)\\
C_{\alpha \alpha}(t,t)&=&E_\alpha (t)^T (A(s)+B(s,t)+a(t,t)) E_\alpha(t).
\end{array}
\]
Note that the last expression does not depend on $s$, since $A(s)+B(s,t)=A(t)$, which is consistent with the first equality. The reason why we introduce $s$ in this expression is to simplify  the following calculations.

The expression $C_{\alpha \alpha}(s,s)C_{\alpha \alpha}(t,t)-C_{\alpha \alpha}(t,s)^2$ is the sum of four sub-expressions:

\[
 \mathcal{E}_1(s,t)=\left(E_\alpha (s)^T A(s)E_\alpha(s)\right)\left(E_\alpha (t)^T A(s)E_\alpha(t)\right)-\left(E_\alpha (t)^T A(s) E_\alpha(s)\right)^2,
\]
which is greater than or equal to 0 because $A(s)$ is a covariance matrix,
\[
 \mathcal{E}_2(s,t)=\left(E_\alpha (s)^T a(s,s)E_\alpha(s)\right)\left(E_\alpha (t)^T a(t,t)E_\alpha(t)\right)-\left(E_\alpha (t)^T a(t,s) E_\alpha(s)\right)^2,
\]
which is also greater than or equal to 0 because $a(t,s)$ is a covariance matrix function,
\begin{multline*}
 \mathcal{E}_3(s,t)=\left(E_\alpha (s)^T A(s)E_\alpha(s)\right)\left(E_\alpha (t)^T a(t,t)E_\alpha(t)\right)+\\
\left(E_\alpha (t)^T A(s)E_\alpha(t)\right)\left(E_\alpha (s)^T a(s,s)E_\alpha(s)\right)-\\
2\left(E_\alpha (t)^T A(s) E_\alpha(s)\right)\left(E_\alpha (t)^T a(t,s) E_\alpha(s)\right)
\end{multline*}
Because $a(t,s)$ is a covariance matrix function we have
\[
 E_\alpha (t)^T a(t,t)E_\alpha(t)+E_\alpha (s)^T a(s,s)E_\alpha(s)-2E_\alpha (t)^T a(t,s) E_\alpha(s) \geq 0,
\]
and , as seen above, $\mathcal{E}_2(s,t) \geq 0$. Because $\mathcal{E}_1(s,t) \geq 0$ we also have
\begin{multline*}
-\sqrt{E_\alpha (s)^T A(s)E_\alpha(s)}\sqrt{E_\alpha (t)^T A(s)E_\alpha(t)} \leq E_\alpha (t)^T A(s) E_\alpha(s) \leq \\
\sqrt{E_\alpha (s)^T A(s)E_\alpha(s)}\sqrt{E_\alpha (t)^T A(s)E_\alpha(t)},
\end{multline*}
and, as it can be readily verified, this implies $\mathcal{E}_3(s,t) \geq 0$.

Therefore we can lowerbound $C_{\alpha \alpha}(s,s)C_{\alpha \alpha}(t,t)-C_{\alpha \alpha}(t,s)^2$ by the fourth subexpression:
\begin{multline*}
 C_{\alpha \alpha}(s,s)C_{\alpha \alpha}(t,t)-C_{\alpha \alpha}(t,s)^2 \geq \left(E_\alpha (s)^T A(s)E_\alpha(s)\right) \left(E_\alpha (t)^T B(s,t)E_\alpha(t)\right)+\\
\left(E_\alpha (s)^T a(s,s)E_\alpha(s)\right)\left(E_\alpha (t)^T B(s,t)E_\alpha(t)\right) \geq \\
\left(E_\alpha (s)^T A(s)E_\alpha(s)\right) \left(E_\alpha (t)^T B(s,t)E_\alpha(t)\right),
\end{multline*}
since $B(s,t)$ and $a(s,s)$ are covariance matrixes. We next have
\[
 E_\alpha (s)^T A(s)E_\alpha(s)=\frac{E_\alpha (s)^T A(s)E_\alpha(s)}{E_\alpha (s)^T E_\alpha(s)}\, \frac{e_\alpha^T \Phi_L(s,t_0)\Phi_L(s,t_0)^T e_\alpha}{e_\alpha^T e_\alpha},
\]
by definition of $E_\alpha(s)$. Therefore
\[
 E_\alpha (s)^T A(s)E_\alpha(s) \geq \lambda^{A(s)}_{\rm min} \lambda^{\Phi_L(s,t_0)\Phi_L(s,t_0)^T}_{\rm min} \geq \lambda_{\rm min}^{\Sigma^{Z_0}} \lambda_{\rm min},
\]
where $\lambda_{\rm min}^C$ is the smallest eigenvalue of the symmetric positive matrix $C$. Similarly we have
\[
 E_\alpha (t)^T B(s,t)E_\alpha(t) \geq \lambda_{\rm min}^{B(s,t)} \lambda_{\rm min}.
\]
Let us write $\Gamma(u)=\mla(u) \mla(u)^T$. We have (assumptions \ref{assumption:all}):
\begin{multline*}
 \lambda_{\rm min}^{B(s,t)} =\min_{\|x\| \leq 1} \int_s^t \frac{x^T \Phi_L(t_0,u) \Gamma(u) \Phi_L(t_0,u)^T x}{x^T x}\,du=\\
\min_{\|x\| \leq 1} \int_s^t \frac{x^T \Phi_L(t_0,u) \Gamma(u) \Phi_L(t_0,u)^T x}{x^T\Phi_L(t_0,u)\Phi_L(t_0,u)^T x}\,\frac{x^T\Phi_L(t_0,u)\Phi_L(t_0,u)^T x}{x^T x}\,du \geq \\
\int_s^t \min_{\|x\| \leq 1} \left(\frac{x^T \Phi_L(t_0,u) \Gamma(u) \Phi_L(t_0,u)^T x}{x^T\Phi_L(t_0,u)\Phi_L(t_0,u)^T x}\,\frac{x^T\Phi_L(t_0,u)\Phi_L(t_0,u)^T x}{x^T x} \right)\,du \geq \\
(t-s) \lambda_{\rm min}  \lambda_{\rm min}^{\Gamma}. 
\end{multline*}
Combining these results we have
\[
 C_{\alpha \alpha}(s,s)C_{\alpha \alpha}(t,t)-C_{\alpha \alpha}(t,s)^2 \geq |t-s| \lambda_{\rm min}^2 \lambda_{\rm min}^{\Sigma^{Z_0}}  \lambda_{\rm min}^{\Gamma}
\]
\end{proof}

\section{Proof of lemma \ref{lemma:integral}}\label{appendix:integral}
\begin{lemma}
The $2n$-dimensional integral 
 \begin{multline*}
I_n=\int_{[t_0,t \vee s]^2} \rho_1(u_1,v_1)\Bigg(\int_{[t_0,u_1 \vee v_1]^2} \cdots \Bigg(\int_{[t_0,u_{n-2} \vee v_{n-2}]^2}\rho_{n-1}(u_{n-1},v_{n-1})\\
\Bigg(\int_{[t_0,u_{n-1} \vee v_{n-1}]^2} \rho_n(u_n,v_n) du_n dv_n\Bigg)du_{n-1} dv_{n-1}\Bigg)\cdots \Bigg) du_1 dv_1,
\end{multline*}
where the functions $\rho_i(u_i,v_i)$, $i=1,\cdots,n$ are either equal to 1 or to $1/\sqrt{\theta(u_i,v_i)}$ (the function $\theta$ is defined in lemma \ref{lemma:lwb2}), is upperbounded by $k^n/(n-1)!$ for some positive constant $k$.
\end{lemma}
\begin{proof}
 First note that the integral is well-defined because of lemma \ref{lemma:lwb2}. Second, note that there exists a constant $K$ such that $K/\sqrt{\theta(u,v)} \geq 1$ for all $(u,v) \in [t_0,t \vee s]^2$, i.e. $K=\lambda_{\rm min} \sqrt{\lambda_{\rm min}^{\Sigma^{Z_0}} \lambda_{\rm min}^\Gamma (T -t_0)}$. Therefore the integral is upperbounded by $K_0^n$, where $K_0=\max(1,K)$ times the integral obtained when
$\rho_i(u_i,v_i)=1/\sqrt{|u_i-v_i|}$ for all $i=1,\cdots,n$. Let us then consider this situation. Without loss of generality we assume $t_0=0$. The
 cases $n=1,2,3$ allow one to understand the process.

\begin{equation}\label{eq:square}
 I_1 \leq  K_0 \int_{[0,t \vee s]^2} \frac{dudv}{\sqrt{|u-v|}}.
\end{equation}
Let us rotate the axes by $-\frac{\pi}{4}$ by performing the change of variables
\begin{align*}
 u &= \frac{U+V}{\sqrt{2}}, \\
 v &= \frac{V-U}{\sqrt{2}}.
\end{align*}

Using the symmetry of the integrand in $s$ and $t$ and the change of variable, the integral in the righthand side of \eqref{eq:square} is equal to (see figure \ref{fig:coordinates}):
 \begin{figure}[!htbp]
\centerline{\includegraphics[width=.7\textwidth]{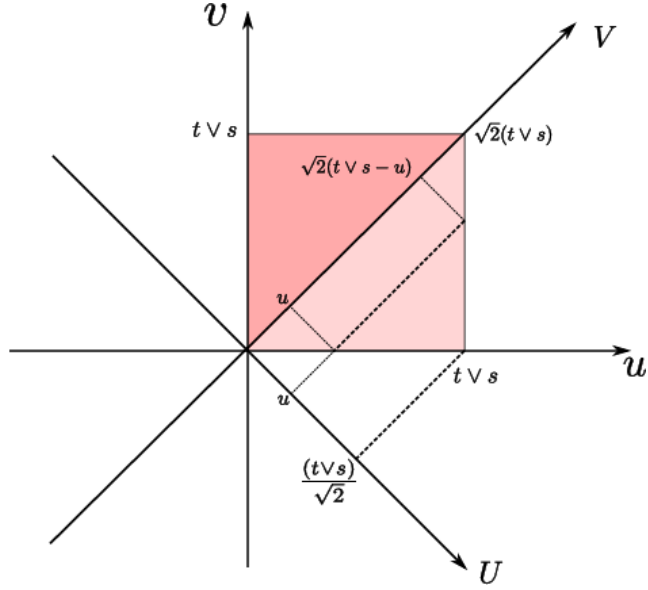}}
\caption{ The change of coordinates.}
\label{fig:coordinates}
\end{figure}

\[
 2 \frac{1}{2^{1/4}} \int_0^{\frac{t \vee s}{\sqrt{2}}} \int_U^{\sqrt{2}(t \vee s)-U} \frac{dVdU}{\sqrt{U}}= 2^{3/4} \int_0^{a/2} \frac{a-2U}{\sqrt{U}} dU=
2^{3/4} \alpha_1 a^{3/2},
\]
where $a=\sqrt{2}(t \vee s)$ and $\alpha_1=\frac{2\sqrt{2}}{3}$.

Let us now look at $I_2$. It is upperbounded by the factor $K_0^2(2^{3/4})^2 \alpha_1$ times the integral
\[
 \int_0^{a/2} \int_U^{a-U} \frac{(\sqrt{2} (u \vee v))^{3/2}}{\sqrt{U}} dU dV.
\]
Since in the area of integration $u \vee v=v=\frac{V-U}{\sqrt{2}}$ we are led to the product of $2/5$ by the one-dimensional integral
\[
\int_0^{a/2} \frac{(a-2U)^{5/2}}{\sqrt{U}} dUdV=\alpha_2 a^{3},
\]
where $\alpha_2=\frac{5 \sqrt{2} \pi}{32}$.

Similarly $I_3$ is upperbounded by the product of $K_0^3 (2^{3/4})^3 \alpha_1 \alpha_2 \frac{2}{5} \frac{2}{8}$ times the integral
\[
 \int_0^{a/2} \frac{(a-2U)^{4}}{\sqrt{U}} dUdV=\alpha_3 a^{9/2},
\]
where $\alpha_3=\frac{128 \sqrt{2}}{315}$. One easily shows then that:
\[
 I_n \leq K_0^nF (2^{3/4})^n 2^n \left(\prod_{i=1}^n \alpha_i\right) \left( \frac{1}{\prod_{j=1}^n (2+3(j-1))}\right).
\]
It can be verified by using a system for symbolic computation that $0< \alpha_i < 1$ for all $i \geq 1$. One also notices that
\[
 \prod_{j=1}^n (2+3(j-1)) \geq \frac{3^{n-1}}{2} (n-1)!,
\]
therefore
\[
 I_n \leq K_0^n (2^{3/4})^n 2^{n-1} 3^{-(n-1)} \frac{1}{(n-1)!},
\]
and this finishes the proof.

\end{proof}

\bibliographystyle{plain}

\end{document}